\documentclass[aps,prd,10pt,nofootinbib,twocolumn,eqsecnum,showpacs,showkeys,superscriptaddress,preprintnumbers,altaffilletter]{revtex4-1}
\usepackage{graphicx}
\usepackage{amsmath}
\usepackage{multirow}
\usepackage{hyperref}
\usepackage{soul}
\usepackage{graphicx}
\usepackage{dcolumn}
\usepackage{amssymb}
\usepackage{amsfonts}
\usepackage{amsbsy}
\usepackage{xcolor}
\usepackage{rotating}
\usepackage[english]{babel}
\usepackage{multirow}
\usepackage{float}
\usepackage{caption}
\usepackage{tabularx}

\usepackage{booktabs,tabularx,threeparttable}
\usepackage{bm}

\usepackage{caption}
\makeatletter
\long\def\@makecaption#1#2{%
  \par\vskip\abovecaptionskip
  \begingroup
    \normalsize\rmfamily
    \samepage
    \flushing
    \let\footnote\@footnotemark@gobble
    \@make@capt@title{#1}{#2}\par
  \endgroup
  \vskip\belowcaptionskip
}
\makeatother

\usepackage{animate}

\newcommand{\be}{\begin{equation}}
\newcommand{\ee}{\end{equation}}
\newcommand{\bea}{\begin{eqnarray}}
\newcommand{\eea}{\end{eqnarray}}
\newcommand{\beal}{\begin{aligned}}
\newcommand{\eeal}{\end{aligned}}
\newcommand{\bi}{\begin{itemize}}
\newcommand{\ei}{\end{itemize}}

\hypersetup{
    colorlinks=true,
    linkcolor=blue,
    filecolor=magenta,
    urlcolor=cyan,
    citecolor=cyan
}

\begin{document}

\title{Updates on dipolar anisotropy in local measurements of the Hubble constant from Cosmicflows-4}

\author{V. Salzano}
\email{vincenzo.salzano@usz.edu.pl}
\affiliation{Institute of Physics, University of Szczecin, Wielkopolska 15, 70-451 Szczecin, Poland}
\author{J. Beltr\'{a}n Jim\'{e}nez}
\email{jose.beltran@usal.es}
\affiliation{Departamento de F\'{ı}sica Fundamental, Universidad de Salamanca, E-37008
Salamanca, Spain}
\affiliation{Instituto Universitario de Física Fundamental y Matemáticas (IUFFyM), Universidad de Salamanca, E-37008
Salamanca, Spain}
\affiliation{Institute of Theoretical Astrophysics, University of Oslo, N-0315 Oslo, Norway.}
\author{D. Bettoni}
\email{dbet@unileon.es}
\affiliation{Instituto Universitario de Física Fundamental y Matemáticas (IUFFyM), Universidad de Salamanca, E-37008
Salamanca, Spain}
\affiliation{Departamento de Matemáticas, Universidad de León,
Escuela de Ingenierías Industrial, Informática y Aeroespacial
Campus de Vegazana, s/n
24071 León, Spain}
\author{P. Brax}
\email{philippe.brax@ipht.fr}
\affiliation{Institut de Physique Th\'{e}orique, Universit\'{e} Paris-Saclay, CEA, CNRS, F-91191 Gif-sur-Yvette Cedex, France}
\author{A. Valade}
\email{avalade@aip.de}
\affiliation{Aix Marseille Universit\'{e}, CNRS/IN2P3, CPPM, Marseille, France}

\date{\today}

\begin{abstract}

We investigate the angular anisotropy of the Hubble constant using the Cosmicflows-4 (CF4) catalogue, with particular emphasis on three issues often treated only implicitly in the literature: the statistical formulation of the Hubble--Lema\^{i}tre relation, the internal consistency of the working sample, and the role of peculiar-velocity corrections. Rather than working in luminosity-distance space, we adopt a logarithmic formulation based directly on distance moduli, thereby preserving the Gaussian error properties of the measured quantities. We first subject the catalogue to internal consistency tests, including the depth dependence of $\langle \log H_0 \rangle$ and the behaviour of residual skewness and kurtosis across radial shells, and use these diagnostics to define conservative subsamples minimally affected by selection effects, namely $\mu \in [31,36]$ and $z \in [0.03,0.06]$. Within these ranges, we reconstruct angular maps of $\log H_0$ and fit them with a spherical-harmonic expansion up to octupole order. We find a statistically significant anisotropic signal in the uncorrected CF4 data, dominated by a dipole and favoured over a monopole-only model by Bayesian evidence. However, when peculiar-velocity-corrected data are used, the anisotropy amplitude is strongly reduced, especially at lower depths, while only a weaker residual signal survives at larger distances. We also test for a monotonic radial evolution of the dipole, as expected in some differential-expansion scenarios, but find no robust evidence for such a trend. These results indicate that the anisotropy seen in CF4 is driven primarily by local velocity flows and catalogue/survey structure, rather than by a large-scale breakdown of isotropic expansion. Finally, we show that although such anisotropy may affect local determinations of $H_0$, its impact on the global Hubble tension is likely limited, since the sky distribution of current SNe~Ia calibrators and Hubble-flow hosts does not show a strong alignment with the detected dipole pattern.

\end{abstract}

\maketitle

\section{Introduction}\label{sec: Intro}

The current era of cosmological tensions is exciting, as it may signal new physics and a paradigm shift in our understanding of the universe \cite{DiValentino:2025sru}. However, ``precision'' cosmology remains in its infancy \cite{Slosar:2019flp}. Although a definitive answer about their nature seems to be far from being reached, fundamental concepts such as dark energy and dark matter are widely accepted, possibly suggesting conventionalist biases in the scientific community \cite{Merritt:2017xeh}. A fruitful and complementary approach might be to challenge mainstream cosmological assumptions, which should be approached with an open and critical mind, as progress may require fundamental revisions.

By ``cosmic tensions” we are primarily referring here to the Hubble tension, the discrepancy between the current expansion rate of the universe, $H_0$, as inferred from the Cosmic Microwave Background radiation (CMB) data by \textit{Planck}, South Pole Telescope (SPT) and Atacama Cosmology
Telescope (ACT) \cite{SPT-3G:2025bzu}, $H_0 = 67.24 \pm 0.35$ km s$^{-1}$ Mpc$^{-1}$, and from the $H0DN$ collaboration and their Local Distance Network, $H_0 = 73.50 \pm 0.81$ km s$^{-1}$ Mpc$^{-1}$ \cite{H0DN:2025lyy}, a $7.1 \sigma$ statistical tension. There is no consensus on its origin, whether theoretical or observational. As Figures 1–3 in \cite{DiValentino:2025sru} illustrate, modifying datasets or extending theoretical models does not show a consistent trend towards solving the issue.

The main criticism of the CMB-based value of $H_0$ is its model dependence, as it assumes a $\Lambda$CDM framework. A similar limitation applies to the recent Dark Energy Spectroscopic Instrument (DESI) results \cite{DESI:2025zgx}, which yield $H_0 = 68.17 \pm 0.28$ km s$^{-1}$ Mpc$^{-1}$ when combined with CMB data, and remain consistent with \textit{Planck} \cite{Planck:2018nkj} even without this assumption. Numerous theoretical models have sought to solve the Hubble tension through model-dependent solutions, but none have been conclusive \cite{Vagnozzi:2023nrq,Poulin:2024ken}. Many revisit old ideas— as cosmic acceleration was itself reinterpreted \cite{Riess:1998cb,Perlmutter:1998np} in terms of the cosmological constant— but no fundamentally new framework has emerged. Observationally, data independent of the CMB rarely match CMB-$H_0$ precision; and all leading alternatives such as the ``Supernovae and $H_0$ for the Dark Energy Equation of State'' (\textit{SH0ES}) project \cite{Riess:2021jrx,Breuval:2024lsv} probe only the local universe. This raises the possibility that our understanding of the local cosmos remains incomplete within the standard cosmological model, a perspective explored in this paper.

One way of challenging established paradigms without introducing new physics is to test the compatibility of the Copernican Principle with observations underlying $H_0$ measurements \cite{Aluri:2022hzs}. Differences between local and global $H_0$ values are expected from local perturbations, as shown in both early \cite{1992AJ....103.1427T,Shi:1997aa} and recent simulations \cite{Odderskov:2017ivg,Hollinger:2025bal}. This variance decreases with distance as peculiar velocities fade, leading to a cosmic rest frame where the nonlinear Hubble flow becomes uniform \cite{McKay:2015nea}. Large-scale bulk flows can produce a dipole, while peculiar velocities add monopole and quadrupole components \cite{Sorrenti:2024ztg}. Therefore, while anisotropy may be detectable, it must be interpreted cautiously, as it can arise from selection effects rather than genuine cosmological features. The literature remains inconclusive, with results depending on both the probes and methodologies used, even when based on the same data. Fig.~1 and Table~1 of \cite{Aluri:2022hzs} summarize several reported anisotropy directions, though comparisons are hindered by large associated uncertainties.

The most prominent anisotropy signal is the CMB dipole, measured by \textit{Planck} \cite{Planck:2018nkj} at galactic coordinates\footnote{Anisotropy directions are typically reported as positive (excess) signals, a convention followed here unless stated otherwise. All directions are given in galactic coordinates.} $l_g = 264.021 \pm 0.011$, $b_g = 48.253 \pm 0.005$, with an amplitude of $v^{dip}_{CMB} = 369.82 \pm 0.11$ km s$^{-1}$, interpreted as our motion relative to the CMB. A recent analysis of \textit{Planck} data \cite{Gimeno-Amo:2025icf} found additional hints of anisotropy, with a dipole in $H_0$ pointing toward $(56^{\circ}, -18^{\circ})$ from temperature-only data and $(81^{\circ}, -20^{\circ})$ when polarization is included. While the amplitude agrees with simulations, directional uncertainties are likely large, as the fit was performed on a low-resolution map\footnote{Private communication from the authors.}. It is important to remember that in analyses using CMB-corrected velocities, any detected anisotropy should not be associated with the CMB dipole, whose effect has already been removed.

Type Ia supernovae (SNeIa) are powerful probes of cosmic anisotropy due to their precise distance estimates across local and cosmological scales. However, their uneven sky distribution can introduce hidden biases \cite{BeltranJimenez:2014otq}. The Pantheon sample \cite{Pan-STARRS1:2017jku}, truncated at $z<0.05$, was analysed  together with the Cosmicflows-3 catalog \cite{Tully:2016ppz} in \cite{Kalbouneh:2022tfw}. These studies found dipoles toward $(285^{\circ} \pm 5^{\circ}, 11^{\circ} \pm 4^{\circ})$ from galaxies and $(334^{\circ} \pm 42^{\circ}, 6^{\circ} \pm 20^{\circ})$ from SNeIa, consistent with the Local Group’s motion relative to the CMB, directed toward $(279^{\circ}, 29^{\circ})$, and nearby mass concentrations such as Hydra–Centaurus and the Shapley Supercluster, located respectively at $(302^{\circ}, 21^{\circ})$ and $(311^{\circ}, 32^{\circ})$. A significant quadrupole component aligned with the dipole was also reported, yielding a maximum $H_0$ variation of $\Delta H_0 = 2.4 \pm 1.1$ km s$^{-1}$ Mpc$^{-1}$ across the sky. The Pantheon+ sample \cite{Brout:2022vxf} has been analyzed by the same authors together with the Cosmicflows-4 catalog \cite{Tully:2022rbj} in \cite{Kalbouneh:2025jnp}, substantially confirming the previous results.

The Pantheon+ sample has also been studied using the hemisphere comparison method \cite{Schwarz:2007wf,Antoniou:2010gw} in a $\Lambda$CDM framework. A maximal variation of $\Delta H_0 \sim 4$ km s$^{-1}$ Mpc$^{-1}$ was found toward $(290^{\circ}, 6^{\circ})$ \cite{McConville:2023xav}. The same data were analyzed in \cite{Perivolaropoulos:2023tdt}, focusing on angular variations in the SNeIa absolute magnitude correlated with $H_0$, yielding an anisotropy at $(136.39^{\circ}, 16.12^{\circ})$. A stronger signal was found within $30$ Mpc, interpreted as evidence for an off-center observer inside a $\sim 20$–$30$ Mpc local bubble. The Pantheon+ sample has also been examined in a series of studies \cite{Sorrenti:2022zat,Sorrenti:2024ztg,Sorrenti:2024ugq}, where the luminosity distance in a flat $\Lambda$CDM model was modified using linear perturbation theory, introducing natural monopole, dipole, and quadrupole terms that allow direct estimation of the observer’s velocity in the CMB frame. For the full sample, the dipole amplitude matches the CMB’s but its direction, $(179.53^{\circ}, 44.34^{\circ})$, deviates by over $3\sigma$ from the \textit{Planck} dipole, implying a bulk flow affecting SNeIa up to $z \simeq 0.0375$. Follow-up analyses confirmed this, identifying a $317$ km s$^{-1}$ bulk flow toward $(310^{\circ}, 9^{\circ})$. A complementary cosmographic approach, expanding the modified luminosity distance, revealed local infall at $z \in [0.04, 0.06]$, corresponding to a $\sim2\%$ overdensity—consistent with $\Lambda$CDM expectations and contrasting with earlier void-based explanations of the Hubble tension \cite{Shanks:2018rka,Riess:2018kzi,Kenworthy:2019qwq,Huterer:2023ldv,Mazurenko:2023sex}.

Another approach to detecting anisotropy uses number-count dipole signals. A purely kinematic CMB dipole should induce a corresponding dipole in radio source counts, aligned in direction and amplitude \cite{1984MNRAS.206..377E}. Extending this to quasars, \cite{Secrest:2020has} finds a dipole directed toward $(238.2^{\circ}, 28.8^{\circ})$, and \cite{Dam:2022wwh} at $(237.2^{\circ +7.9^{\circ}}_{\phantom{\circ}-8.0^{\circ}}, 41.8^{\circ} \pm 5.0^{\circ})$, consistent in direction with the CMB dipole, but with amplitudes up to three times larger, producing a $5.7\sigma$ tension. A more recent analysis accounting for selection effects \cite{Mittal:2023xub} finds full agreement with the CMB dipole in both direction and amplitude. As quasars lie at high redshift (median $z = 1.48$, extending beyond $z \sim 4$), these tests probe (an)isotropy on large scales, so deviations from local measurements are not necessarily problematic.

A recent approach uses galaxy cluster scaling relations to compare cosmology-dependent and independent properties, inferring spatial variations in $H_0$ \cite{Migkas:2020fza,Migkas:2021zdo,Pandya:2024jqg}. In \cite{Migkas:2021zdo} the authors found a variation of $\Delta H_0 = 9.0 \pm 1.7\%$, with a minimum toward $(273^{\circ}{}^{+42^{\circ}}_{-38^{\circ}}, -11^{\circ}{}^{+27^{\circ}}_{-27^{\circ}})$. Updated results \cite{Pandya:2024jqg}, after careful bias analysis, reported an even larger $\Delta H_0 = 27.6 \pm 4.4\%$ toward $(295^{\circ} \pm 71^{\circ}, -30^{\circ} \pm 71^{\circ})$. Interpreting these as bulk flows would require velocities of $\sim 900$ km s$^{-1}$ out to $\sim 500$ Mpc, far beyond $\Lambda$CDM expectations \cite{Migkas:2021zdo}. Recent hydrodynamical simulations \cite{He:2025wly} confirm such signals are rare, though statistical noise and other effects can reduce their apparent significance.

In this study, we explore potential anisotropies in the Hubble constant using the Cosmicflows-4 catalog, focusing on the impact of peculiar velocities and data distribution while minimizing model-dependent assumptions. This work is also motivated by models where dark matter has unconventional properties, such as being charged under dark nonlinear electromagnetism \cite{BeltranJimenez:2020csl,BeltranJimenez:2020tsl,BeltranJimenez:2021imo}. 
In this scenario, the Universe is modelled by a Lema\^{i}tre-Tolman-Bondi (LTB) solution~\cite{Lemaitre:1933gd,Tolman:1939jz,Bondi:1947fta} with non-zero pressure. It is foliated into spherical shells centred around a preferred point, with each shell evolving under the combined influence of gravity and dark nonlinear electrodynamics. Due to non-linear effects, a characteristic scale, the screening radius, emerges: shells outside this radius do not experience the dark repulsive force, while those inside it are accelerated, leading to faster expansion. This mechanism offers a promising way to address the Hubble tension, as an observer located within their shell's screening radius would measure a higher local Hubble rate than for distant shells outside their respective screening regions. If the observer is at a distance $R_O$ from the centre of the configuration, the observed Hubble rate becomes anisotropic and depends on the distance to each emitter as \cite{BeltranJimenez:2021imo}:
\begin{equation}
     H_{\rm obs}=H_{\rm e}(t,r_{\rm e}) \left(1+ \cos \theta \frac{R_{\rm O}}{2d}\sqrt{1-\frac{R_{\rm O}^2}{d^2}\sin^2 \theta}\right),
\end{equation}
where the suffix ``e'' refers to the emitter, $d$ is the distance from the observer to the emitter, and $\theta$ is the polar angle between the emitter and the observer, measured along the line connecting the observer to the centre of the Universe. At very large distances, this expression simplifies to:
\begin{equation}\label{eq:h_angle}
H_{\rm obs}\simeq  H_{\rm e}(t,r_{\rm e})  \left(1+ \cos \theta \frac{R_{\rm O}}{2d}\right).
\end{equation}
At leading order, the observed Hubble rate exhibits a dipole that decreases with distance, with higher-order multipoles aligned with this dipole. Two key features emerge: first, the Hubble rate is anisotropic; second, it asymptotically approaches the standard Hubble flow at large distances. Although this serves as a motivation, we would like to emphasise that the aim of this work is not to test the models in \cite{BeltranJimenez:2020csl,BeltranJimenez:2020tsl,BeltranJimenez:2021imo} specifically, but rather to verify whether the Hubble rate is anisotropic and whether there is any possible distance dependence of the dipole. Only once this has been confirmed (or refuted) a discussion about the theoretical explanation can be opened. This is left for future work. 

This paper is structured as follows: in Sec.~\ref{sec:data} we introduce the dataset used for our analysis; in Sec.~\ref{sec:HL} we provide a brief review of the Hubble-Lema\^{i}tre law and possible caveats in its application; Sec.~\ref{sec:preliminary} describes a preliminary assessment check to validate the use of the sample; Sec.~\ref{sec:statistical} enlists the key aspects of our analysis; in Sec.~\ref{sec:results} we present the main results; and in Sec.~\ref{sec:discussion} we compare our findings with existing literature and draw conclusions.

\section{Data: Cosmicflows-4}
\label{sec:data}

We use data from the Cosmicflows-4 (CF4) project, which provides many observational quantities in various complementary forms. The reference catalogue is described in \cite{Tully:2022rbj}; in the following, we will refer to it as CF4.

The CF4 catalog compiles distances for $55877$ galaxies in $38065$ groups. Multiple galaxies per group and diverse methodologies reduce distance uncertainties (specifically in distance moduli). Primary methods include the Tully–Fisher (TF) relation for spiral and the Fundamental Plane (FP) for elliptical galaxies, with smaller contributions from surface brightness fluctuations (SBF) and Type II supernovae (SNeII). These methods are complemented by SNeIa, Cepheid measurements, the Tip of the Red Giant Branch (TRGB) data and maser parallaxes. 

An important point concerns the value of the Hubble constant $H_0$. Using the standard CF4 pipeline, which combines SNeIa, FP, TF, SBF, and SNeII measurements and calibrates them with Cepheids and TRGB, the team finds $H_0 = 74.6 \pm 0.8 \pm 3.0$ km s$^{-1}$ Mpc$^{-1}$. Direct calibration of SNeIa with TRGB, Cepheids, and masers yields $H_0 = 72.1$ km s$^{-1}$ Mpc$^{-1}$, which matches \textit{SH0ES} results \cite{Riess:2021jrx,Breuval:2024lsv}. This is notable for two reasons: first, the \textit{Chicago Carnegie Hubble Program} (CCHP) finds a lower $H_0$ than \textit{SH0ES} \cite{Freedman:2024eph} using Cepheids and TRGB, though selection biases may play a role \cite{Riess:2024vfa,Li:2025lfp}; second, as the Hubble tension could be seen as a calibration issue \cite{Perivolaropoulos:2024yxv, Poulin:2024ken}, this shows that changing the calibration can shift $H_0$ by at least 2.5 km s$^{-1}$ Mpc$^{-1}$. For this reason, we focus on trends and relative variations rather than the absolute value of $H_0$.

The CF4 catalog provides $22$ quantities per galaxy group, but for our analysis we use only a subset: the distance moduli $\mu$ with their errors $\sigma_{\mu}$; positions in galactic coordinates; and observed systemic group radial velocities in the CMB frame \cite{Tully:2007ue}. These velocities are observationally based, derived from spectroscopic redshifts obtained in large surveys (e.g. SDSS, 6dFGSv), and combined at the group level to estimate systemic velocities \cite{Tully:2022rbj}. Importantly, distance moduli and velocities in CF4 are measured independently, as the $\mu$ are obtained from redshift-independent methods and calibrated using geometric anchors, without assuming $H_0$. Finally, we acknowledge that distance moduli are inferred quantities and that the CF4 catalogue does not provide a full covariance matrix for them reflecting the heterogeneous nature of the underlying distance indicators and their shared calibration anchors. In the absence of this information, we treat the reported distance moduli as conditionally independent measurements with Gaussian uncertainties, and restrict our analysis to angular modulations and relative variations rather than absolute scale determination. Any unmodelled covariance is therefore expected primarily to rescale the inferred parameter uncertainties (that could be underestimated) rather than to generate or erase coherent large-scale angular patterns: correlated calibration uncertainties would most naturally project onto monopole-like modes, while higher-order angular modulations would require a specific alignment between calibration systematics and sky position. Nevertheless, we interpret the statistical significance of the detected modulation conservatively (small-amplitude signals especially should be interpreted cautiously).

Other quantities—luminosity distances and their errors, individual $H_0$ estimates (from the ratio of the observed radial velocity to the luminosity distance), and radial peculiar velocities—are not used. Furthermore, errors in distance moduli are Gaussian, whereas luminosity distances are log-normally distributed, and this can bias peculiar velocity estimates if unaccounted for \cite{2024MNRAS.527.3788H}. 

Indeed, the peculiar velocities provided in the CF4 catalog are biased in this sense. The analysis of \cite{Valade:2024riw} built an alternative CF4 catalog based on the original compilation of galaxy distances and redshifts which are treated as noisy constraints on an underlying continuous velocity and density field. Peculiar velocities are inferred through a Bayesian forward-modelling approach that reconstructs the three-dimensional matter density and velocity fields consistent with the CF4 data. The authors assume a $\Lambda$CDM cosmological framework (of interest here, the Hubble constant is fixed at the value $H_0 = 74.6$ km s$^{-1}$ Mpc$^{-1}$ as derived by \cite{Tully:2022rbj}) and linear gravitational instability theory, linking the velocity field to the density field via the continuity and Poisson equations. A Hamiltonian Monte Carlo sampler is used to explore the posterior distribution of the initial density field, evolved to the present epoch, such that the predicted radial peculiar velocities reproduce the observed distance–redshift relation from the original CF4. In this framework, peculiar velocities are therefore not direct observables, but model-dependent quantities obtained by combining the measured distance moduli and redshifts with a dynamical prior on structure formation. The resulting peculiar-velocity field is a smoothed, probabilistic reconstruction that suppresses small-scale noise while retaining large-scale coherent flows. From this reconstructed velocity field, the authors derive secondary dynamical quantities, such as the gravitational potential and its local minima, which define ``basins of attraction''. These basins correspond to regions of convergent flow toward dominant attractors and provide a physically motivated partition of the local Universe. Importantly, the inferred peculiar velocities inherit both the assumptions of the $\Lambda$CDM model (we highlight that the role played by $\Lambda$CDM in the construction of the catalog makes it unfit to test alternative cosmological models, though this is not of matter for this work) and the statistical regularization imposed by the reconstruction, and are therefore best interpreted as model-dependent estimates of large-scale flows, rather than as raw velocity measurements tied one-to-one to individual galaxies. We refer to the CF4 dataset integrated with peculiar velocities from \cite{Valade:2024riw} as CF4$_{pec}$.

\section{The Hubble-Lema\^{i}tre law: an historical and technical excursus}
\label{sec:HL}

The purpose of this section is to illustrate some possible ambiguities in the way the Hubble-Lema\^{i}tre (HL) law can be used and \textit{is commonly used} in the literature.

The HL law is now almost a century old \cite{Hubble:1929ig}, and its crucial role in the building of modern cosmology is an undisputed fact. It states that the recessional (from us) radial velocity, $v_{rec}$, of the galaxies is proportional to their distance, $d$, from us, and that the proportionality factor is the Hubble constant, $H_0$, namely:
\begin{equation}\label{eq:Hubble_law_original}
v_{rec} = H_0\, d \, .
\end{equation}

\begin{figure}
\centering
\includegraphics[width=\columnwidth]{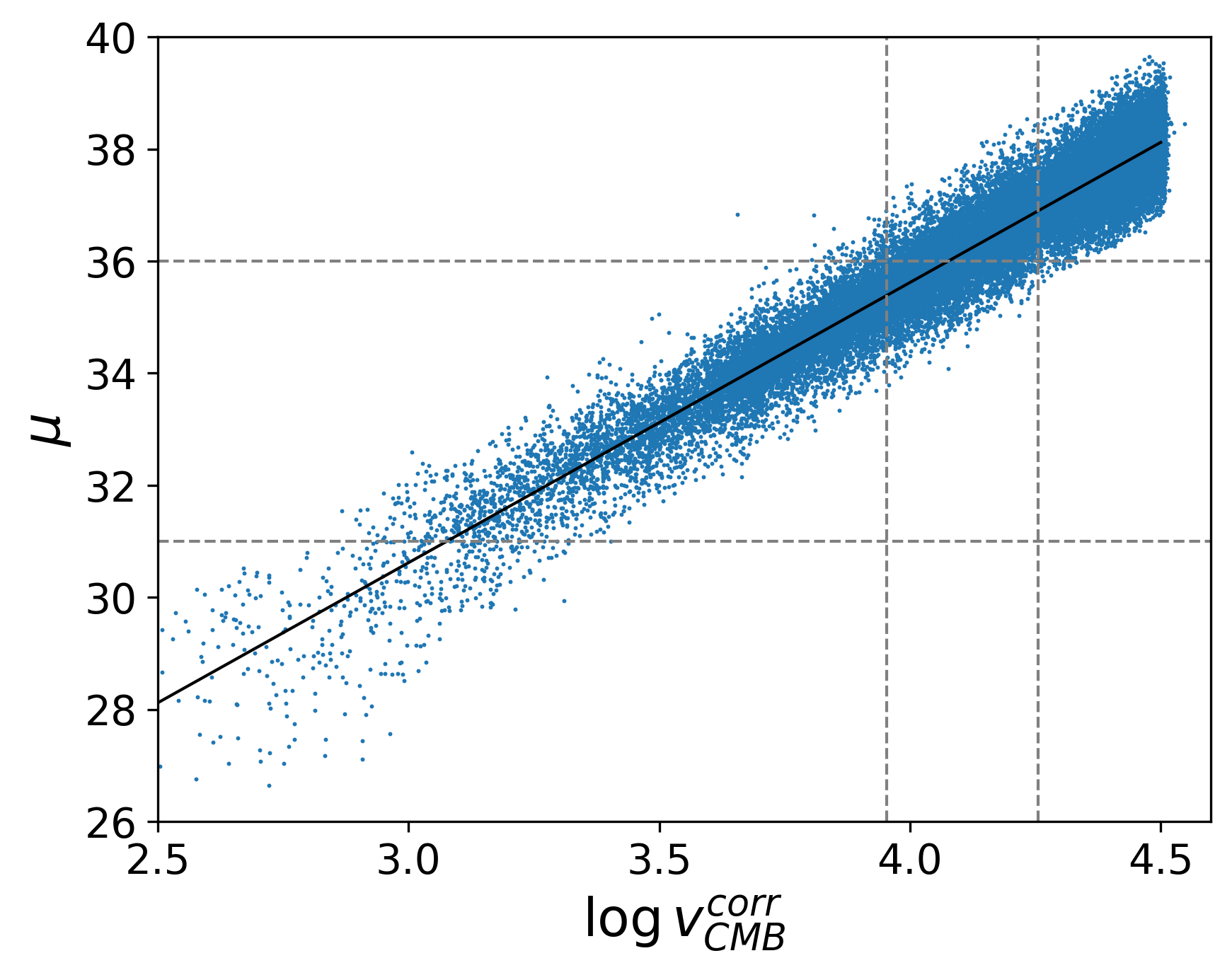}
\caption{Hubble--Lema\^{i}tre law from the Cosmicflows-4 data. Horizontal (vertical) dashed lines indicate the intervals of distance in $\mu$ ($z$) on which our nominal analysis is based.}\label{fig:HL_plot}
\end{figure}

When first formulated, this law provided clear evidence that the universe is expanding, despite limited observational quality and coverage \cite{1927ASSB...47...49L,Trimble:1996jqa}. Today, some ambiguity remains, as cosmology allows multiple distance definitions: proper, comoving $(d_C)$, luminosity $(d_L)$, and angular diameter $(d_A)$. For nearby objects, the choice is largely irrelevant since these distances coincide at lowest order, but differences become important at larger scales, even within the lower bound of the Hubble flow.

Ideally, the distance $d$ in the HL law, Eq.~(\ref{eq:Hubble_law_original}), should correspond to the proper distance to each galaxy, which can be considered equivalent to the present-time comoving distance $d_C$. However, at larger scales, it becomes model-dependent. Observationally, we measure the luminosity distance $d_L$ and the angular diameter distance $d_A$, though here we focus on $d_L$. Even this statement is not entirely accurate and can introduce biases. In fact, in most cases we directly measure the distance modulus $\mu$, which is related to the luminosity distance via the well-known relation:
\begin{equation}\label{eq:distance_modulus}
\mu = m - M = 25 + 5 \log d_L\, ,
\end{equation}
when the luminosity distance is expressed in Mpc, and where $m$ and $M$ are respectively the apparent and the absolute magnitude of the observed object.

In the CF4 catalogue the provided Hubble constant value for each object is calculated using:
\begin{equation}\label{eq:Hubble_Cosmicflows}
H_0 = \frac{\, cz f_{d_L}(z)}{d_L}\,, 
\end{equation}
where $z$ is the redshift of the object derived from the \textit{observed} radial velocity, and $f_{d_L}(z)$ is a function that will be defined shortly. The path to finding Eq.~(\ref{eq:Hubble_Cosmicflows}) starts from the luminosity distance $d_L$ which, within the redshift range covered by the data, $z \lesssim 0.1$, can be safely described by its cosmographic series \cite{Weinberg:1972kfs,Visser:2003vq,Dabrowski:2005fg,Cattoen:2007sk,Capozziello:2011tj}:
\begin{equation}\label{eq:Hubble_Cosmicflow_2}
d_L(z) = \frac{c}{H_0} D_L(z) 
       = \frac{c z}{H_0}  f_{d_L}(z,q_0,j_0)\, ,
\end{equation}
where $D_L$ is the dimensionless luminosity distance and $f_{d_{L}}(z,q_0,j_0)$ is the third-order truncated cosmographic series after factoring out the $c z$ term \cite{Capozziello:2011tj}:
\begin{equation}\label{eq:cosmography}
f_{d_L}(z,q_0,j_0) = 1 + \frac{1-q_0}{2} z - \frac{1-q_0-3 q^{2}_{0} + j_0}{6} z^2\, ,
\end{equation}
with $q_0$ the deceleration parameter and $j_0$ the jerk parameter. It is commonly stated that this approach avoids assuming a cosmological model, but actually there is an implicity assumption of a Friedmann–Lema\^{i}tre–Robertson–Walker metric. A detailed treatment of the operational definition of the generalized Hubble function can be found in \cite{Maartens:2023tib}. Although one can relate the cosmographic parameters to any cosmological model, the idea behind cosmography is that one should \textit{first} determine the cosmographic parameters, $q_0$ and $j_0$ in this case, and \textit{then} compare any given cosmological model with them. In \cite{Tully:2022rbj}, the authors assume a flat $\Lambda$CDM model with $\Omega_m = 0.27$, yielding $q_0 = -0.595$ and $j_0 = 1$ (neglecting radiation). 

Eventually, Eq.~(\ref{eq:Hubble_Cosmicflow_2}) can be reshaped as the HL law,
\begin{align}
d_{L}  &= \frac{v^{obs}_{CMB}}{H_0} f_{d_L}(z,q_0,j_0)\, ,
\end{align}
where we use the fact that $c z$ is the \textit{observed} radial recessional velocity in the CMB reference frame, leading to the final expression:
\begin{align}\label{eq:Hubble_lin}
d_{L} =  \frac{v_{corr}}{H_0}\, ,
\end{align}
where we use the nomenclature of \cite{Tully:2022rbj}. Actually, CF4 provides also $v_{corr}$ in the final data release, where it is called the ``cosmological curvature-adjusted'' velocity.

However, there are several issues with Eq.~(\ref{eq:Hubble_lin}). The least significant of these is its dependence on $q_0$ and $j_0$, which could be treated as free parameters. We have fitted the distance moduli provided by the CF4 catalogue to the corresponding cosmographic expansion, Eq.~(\ref{eq: f_mu}) that will be introduced and motivated below. Firstly, we lack the absolute calibration magnitude, meaning we cannot constrain them alongside $H_0$. Moreover, given the limited redshift range and the dispersion of the data, the jerk parameter is completely unconstrained. Consequently, when focusing only on the deceleration parameter, we obtain $q_0 = -1.22^{+0.06}_{-0.05}$ which, when assuming a $\Lambda$CDM model, would correspond to an unphysical negative value of $\Omega_m$. Excluding very low redshift objects, which do not provide information about cosmological dynamics and are likely dominated by local motion, and focusing on those with $v_{CMB}> 4000$ km s$^{-1}$, as suggested in \cite{Tully:2022rbj}, yields $q_0 = -0.52^{+0.07}_{-0.07}$, which is perfectly consistent with the CF4 default value. Notably, the \textit{SH0ES} team has historically fixed $f_{d_L}(z)$ with a specific $q_0$ (and $j_0 = 1$ for flat $\Lambda$CDM) \cite{Riess:2016jrr}, only recently verifying that allowing $q_0$ to vary has no impact on their $H_0$ determination \cite{Riess:2021jrx}. Thus, $q_0$ and $j_0$ are statistically negligible for the estimation of $H_0$ and the issue of letting them vary can be ignored here. We have therefore chosen to fix them at the values used by the CF4 team.

A related problem is that we assume the distance $d$ in the original HL relation corresponds to the luminosity distance $d_L$. Extending the HL relation beyond the linear regime requires $d \equiv d_L$, since using $d_C$ would yield a different $v_{corr}$ due to $f_{d_C} \neq f_{d_L}$ \cite{Capozziello:2011tj}. While the difference may be negligible within observational errors and in our redshift range, this would not ease the methodological and theoretical interpretation  of the HL relation.

Another, and arguably more important, issue concerns the use of the observed CMB-frame radial velocity. The HL law requires the recessional velocity; thus, the observed velocity should first be corrected for peculiar motions, i.e. $v_{rec} = v^{obs}_{CMB} - v_{pec}$. This correction naturally affects the inferred redshift as well. Most studies assume $v_{pec} \ll v^{obs}_{CMB}$ and do not explicitly test the validity of this approximation. In our analysis, however, this check is essential. We therefore consider both cases: one where $v_{corr} \equiv v^{obs}_{CMB}$, and another where $v_{corr} \equiv v^{obs}_{CMB} - v_{pec}$. Note that in this case, the cosmological redshift for each group, i.e. the CMB-frame redshift cleaned of the peculiar velocities (in the notation of Pantheon+ team \cite{Brout:2022vxf}, this is equivalent to their ``Hubble diagram'' redshift, $z_{HD}$) is defined as:
\begin{equation}
1 + z_{cosmo} = \frac{1+z^{obs}_{CMB}}{1+z_{pec}}\, , 
\end{equation}
where $z_{CMB} = v^{obs}_{CMB}/c$ and $z_{pec} = v_{pec}/c$.

A further, and more fundamental, issue does not appear to have  clear solution. Equation~(\ref{eq:cosmography}) is derived from a series expansion of the luminosity distance \textit{after assuming} a Friedmann--Lema\^{i}tre--Robertson--Walker metric, which rests on the assumptions of \textit{isotropy and homogeneity}. Strictly speaking, it should therefore not be applied simultaneously to local, anisotropic regions and to distant, statistically isotropic scales. Recent attempts to generalize cosmography to inhomogeneous settings \cite{Heinesen:2020bej,Dhawan:2022lze, Cowell:2022ehf,Koksbang:2024nih,Kalbouneh:2024szq,Macpherson:2025qec} reduce this reliance but at the cost of introducing numerous   parameters. The abundance of upcoming data (from DESI, the Zwicky Transient Facility and the Vera C. Rubin Observatory) will make this less of an issue and even a promising avenue for the field, as preliminary shown very recently in \cite{Kalbouneh:2025jnp}.

Finally, as noted earlier, for nearly all objects in the catalogue the distances are not measured  but their distance moduli. It is therefore helpful to rewrite the HL law in Eq.~(\ref{eq:Hubble_lin}) in a logarithmic version as well:
\begin{align}\label{eq:Hubble_log}
\frac{\mu}{5} - 5 &= \log v_{corr}  - \log H_0\, .
\end{align}
While many studies in the literature rely on Eq.~(\ref{eq:Hubble_lin}), there are several compelling reasons to adopt Eq.~(\ref{eq:Hubble_log}). The first concerns the statistical properties of the errors: uncertainties in distance moduli are Gaussian, whereas those in luminosity distances are log-normally distributed. Using the latter can therefore introduce biases. Consequently, several analyses that search for possible anisotropic signals, while employing distances indirectly derived from distance moduli may be methodologically inconsistent, as they typically use the standard $\chi^2$ definition, $\chi^2 = \sum_i (obs_i - theo_i)^2 / err_i^2$, which assumes Gaussian errors. Indeed fitting the full CF4 dataset using Eq.~(\ref{eq:Hubble_lin}) or Eq.~(\ref{eq:Hubble_log}) within this same $\chi^2$ framework yields different values of the Hubble constant: $H_0 = 78.28^{+0.08}_{-0.08}$ and $H_0 = 75.84^{+0.06}_{-0.06}$, respectively, a discrepancy as large as $3.0$ km s$^{-1}$ Mpc$^{-1}$. This difference arises from both the data distribution (since a linear spacing in $\mu$ corresponds to a logarithmic spacing in $d_L$) and the incorrect assumption of Gaussianity in the distance errors. Both factors alter the statistical weight of each data point in the fit, thereby affecting the slope of the HL relation and, ultimately, the inferred value of $H_0$.


If the distance modulus is used instead of the luminosity distance, the correction to the velocity applied by the CF4 team must be modified accordingly, since it currently employs the correction term $f_{d_L}(z,q_0,j_0)$ derived from the series expansion of the luminosity distance. Following the same reasoning that leads from Eq.~(\ref{eq:Hubble_Cosmicflow_2}) to Eq.~(\ref{eq:Hubble_lin}), but starting directly from the distance modulus, the appropriate correction should be $f_{\mu}(z,q_0,j_0)$, which stems from the series expansion of the distance modulus and differs from $f_{d_L}(z,q_0,j_0)$ (see, e.g., Eqs.~(4) and (A5) of \cite{Capozziello:2011tj}):
\begin{align}
5 \log D_{L} &= 5 \log z + f_{\mu}(z, q_0, j_0)\, , \\
f_{\mu}(z, q_0, j_0) &= \frac{5}{\ln 10} \left[ \frac{1}{2}(1-q_0)z  \nonumber \right. \\ 
&\left. - \frac{1}{24} (7-10 q_0 - 9 q^{2}_{0} + 4 j_0)z^{2} \right]\, . \label{eq: f_mu}
\end{align}
More specifically, the \textit{correct} logarithmic version of the HL law, after the cosmographic expansion of the distance modulus, is:
\begin{equation}\label{eq:Hubble_mucorr}
\frac{\mu}{5} -5 = \log \left(v^{obs}_{CMB} - v_{pec}\right)+ \frac{f_{\mu}(z,q_0,j_0)}{5} - \log H_0 \, .
\end{equation}
We emphasize that Eq.~(\ref{eq:Hubble_mucorr}) represents the  formulation used in this paper in a self-consistent manner. Indeed, we have recently noted several instances in the literature where the cosmographic expansions for $d_L$ and $\mu$ have been incorrectly applied. Both the \textit{SH0ES} \cite{Riess:2021jrx} and the \textit{CCHP} \cite{Freedman:2024eph} teams, for instance, employ $\log d_L$ with $d_L$ expressed through its cosmographic expansion, Eq.~(\ref{eq:cosmography}). This procedure is problematic, as the logarithm of a series is not equivalent to the series of the logarithm.

It is straightforward to verify that $f_{\mu}(z)$ is systematically larger than the corresponding expression $5 \log f_{d_L}(z)$, with a relative difference of $\sim 0.06\%$ at $z \simeq 0.15$, the upper limit of the Hubble flow SNeIa sample used by \textit{SH0ES} to infer $H_0$. It is therefore worth examining how this discrepancy might affect their $H_0$ estimate. From Eqs.~(4) and (5) in \cite{Riess:2021jrx}, adopting $f_{\mu}$ instead of $5 \log f_{d_L}$ leads to a decrease in $a_B$, and consequently in $\log H_0$. The effect is negligible for very nearby objects but increases slightly with redshift. Using the \textit{SH0ES} calibration $M^0_B = -19.25$, we find that the average $H_0$ inferred for the Hubble flow SNeIa sample would decrease by only $\Delta H_0 \simeq 0.02$ km s$^{-1}$ Mpc$^{-1}$ (about $0.03\%$). While clearly insignificant in the context of the Hubble tension, this nonetheless illustrates how small, often overlooked systematics can pervade even the most carefully calibrated analyses.

\begin{figure}
\centering
\includegraphics[width=\columnwidth]{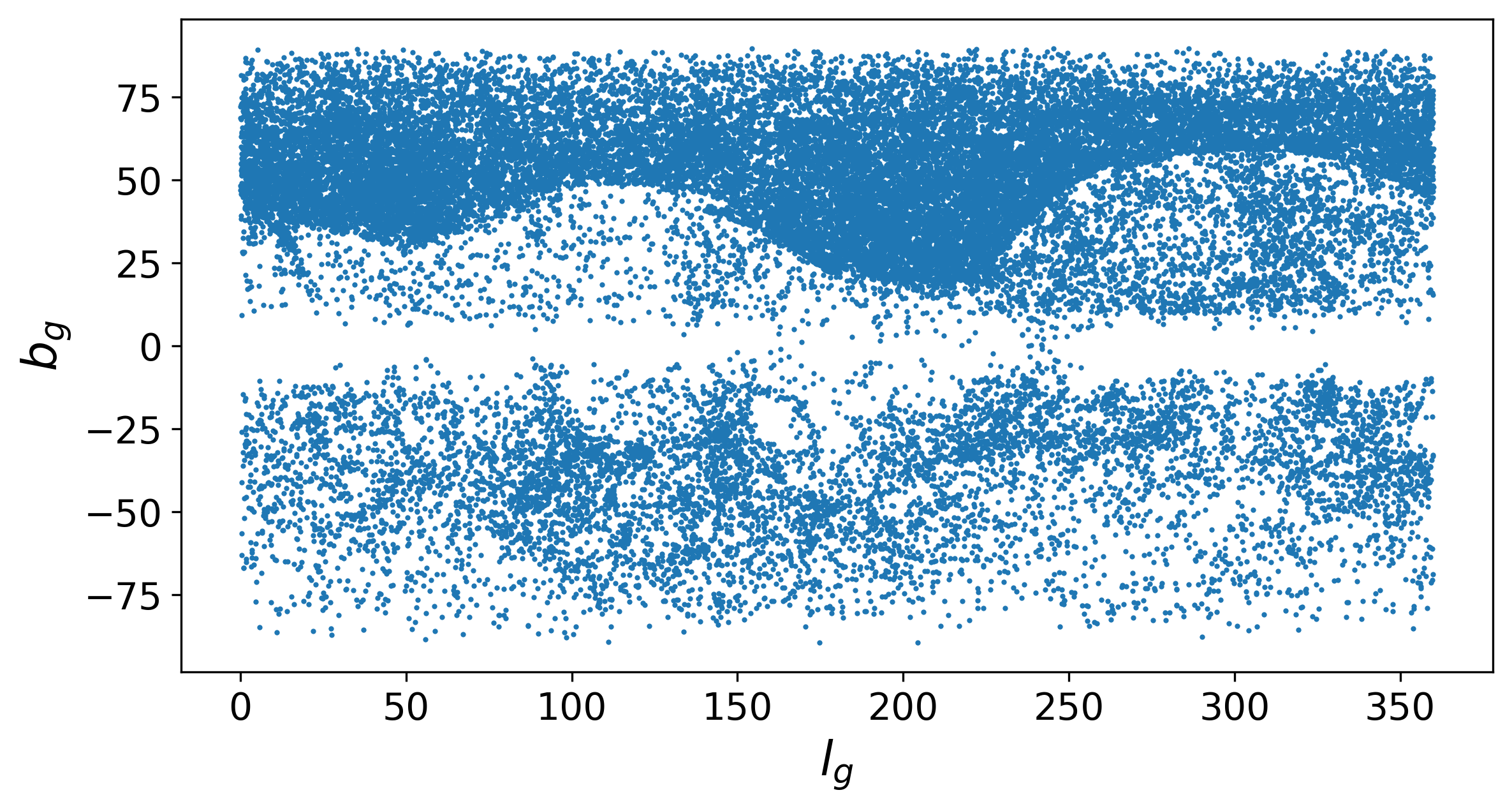}
\caption{Positions of the objects in the CF4 catalog in galactic coordinates.}\label{fig:Full_CF4_gal}
\end{figure}

\section{Preliminary analysis}
\label{sec:preliminary}

\begin{figure*}[!htbp]
~~~~~~~~~~\includegraphics[width=0.95\columnwidth]{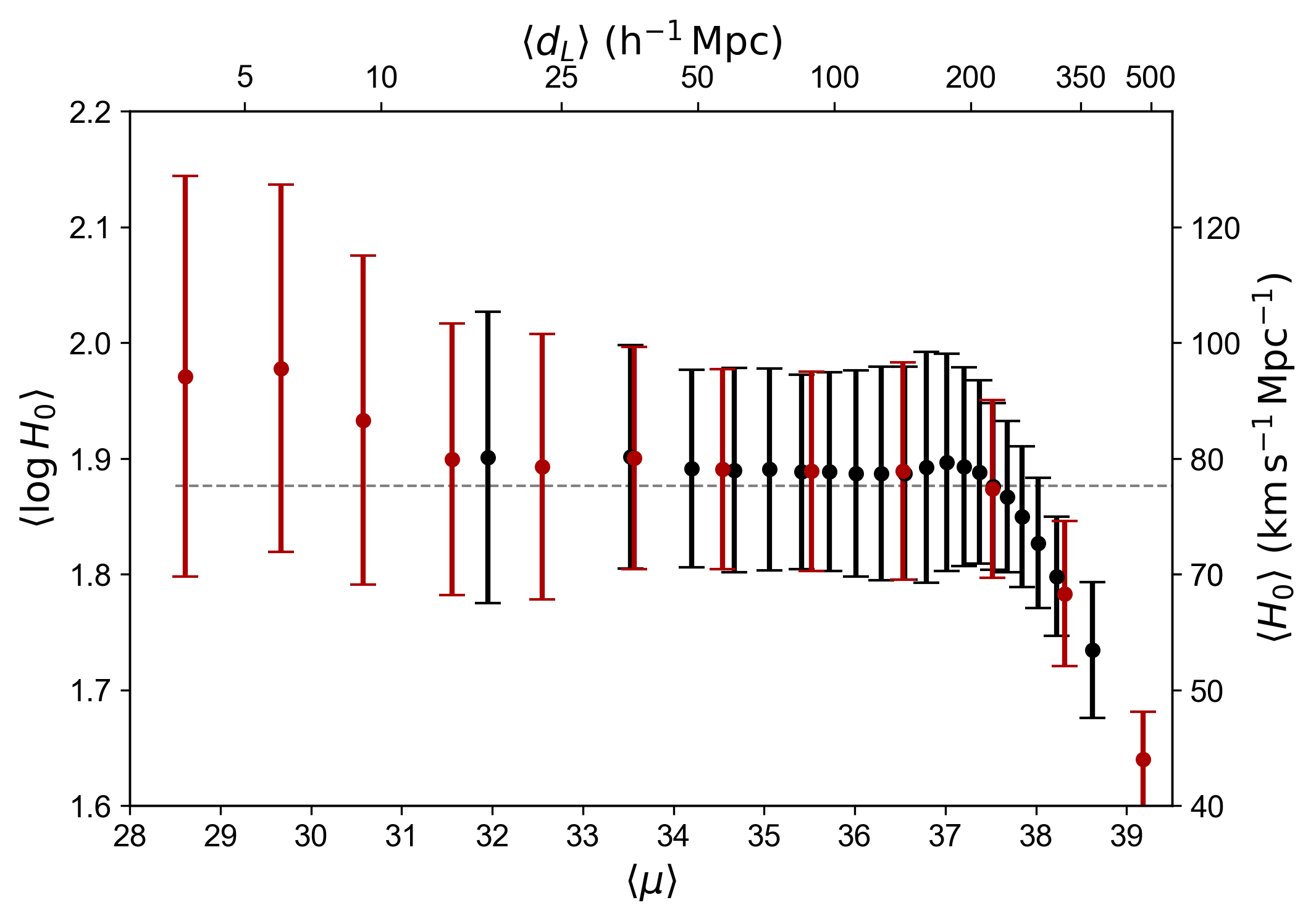}~~~~~~
\includegraphics[width=0.95\columnwidth]{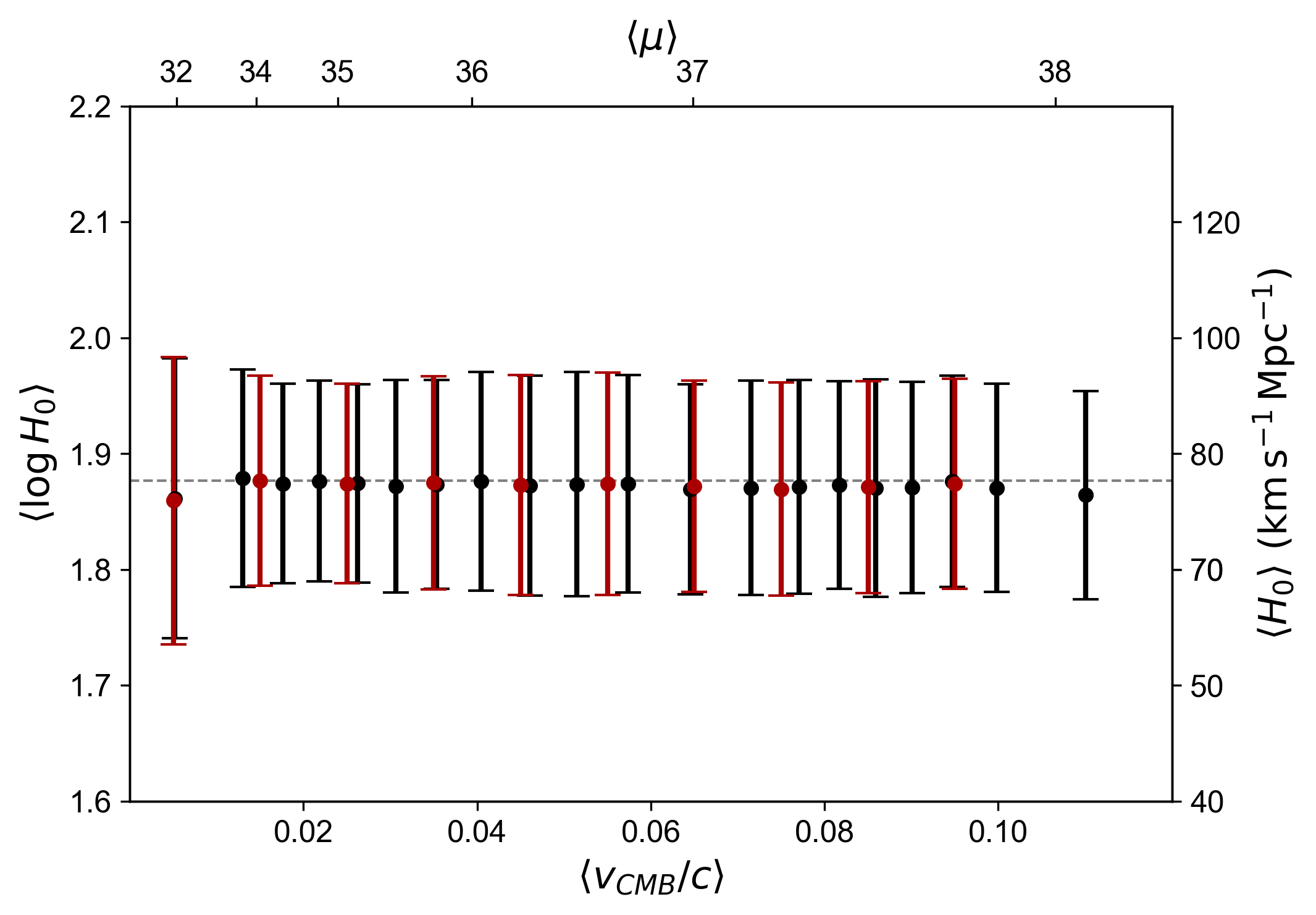}\\
~~~\\
\includegraphics[width=\columnwidth]{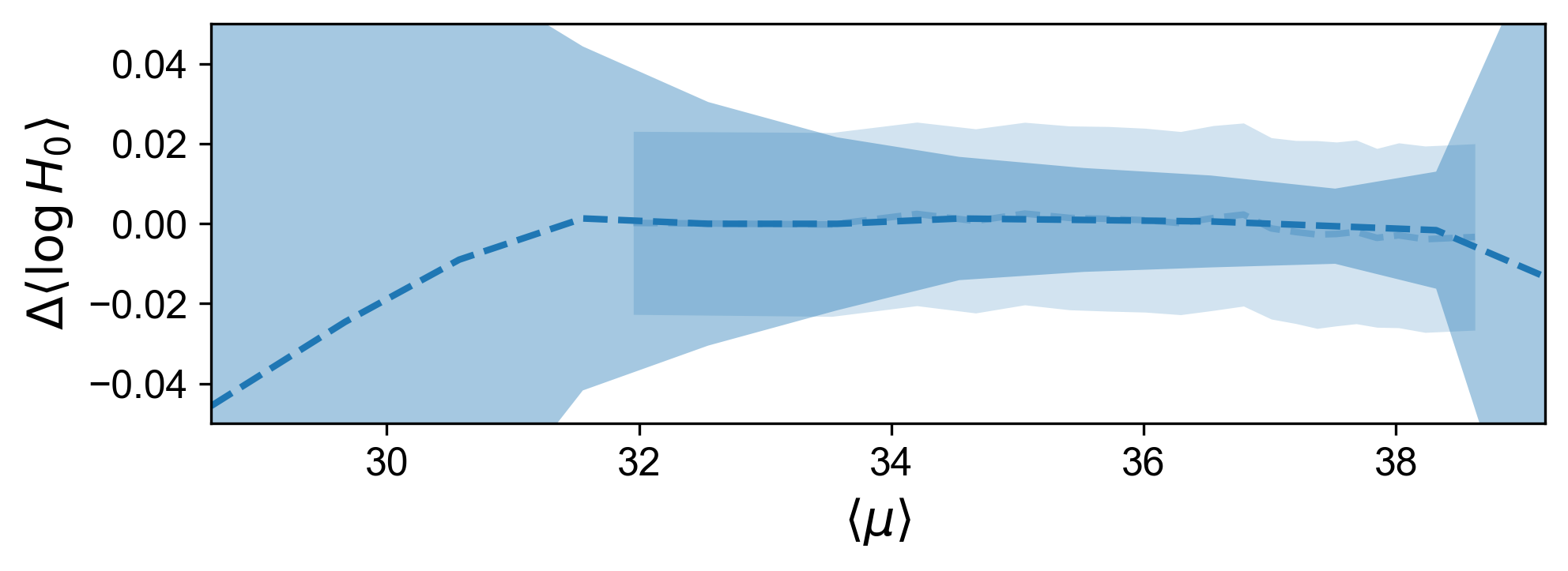}~~
\includegraphics[width=\columnwidth]{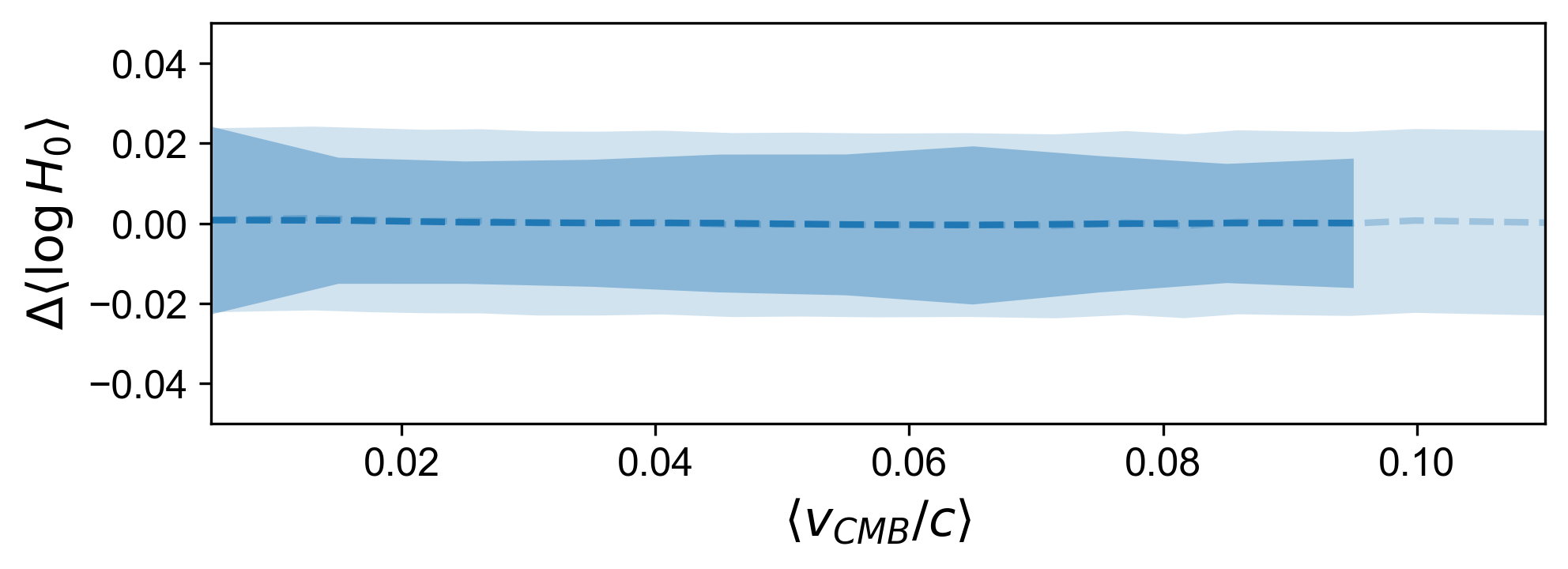}\\
~~~\\
\includegraphics[width=\columnwidth]{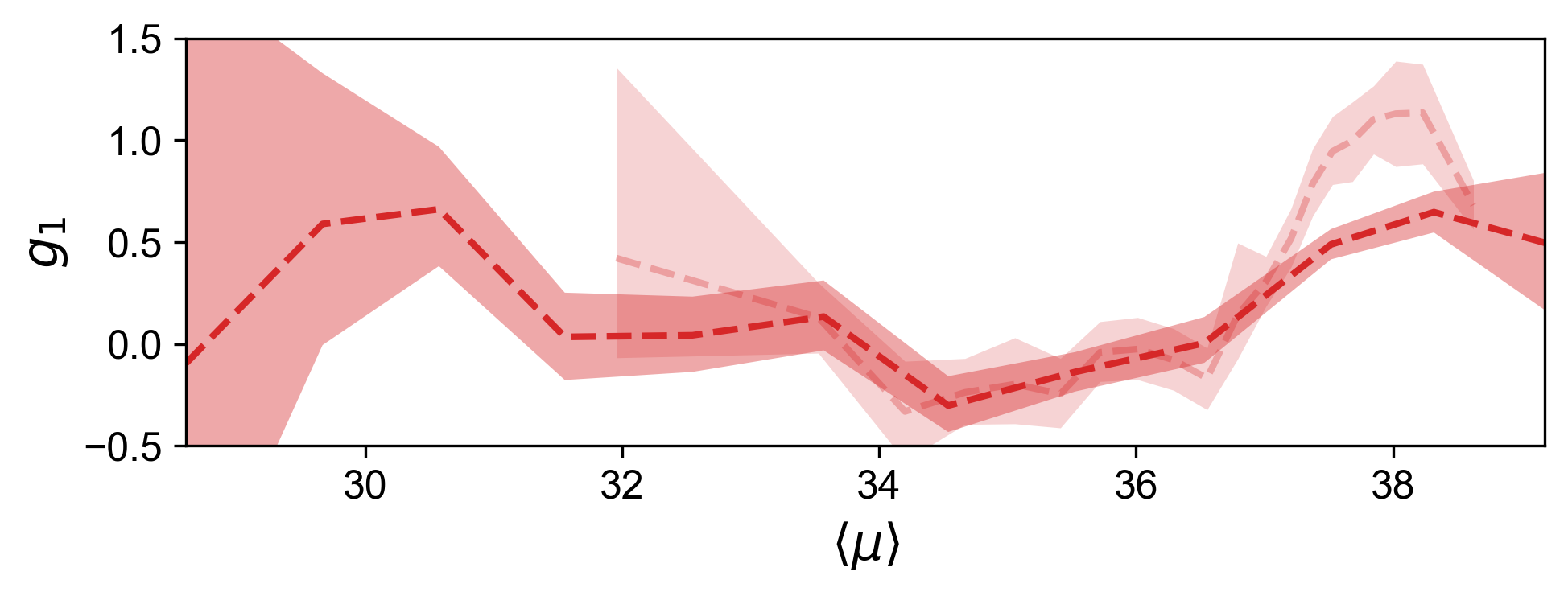}~~
\includegraphics[width=\columnwidth]{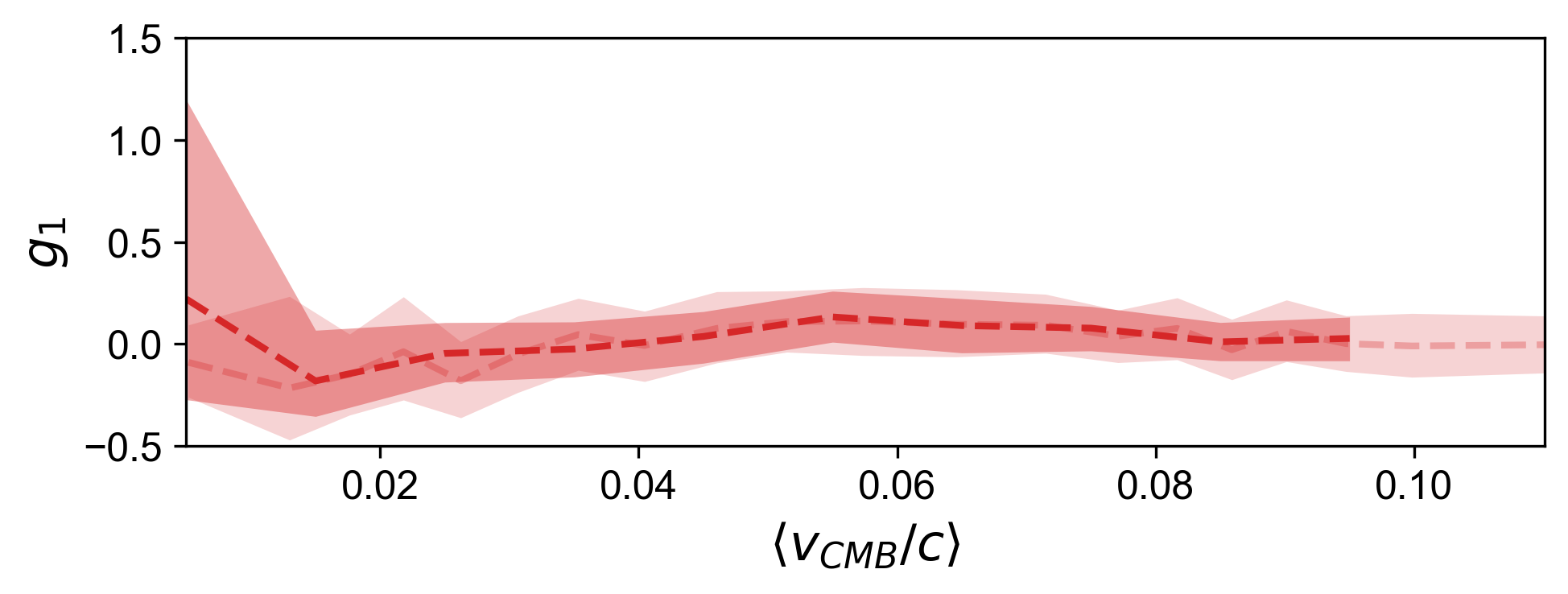}\\
~~~\\
\includegraphics[width=\columnwidth]{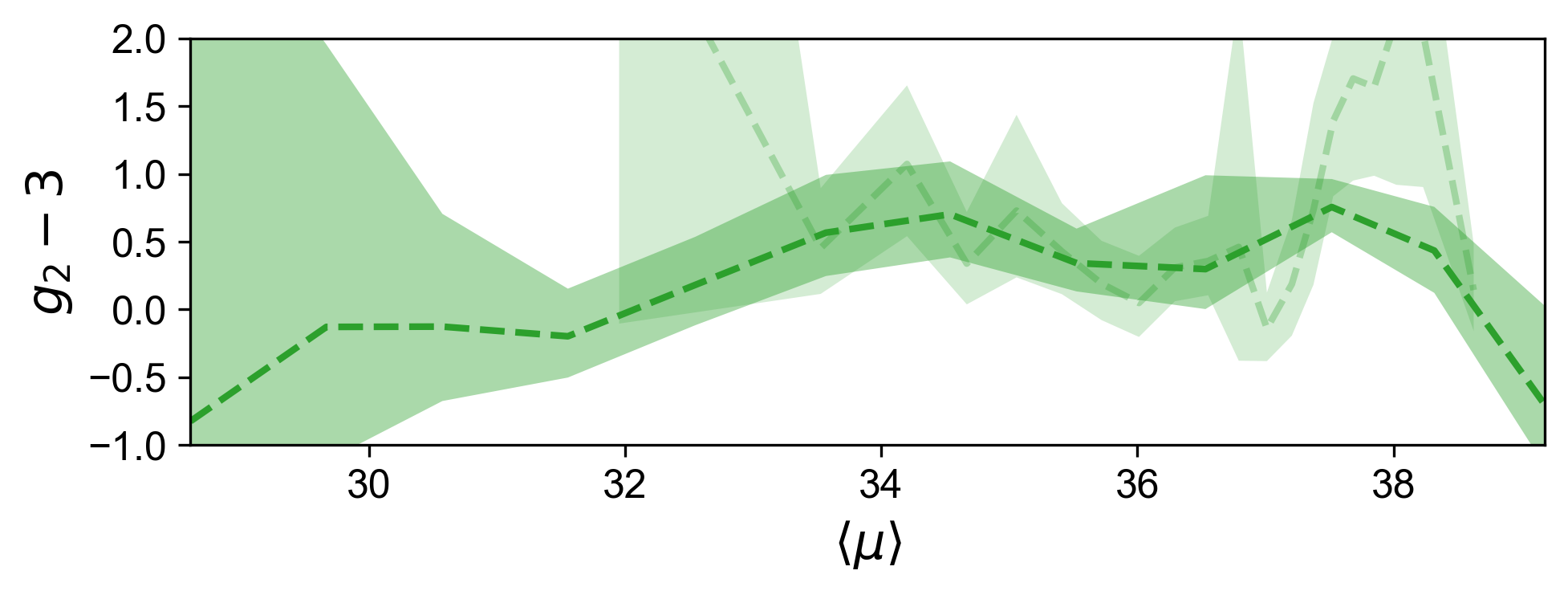}~~
\includegraphics[width=\columnwidth]{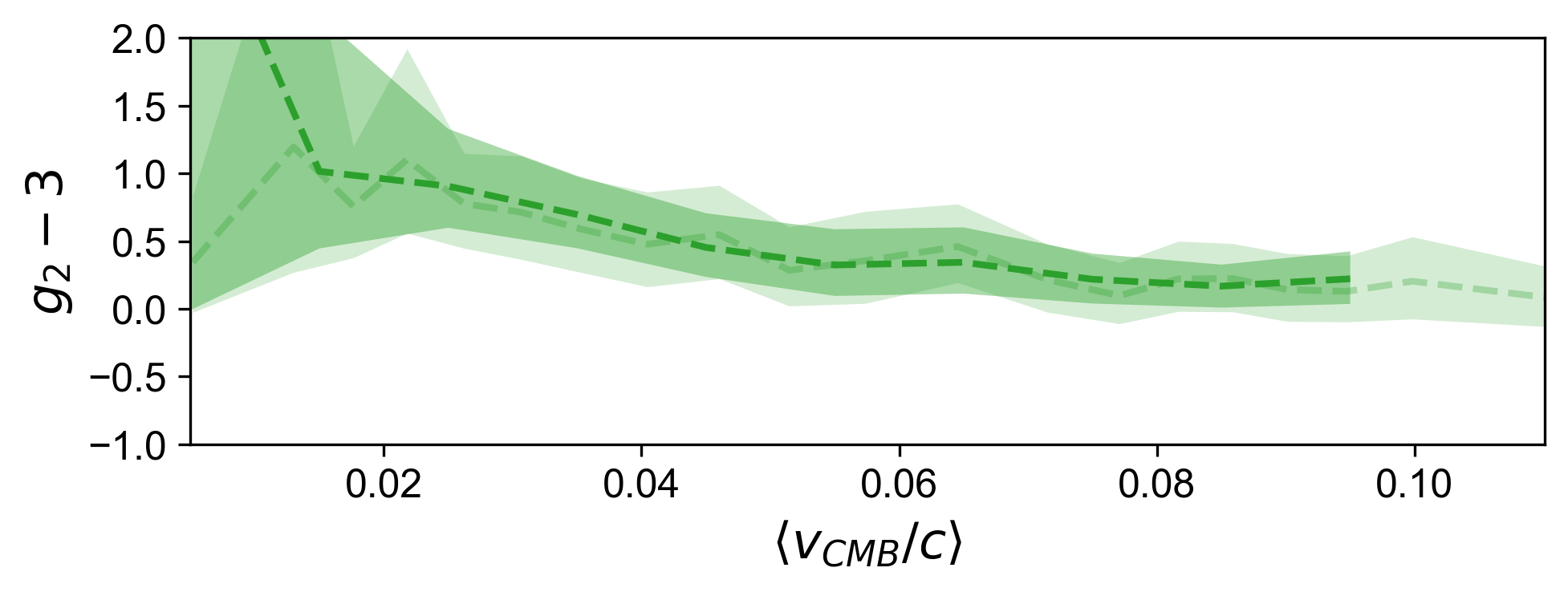}\\
\caption{Hubble constant average, $\langle \log H_0 \rangle $, as a function of distance modulus, $\mu$, and redshift, $v_{CMB}/c$.}\label{fig:Residual}
\end{figure*}

The main objective of this section is to perform a qualitative and quantitative analysis of the CF4 catalogue in order to assess the viability of its use for the purposes of this work. Indeed, the CF4 catalogue compiles distance estimates from multiple methodologies and data sets, which makes it a powerful general tool, but also complicates the identification and quantification of intrinsic selection effects and biases due to the heterogeneity of the sources. A heterogeneous compilation generally does not have a single, clean selection function, different probes come with different selection mechanisms (magnitude limits, host selection, targeted follow-up, cluster selection), and that indeed complicates forward modelling of selection and Malmquist-type effects in a homogeneous way. Although the CF4 team applies an overall cross-calibration/zero-point homogenization to bring heterogeneous distance estimates onto a common scale \cite{Tully:2022rbj}, it remains challenging to account for all the different contributions in a homogeneous manner. Addressing this issue is beyond the scope of this work, and
recent literature has emphasized and quantified how distance priors and selection effects can bias inferred distances and derived parameters, and therefore should be modelled from the earliest stages of constructing these types of catalogues, particularly for cosmological inference \cite{Desmond:2025ggt,Nusser:2025hob,Hogas:2026urs}.

As our goal is to test for possible anisotropic signals and their variation moving farther from us — which will be accounted for by dividing the full sample into radial shells — particular care must then be taken when working with the CF4 catalogue. As can be seen in Fig.~\ref{fig:Full_CF4_gal}, which plots the catalogue objects in galactic coordinates, and in Fig.~21 of \cite{Tully:2022rbj}, which uses Cartesian coordinates for the observed radial recessional velocities, the full sample is not uniformly distributed across the sky. In particular, data are missing in the Zone of Avoidance around the galactic plane, and there is an obvious excess of objects in the northern hemisphere due to SDSS coverage.

The CF4 team restricts its analysis to objects with observed radial velocities $v^{obs}_{CMB} > 4000$ km s$^{-1}$, corresponding to approximately $93\%$ of the full catalog. As qualitatively noted in \cite{Tully:2022rbj}, for $v^{obs}_{CMB} > 4000$ km s$^{-1}$, the observed velocity can be regarded as the dominant component of the total source velocity, assuming an average peculiar bulk-flow velocity of $v_{pec} \sim 300$ km s$^{-1}$. It is worth noting that the ``fiducial'' best estimate of the Hubble constant reported in \cite{Tully:2022rbj}, $H_0 = 74.6 \pm 0.8$ km s$^{-1}$ Mpc$^{-1}$, is derived from this very subsample. When we fit the same subsample using Eq.~(\ref{eq:Hubble_mucorr}) without applying statistical weights, we obtain a consistent result of $H_0 = 74.55^{+0.08}_{-0.08}$ km s$^{-1}$ Mpc$^{-1}$, with an intrinsic scatter of $\sigma_{intr} = 0.46$ km s$^{-1}$ Mpc$^{-1}$. Applying a weighted fit instead yields $H_0 = 75.23^{+0.07}_{-0.07}$ km s$^{-1}$ Mpc$^{-1}$, with a reduced intrinsic scatter of $\sigma_{intr} = 0.05$ km s$^{-1}$ Mpc$^{-1}$. 

From these premises, it is clear that we will need a stronger and more valid motivation for any cut from our side. We thus aim to perform a set of basic tests to identify whether residual Malmquist signatures remain in the working sample (in \cite{Tully:2022rbj} the authors explicitly discuss bias issues and apply bias-related corrections in at least some contributing subsamples e.g., FP pipelines, and update to 6dFGSv-related bias handling) and eventually, which subset of the full catalogue can be used to minimize, or render negligible, the impact of potential selection effects and biases. A similar investigation, focusing on anisotropy, was performed in \cite{McKay:2015nea} using Cosmicflows-2. When compared to the COMPOSITE sample, a specific issue was identified: {the variation of the Hubble constant with respect to its asymptotic mean value was found to decrease more slowly than expected} and failed to reach the expected uniformity at large scales. This systematic offset was attributed to the treatment of Malmquist biases, which were explicitly corrected in the COMPOSITE sample but not in Cosmicflows-2. The subsequent Cosmicflows-3 catalogue—used, for instance, in \cite{Kalbouneh:2022tfw}—also lacks explicit Malmquist-bias corrections and is redshift-limited, which affects the estimation of bulk-flow velocities for objects with large distance measurements \cite{Peery:2018wie}. From related studies, we can infer that the CF4 sample should not have been treated and corrected for Malmquist bias  (see Sec.~4.1 of \cite{Tully:2022rbj}). In order for it to be successfully employed in several recent studies of bulk-flow measurements \cite{Hoffman:2023pac,Watkins:2023rll,Whitford:2023oww} and in anisotropy searches using the Tully–Fisher relation \cite{Boubel:2024cmh} this has to be taken into account.

The quickest ``smoke detector'' is the Hubble parameter flatness test. CF4 explicitly uses the \textit{``rough constancy of the Hubble parameter with redshift''} \cite{Tully:2022rbj} as a necessary test for distances beyond the nearby regime. To test this trend, we first bin the data in radial shells and compute the weighted average $\langle \log H_{0} \rangle$ within them. Here, the term ``radial'' refers to the distance of the objects from us. We will use two different quantities to measure this distance: the distance modulus, $\mu$, and the observed radial velocity, $v^{obs}_{CMB}$, both of which are used in the literature as distance indicators, but which have different impacts and roles, especially in connection to possible biases \cite{Nusser:2025hob}. The weighted average $\langle \log H_{0} \rangle$ and the corresponding error in each shell will be calculated, as per standard procedure, as
\begin{equation}\label{eq:average_H0}
\langle \log H_0 \rangle =
{\frac{\sum_{i=1}^\mathcal{N} w_i \log H_{0,i}} {\sum_{i=1}^N w_i}}\, ,
\end{equation}
\begin{equation}\label{eq:average_sH0}
\sigma_{\langle \log H_0 \rangle}^2
= \left( \sum_{i=1}^N w_i \right)^{-1}\, ,
\end{equation}
where $\mathcal{N}$ is the number of objects per shell, $\log H_{0,i}$ is the logarithm of the Hubble constant for each object in the shell derived from Eq.~(\ref{eq:Hubble_mucorr}), and the corresponding weights are defined as
\begin{equation}\label{eq:weight_stat}
w_i^{-1} = \sigma_{\log H_{0,i}}^2 + \sigma_{\rm int}^2\, , 
\end{equation}
with the statistical errors, $\sigma_{\log H_{0,i}} = \sigma_{\mu,i}/5$, obtained propagating from the error on the distance moduli, available in the catalogue, while errors on the velocities are assumed to be much smaller and omitted. The intrinsic scatter $\sigma_{int}$ is derived per each shell and added in quadrature to the statistical errors so that the reduced chi-square (per shell)
\begin{equation}\label{eq:int_err}
\chi^2_{\rm red}(\sigma_{\rm int})
= {\frac{1}{\nu}}
\sum_{i=1}^N
{\frac{(\log H_{0,i}- \langle \log H_0 \rangle)^2} {\sigma_{\log H_{0,i}}^2+\sigma_{\rm int}^2}}\, ,
\end{equation}
with $\nu$ the number of degrees of freedom, becomes equal to $1$.

The Malmquist bias is fundamentally a selection and scatter effect so we want to test also whether residuals correlate with the variable that drives inclusion, in this case the distance modulus and the radial velocity. In \cite{Tully:2022rbj}, the CF4 team explicitly discusses bias trends with apparent magnitude for at least some inputs and shows that changing the effective magnitude limit used in the probability distribution function normalization reduces bias in mocks. A generic, indicator-agnostic normalized residual is:
\begin{equation}
\Delta \log H_{0,i} = \frac{\log H_{0,i} - \langle \log H_0 \rangle\,}{\sqrt{\sigma_{\log H_{0,i}}^2+\sigma_{\rm int}^2}} , 
\end{equation}
and we have analyzed the behaviour of the residuals in the same shells, following their distribution, standard deviation, and more importantly their skewness and kurtosis. In order to assure that the statistics (in particular the estimation of the skewness and of the kurtosis) is well based and fully reliable, we apply bootstrap in each shell, resampling the data $5000$ times.

We start with binning in distance moduli for a total of $20$ bins over the full range covered by CF4, $\mu \in [26;40]$, and containing $\sim 1900$ objects each. Then, we look for a statistically significant slope of $\langle \log H_0 \rangle$ vs the distance modulus, because residual Malmquist or selection biases in distance indicators tend to create a distance-dependent bias in inferred distances. From a simple linear fit, we find
\begin{equation}
\frac{\mathrm{d}\, \langle \log H_0 \rangle}{\mathrm{d}\, \langle \mu \rangle} = -0.022 \pm 0.005\, , 
\end{equation}
which implies a variation at $\approx 4\sigma$ confidence level.

When we look at the residual statistics for this binning cases, in light colour plots in the right panels of Fig.~\ref{fig:Residual}, we have clear indications about what a healthy range/subsample to be used in our analysis could be. From the binning in distance moduli, we see how the skewness $(g_1)$ is fully consistent with zero, or just mildly even when the confidence intervals are wide, up to $\mu \approx 36.5$, which is telling that no strong asymmetry is present yet in the data. For $\mu \gtrsim 36.5$ it is important to note that we have a monotonically increasing skewness which cannot be simply noise, but very likely is connected to selection or truncation biases. Indeed we expect a wedge in $\mu$-space given that the catalog is cut in redshift space ($z=0.055$ for 6dF and $z=0.1$ for SDSS). These conclusions are reinforced when looking at the kurtosis $(g_3)$: at the lower end, up to $\mu \sim 33$, the consistency is due to large error bars; then, there is a general consistency with the expected value of $3$ up to $\mu \sim 36.5-37$; at larger distance moduli it raises up to $5$, clearly signalling heavy tails and selection domination.

For our later statistical investigation of a possible anisotropy, although, a different binning is more meaningful: we bin data requesting that their width is approximately twice the average error on $\mu$ in that shell. The data distribution in the shells so obtained is shown in Fig.~\ref{fig:Shells_mu}. It is clear that below $\mu < 31$ there is a scarcity of objects, accounting for only the $0.71 \%$ of the total sample, meaning that any statistical inference will be weak. The SDSS becomes the dominant contributor at $\mu > 37$, constituting at least the $40\%$ of the full sample. Thus, it is clear that these ranges will be highly anisotropically populated, which is an important issue for our analysis concerning anisotropy detection. Looking carefully at Fig.~\ref{fig:Residual} and the dark colour plots, we also see how the general conclusions do not qualitatively change with respect to the equal-statistics binning discussed before. In particular, the asymmetry trend at large distances, as well as the very large uncertainty regions at the lower end are confirmed. For this reason, we think that when performing our analysis with respect to a $\mu$ binning, the safest conclusions can be drawn from the shells with $\mu \in [31;36]$. We finally note how the variation of the average Hubble constant with the distance can be influenced by the data at both ends of the full interval. If we restrict the fit to the healthy range, $\mathrm{d} \langle \log H_0 \rangle / \mathrm{d} \langle \mu \rangle = -0.0022 \pm 0.0009$, so that the variation is reduced both by absolute value and by confidence, down to $\sim 2\sigma$.

When we consider the binning in terms of radial velocity, shown in the left panels of Fig.~\ref{fig:Residual}, we get much better results. First, we consider the case of binning requesting an equal number of objects per shell (light colour plots): the trend in $\langle H_0 \rangle$ is basically consistent with a zero signal:
\begin{equation}
\frac{\mathrm{d}\, \langle \log H_0 \rangle}{\mathrm{d}\, \langle \mu \rangle} = -0.0006 \pm 0.0005 .
\end{equation}
If we look at the skewness, every confidence interval includes zero and no monotonic trend with velocity is found, so that we can conclude that the residual distributions are statistically symmetric at all depths and no sign of distance–estimator saturation effects. In contrast to the distance modulus binning, the asymmetry problem disappears entirely when binning in observed CMB-frame radial velocity, which might point to indicate that a large part of the skewness seen before was more a binning artefact, than an intrinsic physics effect. As for the kurtosis, for $v_{CMB} \gtrsim 15000$ km s$^{-1}$, or equivalently $z \gtrsim 0.05$, almost all confidence intervals include $3$, the upper bounds steadily decrease and the distributions become more Gaussian with depth: if Malmquist bias or truncation were dominating, kurtosis would increase, not decrease, with velocity. For lower velocities regimes, the first few bins reach an upper bound of $4$-$6$ and are not consistent with $3$ at $3\sigma$ level, while for $v_{CMB} \gtrsim 10000$ km s$^{-1}$ $(z \gtrsim 0.03$) we only have mild leptokurtosis.

Considering that the errors on the velocities are very small, binning based on the average error would result in quite narrow intervals. We thus opt for a second binning modality in which we have $10$ bins of $\Delta z = 0.01$, assuring that enough data are inside each shell. Even in this case, the same general considerations hold (dark colours plots in Fig.~\ref{fig:Residual}). The problem would be that, as it can been deducted by looking at Fig.~\ref{fig:Shells_v}, SDSS dominates at $z>0.06$, which makes this range unimportant for us. Thus, we should finally conclude that a healthy interval in this case corresponds to $v_{CMB} \in [10000, 15000]$ km s$^{-1}$, or equivalently $z \in [0.03,0.06]$.

In summary, a coherent picture emerges from these complementary and preliminary tests. When the CF4 catalogue is analysed in bins of distance modulus, statistically significant depth-dependent trends in the mean Hubble parameter, together with the onset of skewed and heavy-tailed residual distributions beyond $\mu \simeq 36.5$–$37$, indicate the increasing impact of selection and truncation effects at large distances, compounded by strong angular inhomogeneities associated with survey dominance. At the opposite extreme, very nearby distances suffer from limited statistics and poor sky coverage. In contrast, binning in observed CMB-frame radial velocity yields a substantially cleaner behaviour: the mean $\langle \log H_0 \rangle$ is consistent with depth independence, residuals remain statistically symmetric at all velocities, and kurtosis decreases rather than increases with depth, as expected in the absence of strong Malmquist-type distortions. Taken together, these results justify restricting the baseline anisotropy analysis to an intermediate-depth subsample, corresponding to $\mu \in [31;36]$ and $z \in [0.03,0.06]$, where the Hubble flow is stable, residual distributions are close to Gaussian, and selection effects are demonstrably minimized. This choice is therefore not ad hoc, but is directly motivated by internal consistency tests of the data themselves, and provides a statistically robust and conservative foundation for the subsequent anisotropy analysis.

\section{Statistical analysis}
\label{sec:statistical}

\begin{figure*}[!ht]
\centering
\includegraphics[width=0.65\columnwidth]{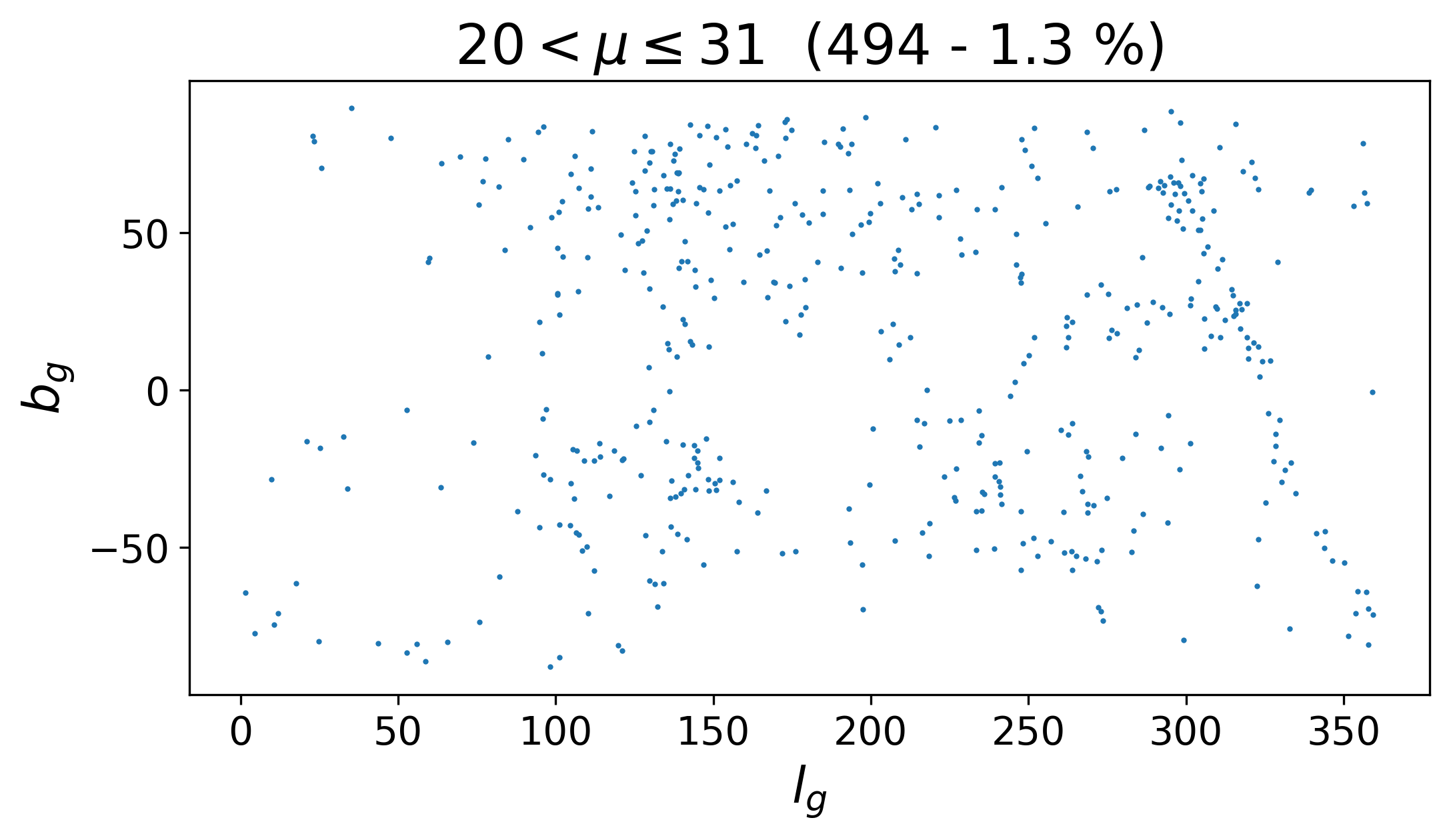}~~~
\includegraphics[width=0.65\columnwidth]{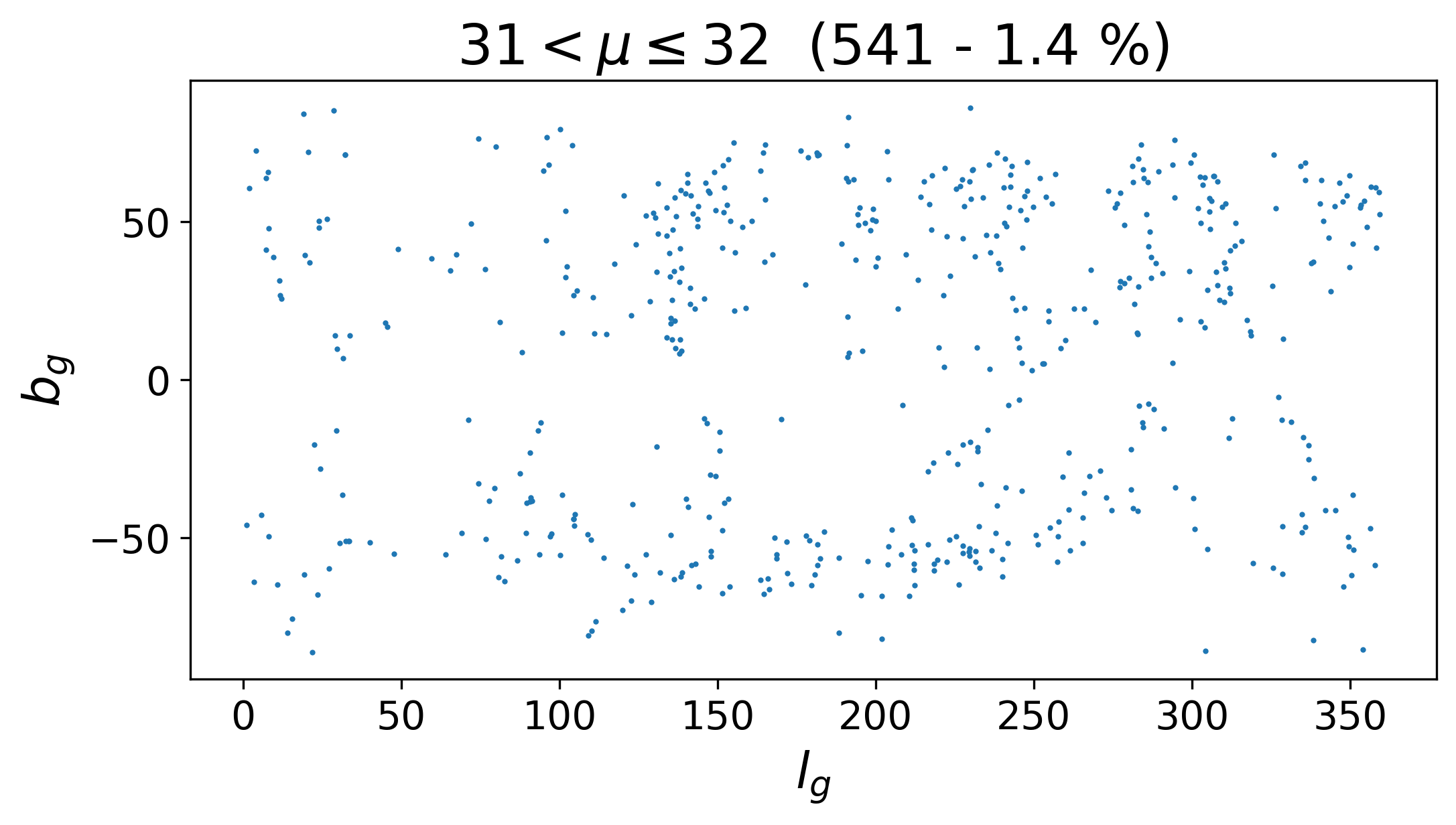}~~~
\includegraphics[width=0.65\columnwidth]{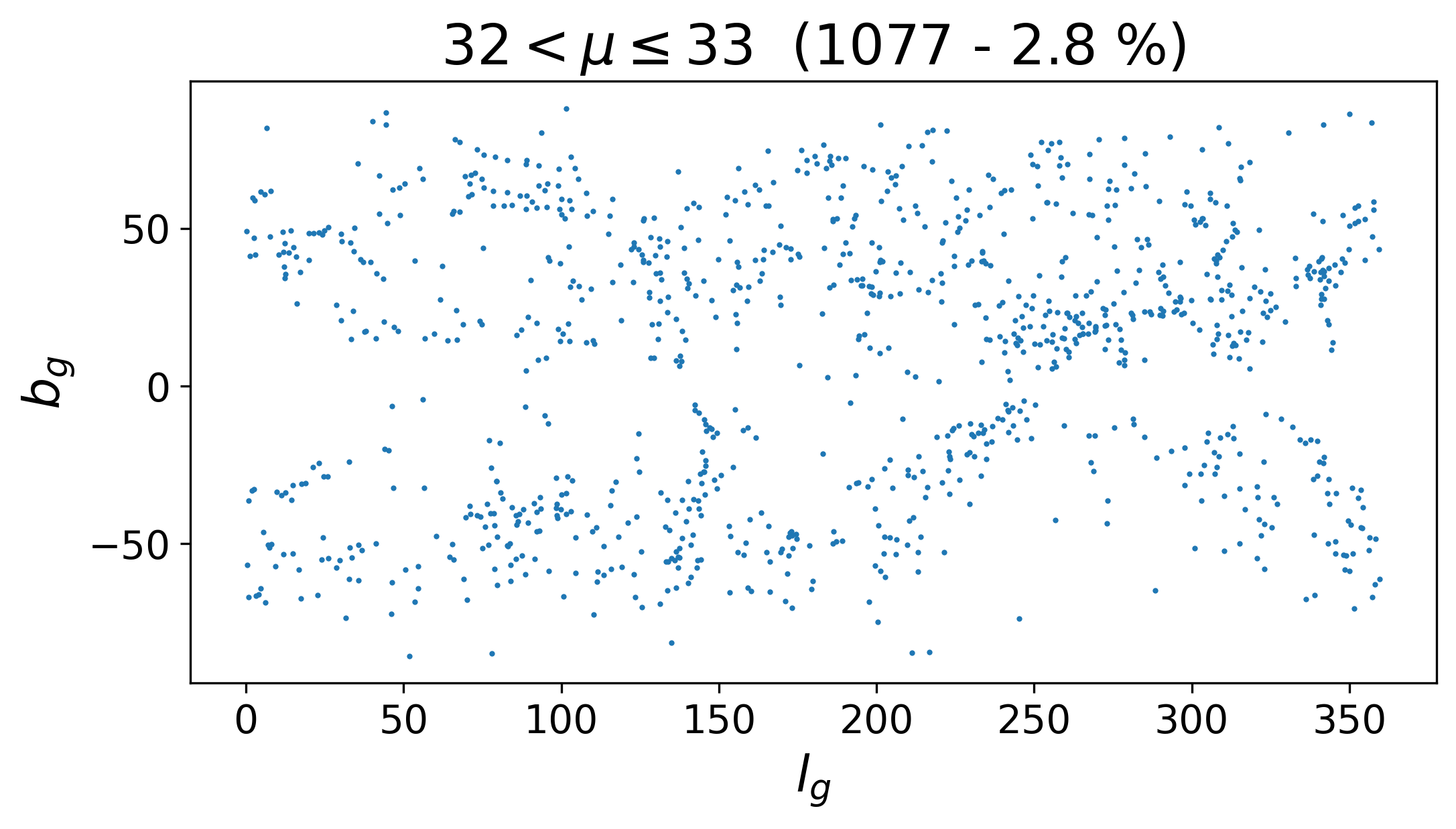}\\
~~~\\
\includegraphics[width=0.65\columnwidth]{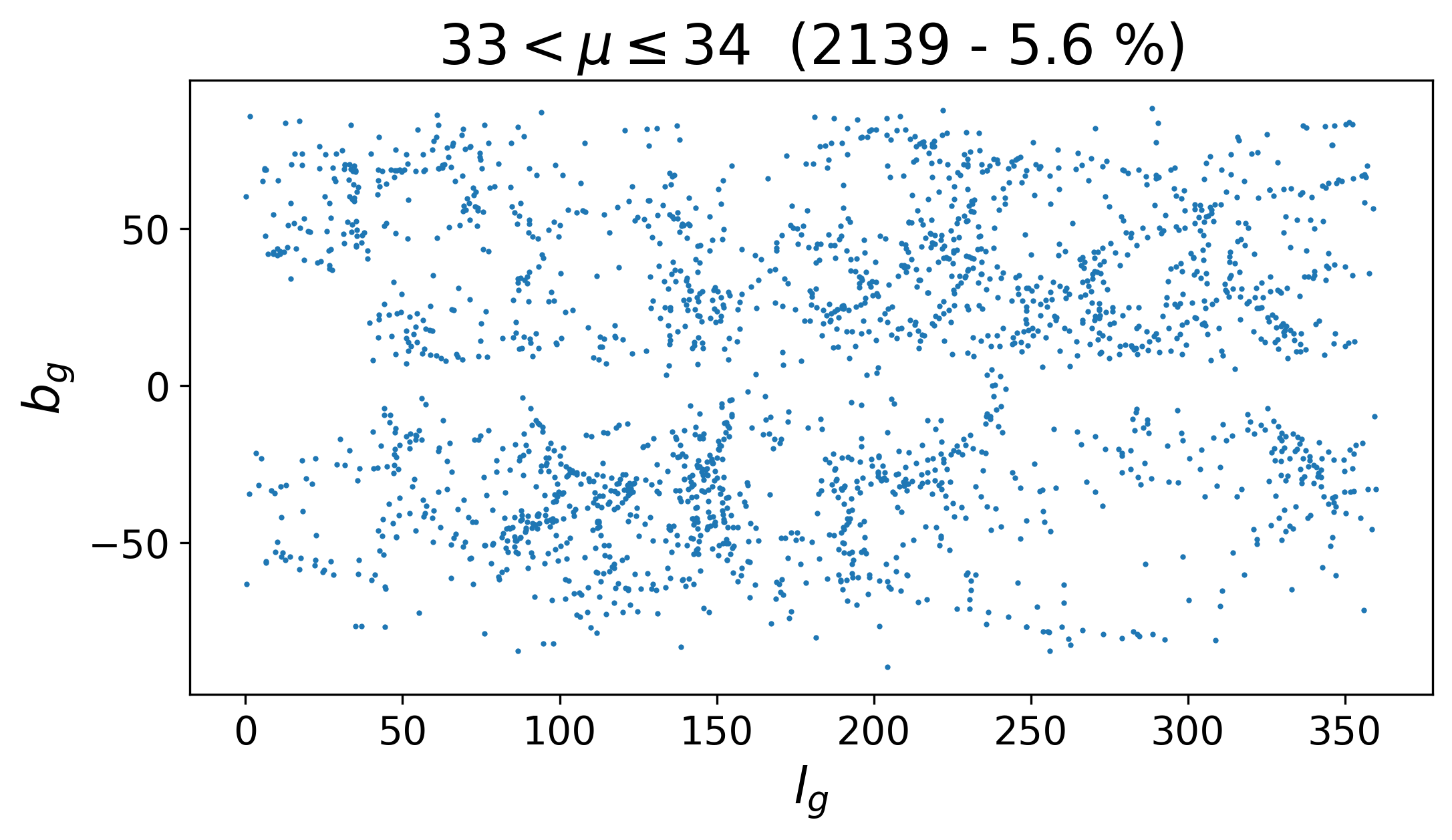}~~~
\includegraphics[width=0.65\columnwidth]{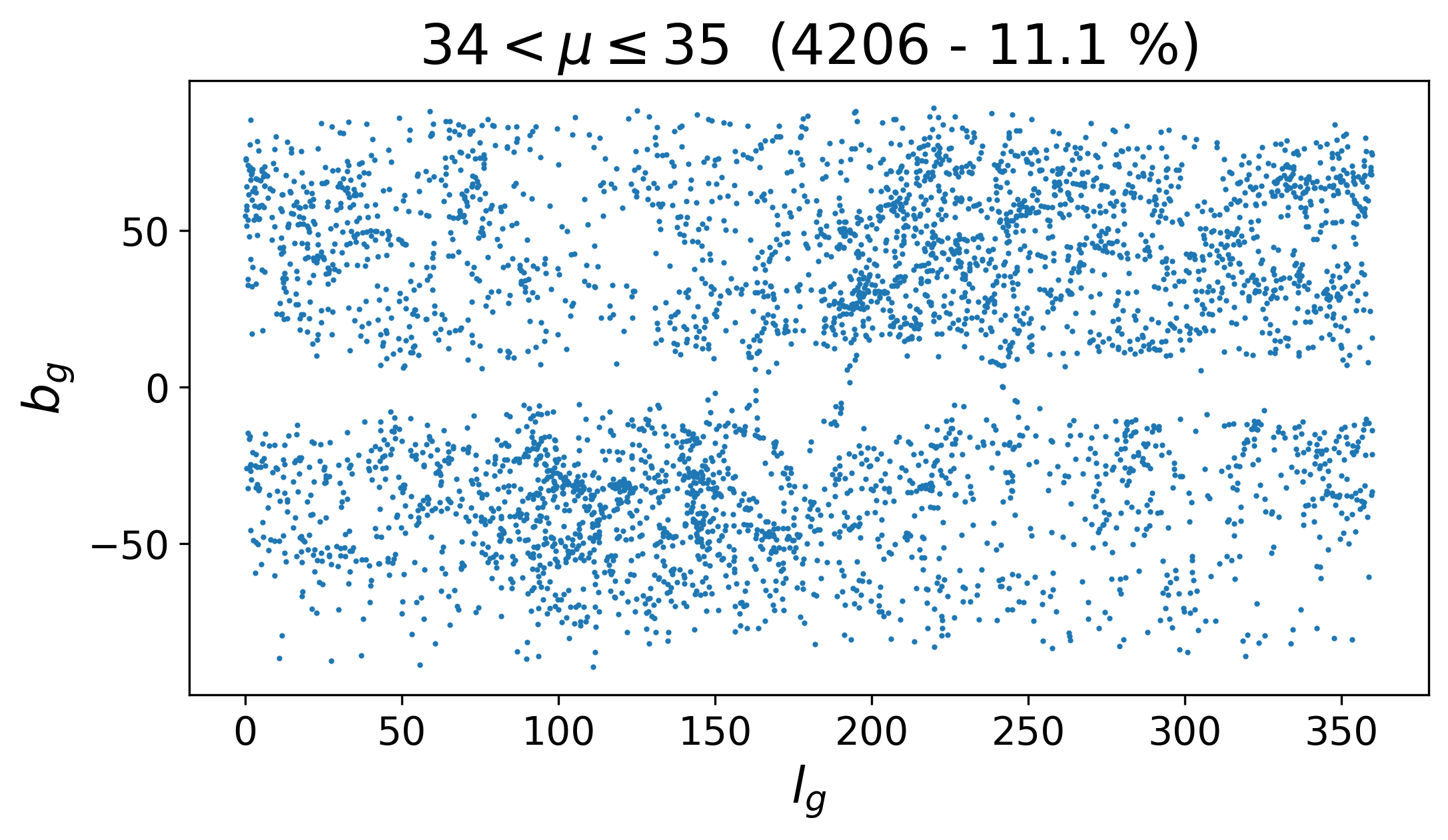}~~~
\includegraphics[width=0.65\columnwidth]{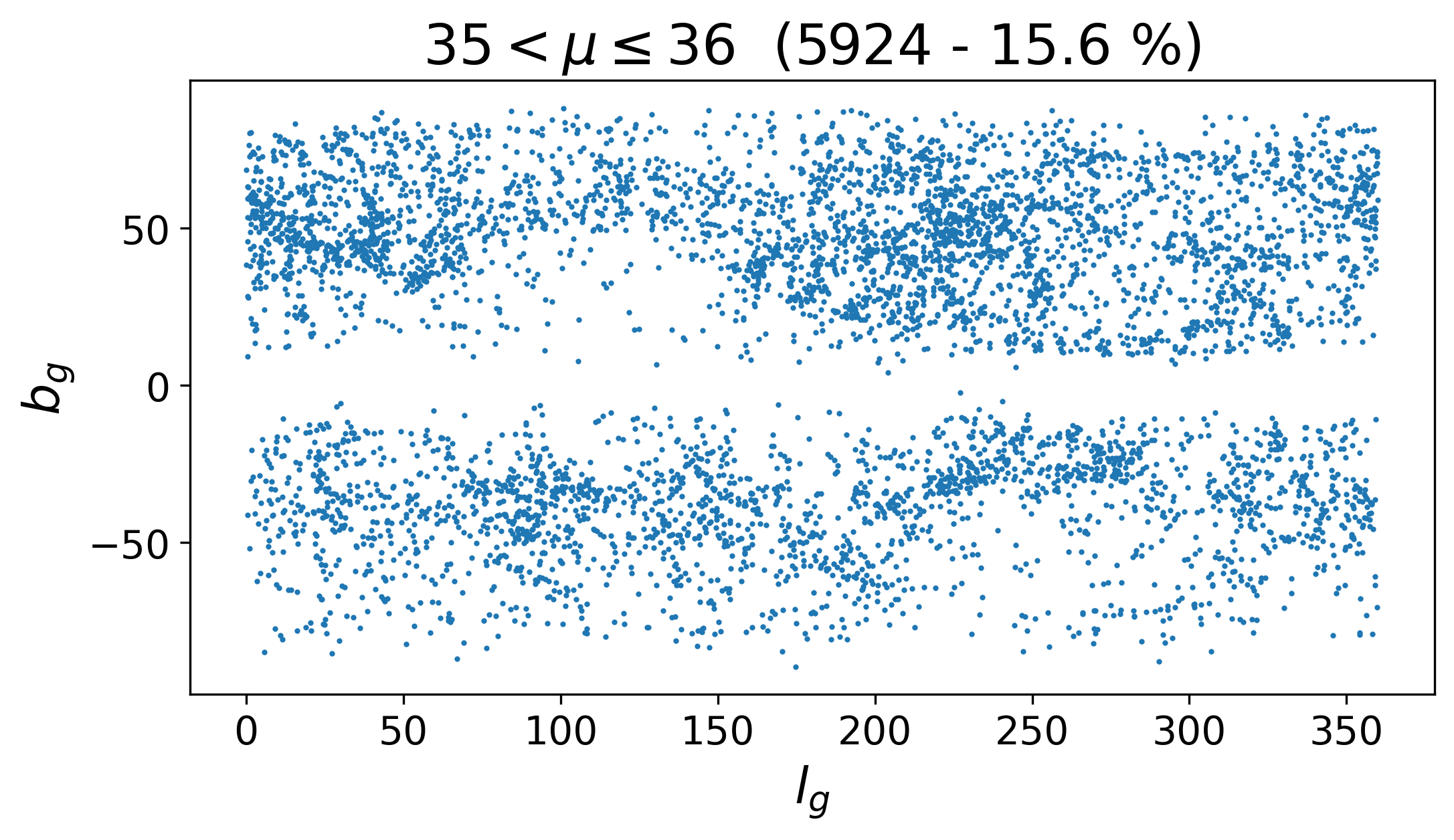}\\
~~~\\
\includegraphics[width=0.65\columnwidth]{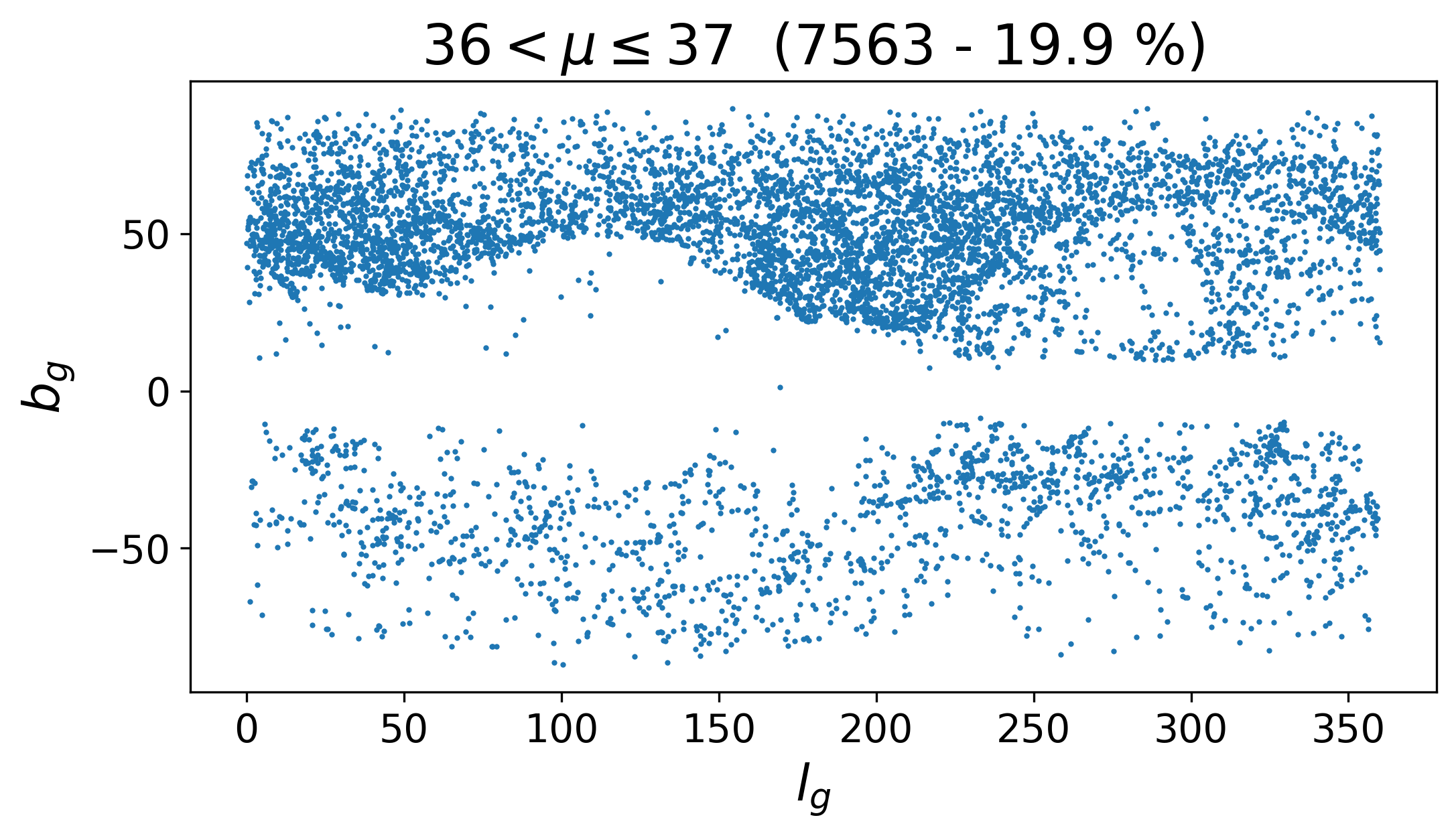}~~~
\includegraphics[width=0.65\columnwidth]{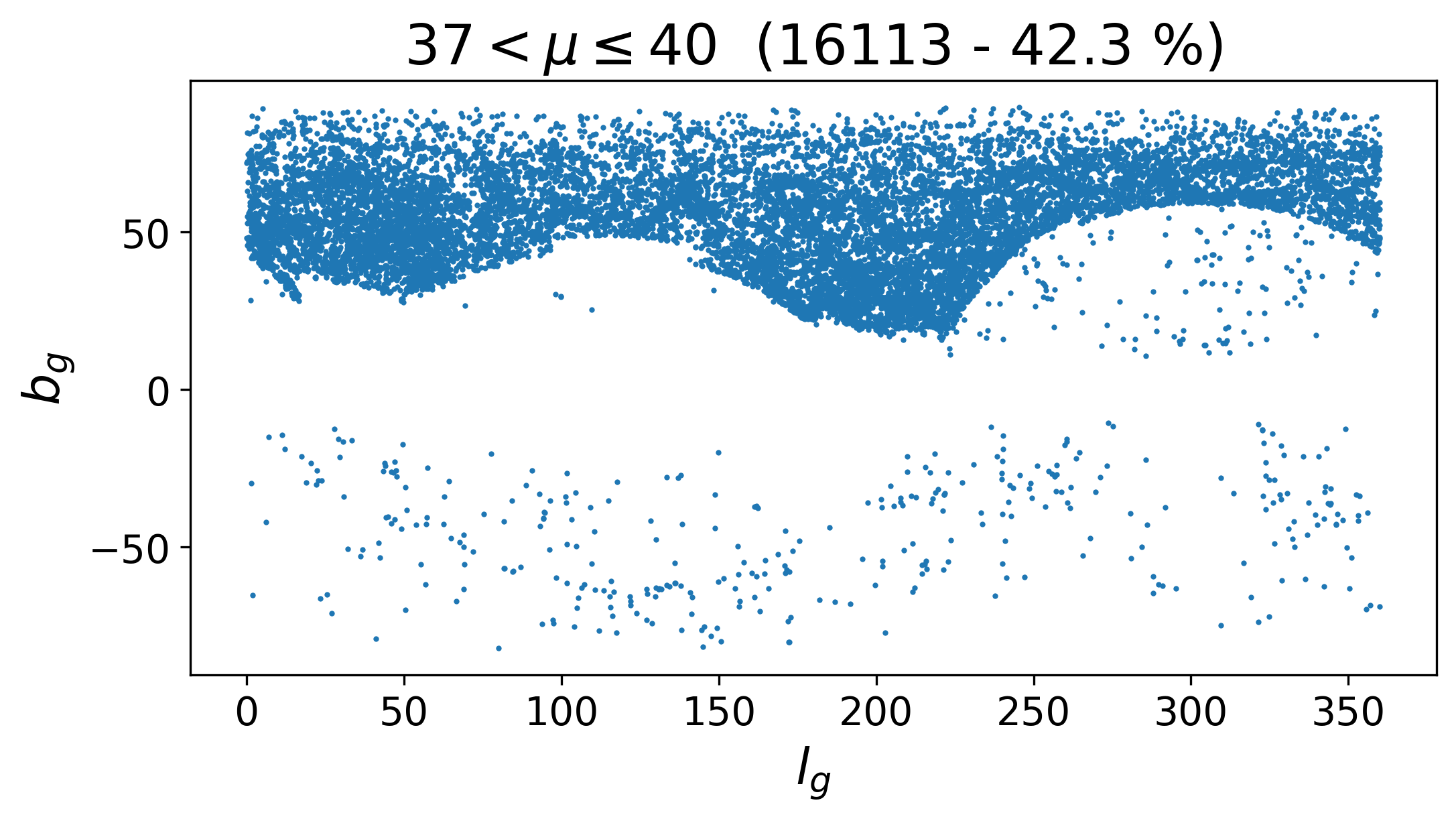}
\caption{CF4 angular distribution in galactic coordinate system and radial shells  in distance modulus, $\mu$.}\label{fig:Shells_mu}
\end{figure*}

\begin{figure*}[!ht]
\centering
\includegraphics[width=0.65\columnwidth]{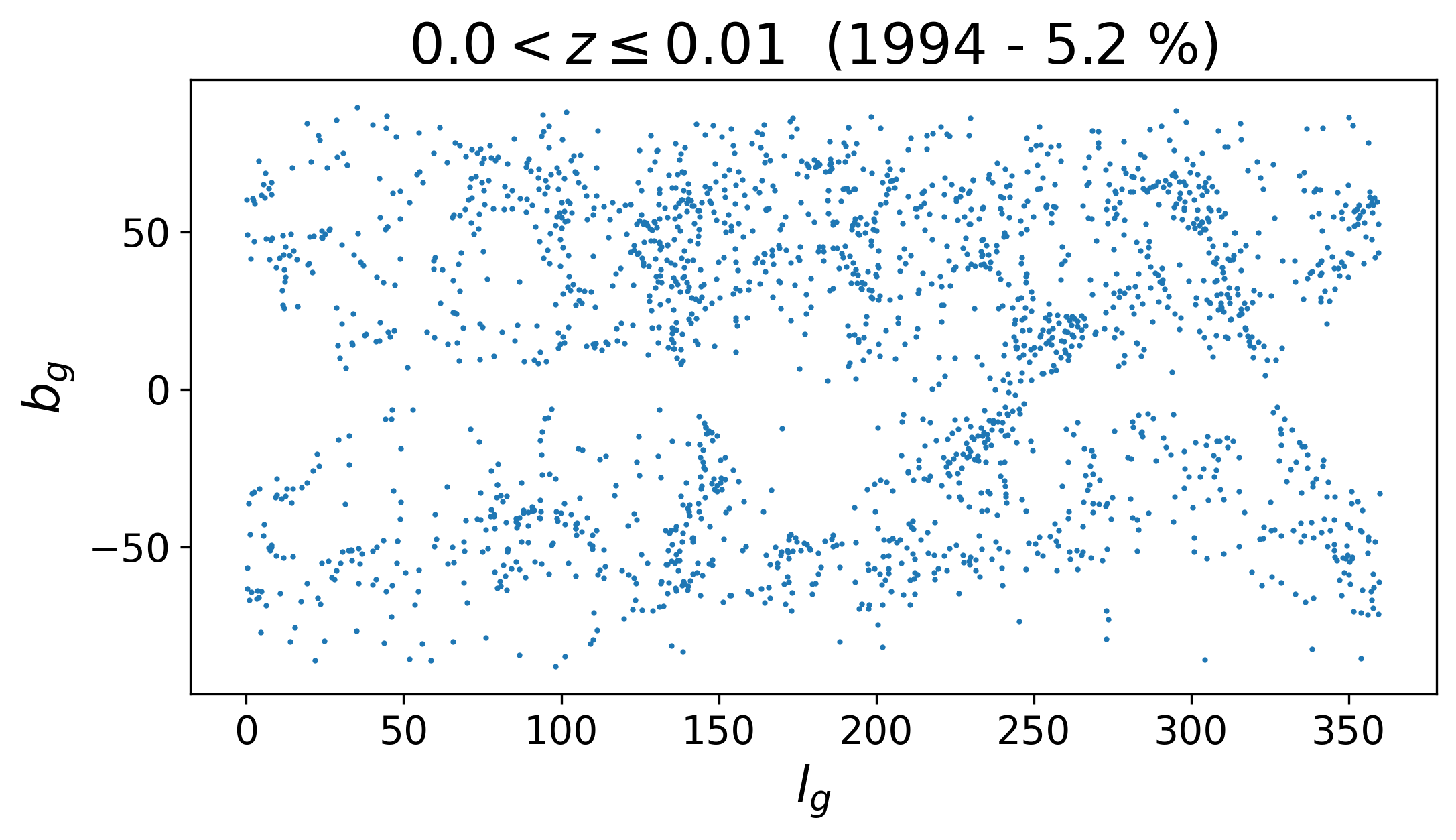}~~~
\includegraphics[width=0.65\columnwidth]{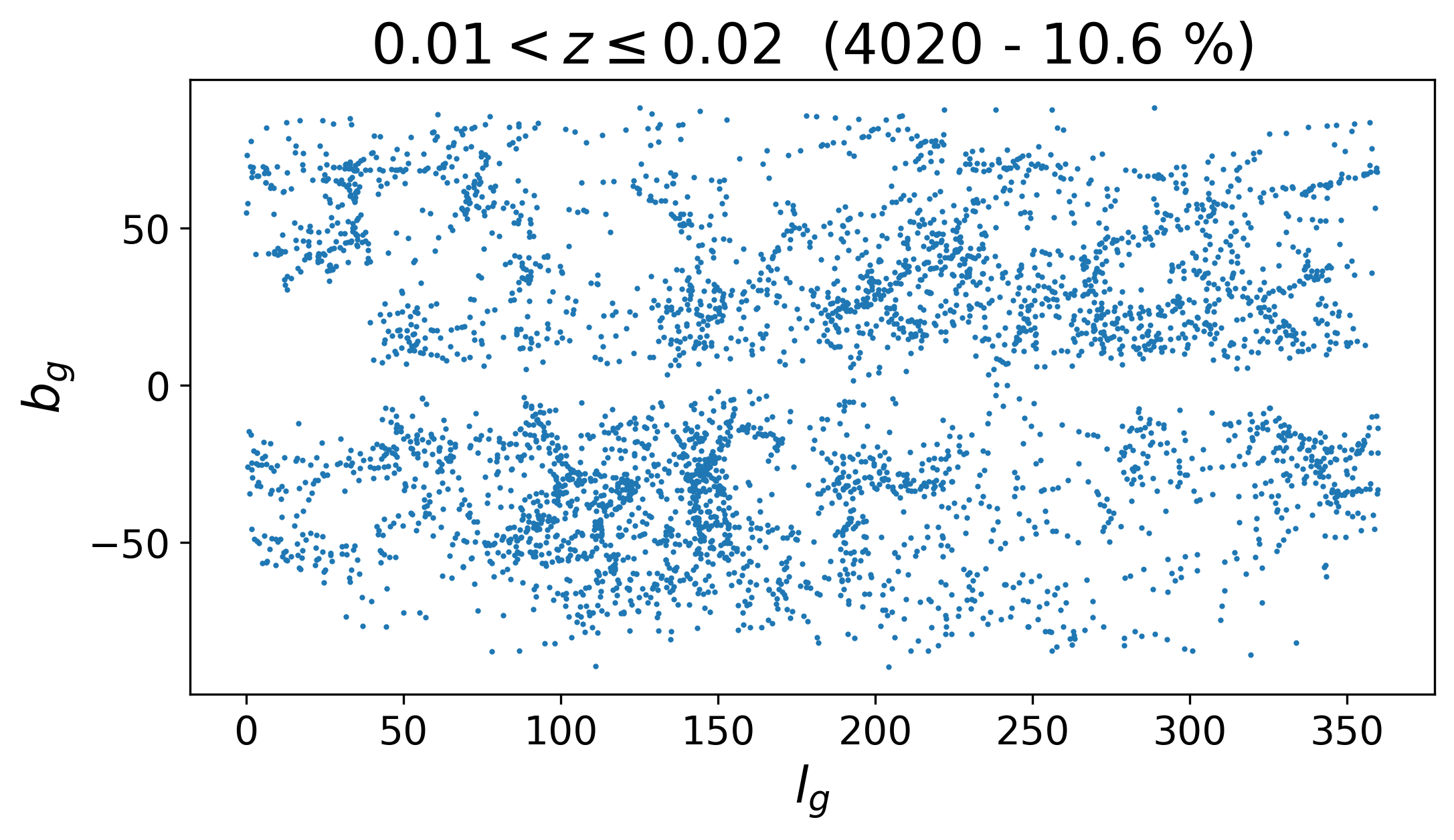}~~~
\includegraphics[width=0.65\columnwidth]{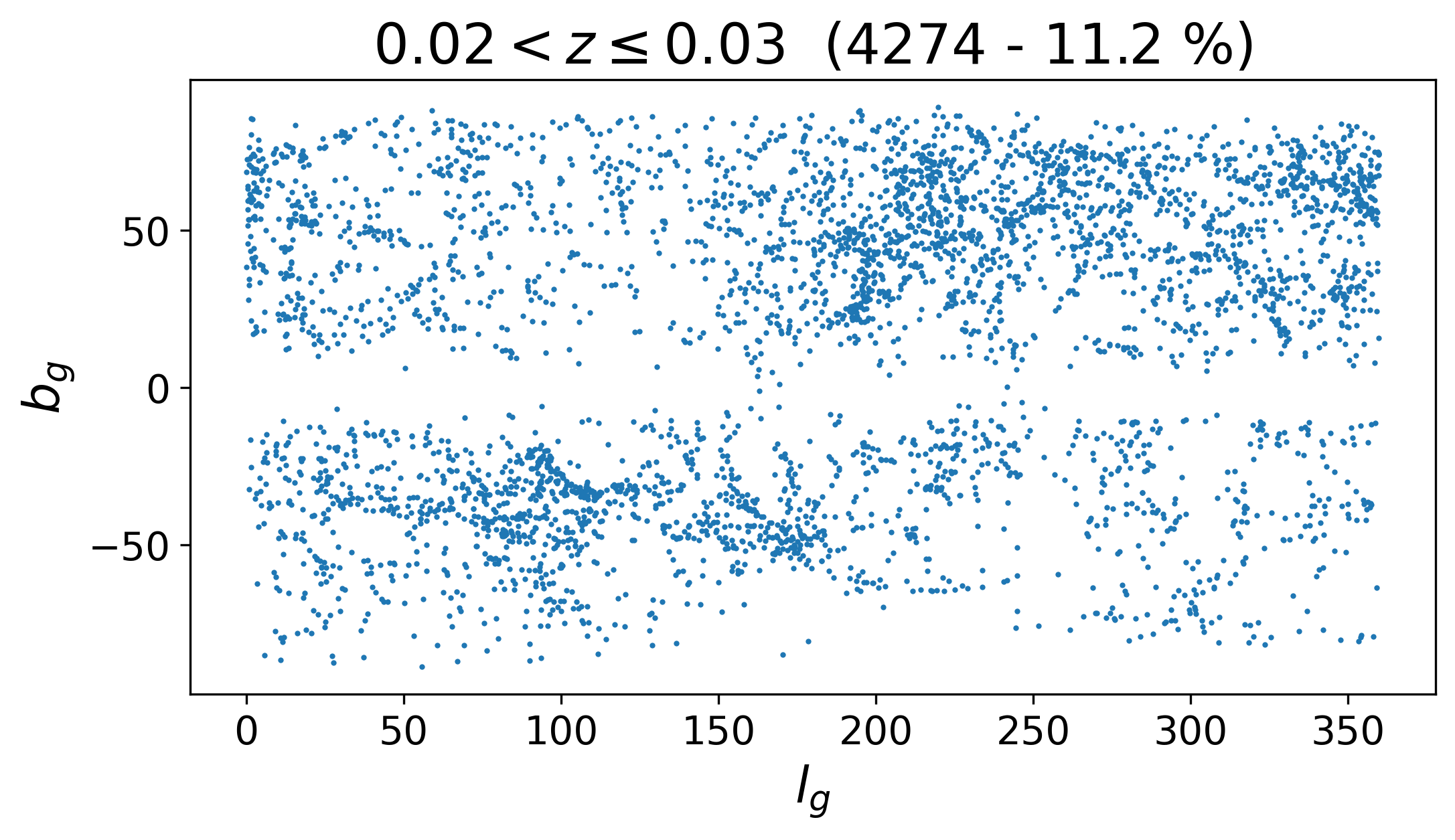}\\
~~~\\
\includegraphics[width=0.65\columnwidth]{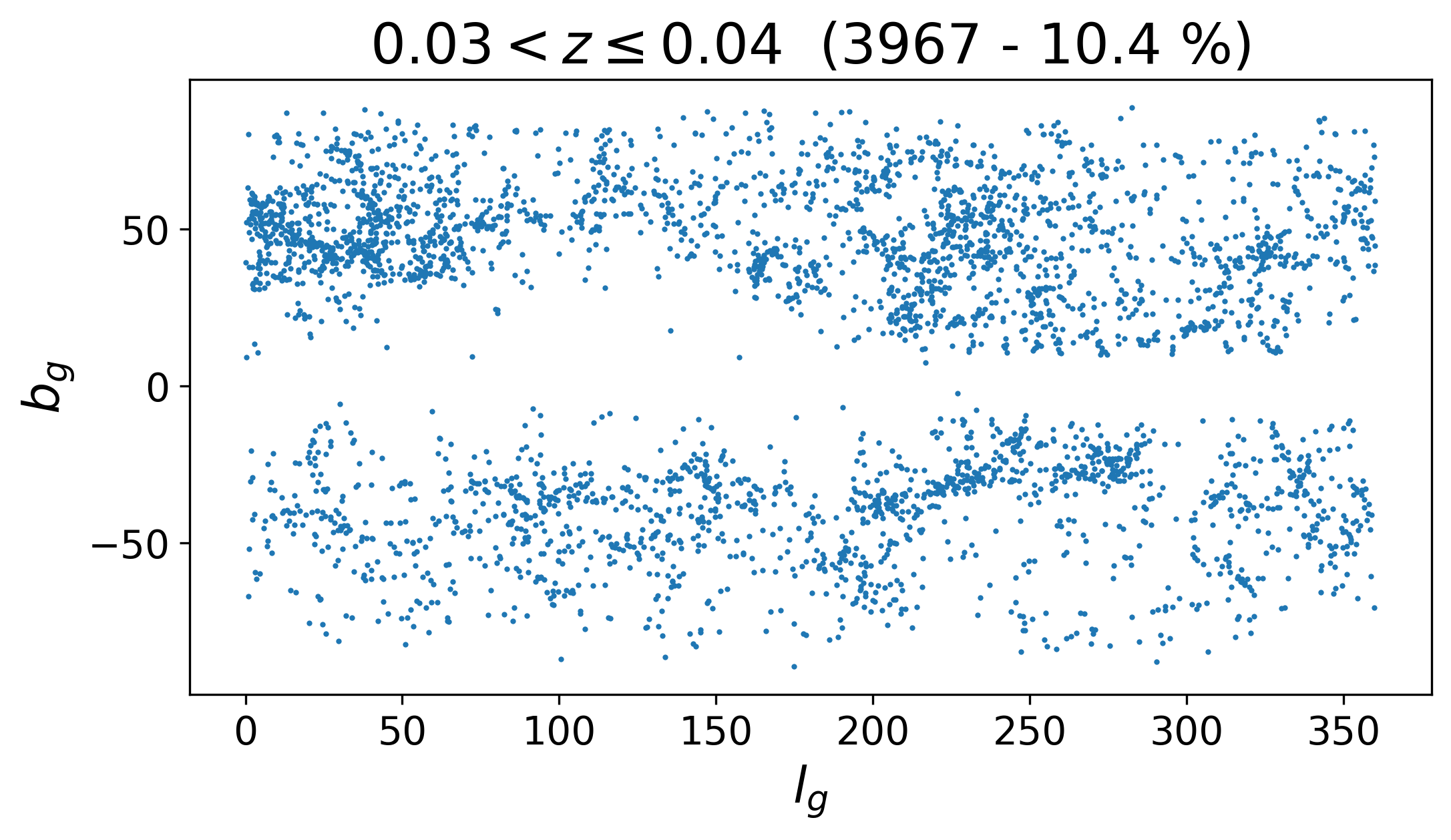}~~~
\includegraphics[width=0.65\columnwidth]{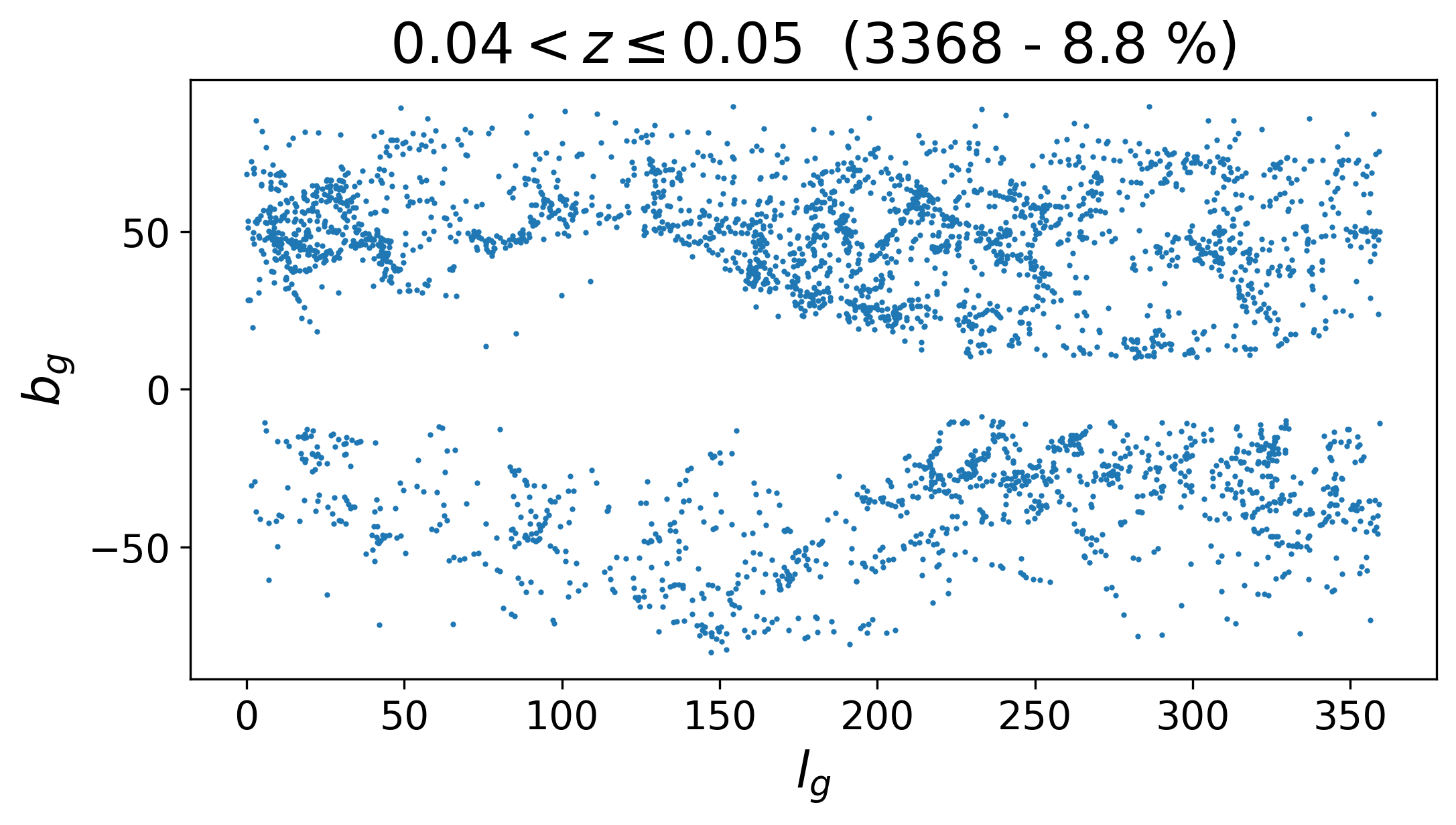}~~~
\includegraphics[width=0.65\columnwidth]{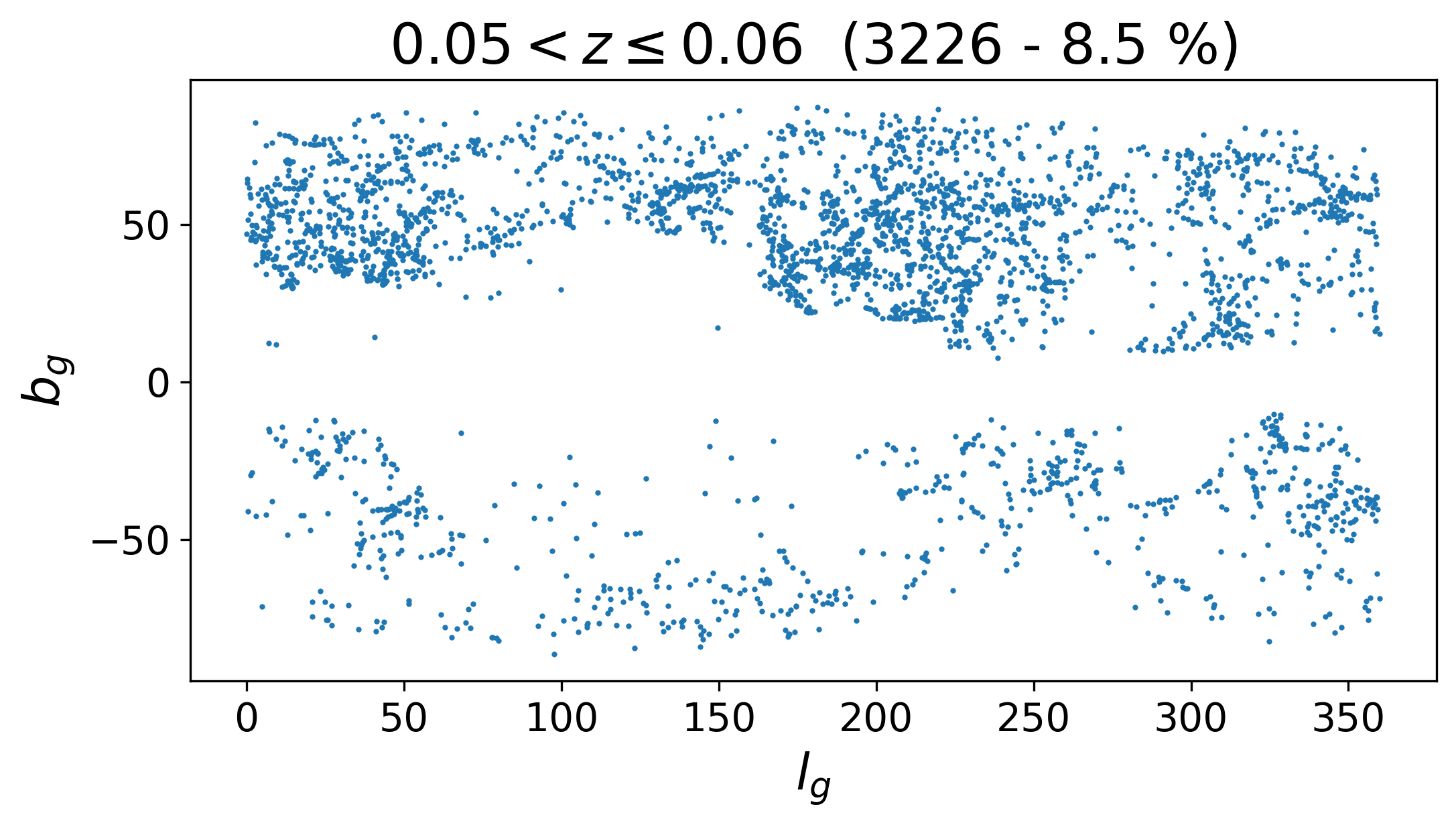}\\
~~~\\
\includegraphics[width=0.65\columnwidth]{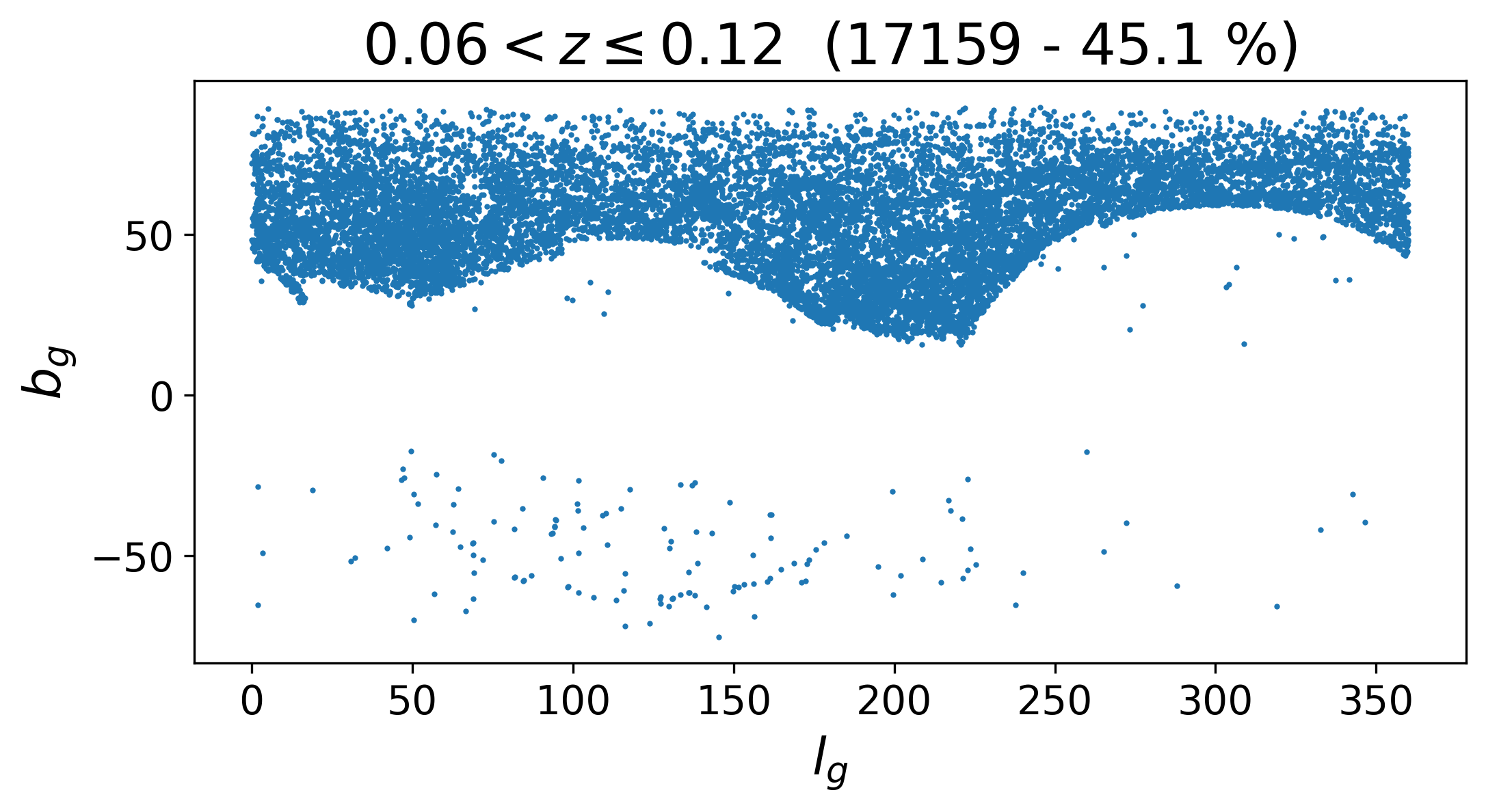}
\caption{CF4 angular distribution in galactic coordinate system and radial shells  in radial observed velocity, $v_{CMB}$.}\label{fig:Shells_v}
\end{figure*}

After this preliminary consistency check on the sample, we proceed now to elucidate the theoretical model we want to analyse. From the CF4 and CF4$_{pec}$ we can derive the $\log H_0$ for each object in the catalogue, using Eq.~(\ref{eq:Hubble_mucorr}) with $v^{obs}_{CMB}$ and $v^{obs}_{CMB} - v_{pec}$ when using CF4 and CF4$_{pec}$ respectively. Furthermore, we are interested in detecting a possible distance-related dependence of an anisotropy signal (if any) so that we are going to analyse not only the full consistent ranges delineated in the previous section, namely $\mu \in [31;36]$ and $z \in [0.03;0.06]$, but also separate radial shells. As stated in the previous section, the width of each radial shell was chosen to be roughly twice the mean uncertainty in the chosen distance variable of the objects it contains. The yielded shells are shown in Figs.~\ref{fig:Shells_mu} and \ref{fig:Shells_v} when using respectively, binning in the distance modulus and in the  redshift (radial velocity).

We fitted the maps of $\log H_0$ in each considered radial shell using a spherical harmonics expansion for $H_0$,
\begin{equation}\label{eq:spherical_harm}
H_0 = \sum^{\ell_{max}}_{\ell=0} \sum^{\ell}_{m=-l} a_{\ell m} Y_{\ell m}(\theta,\phi)\, ,
\end{equation}
where we truncated the expansion at the octupole ($\ell_{\max}=3$), $\theta$ and $\phi$ are the polar (co-latitude) and azimuthal angles, and $Y_{\ell m}$ are the real spherical harmonics defined from the complex $Y_{\ell}^{m}$ as\footnote{Different conventions exist in the literature, but constant or sign differences can always be absorbed into the coefficients $a_{\ell m}$.}:
\begin{equation}\label{eq:harmonics_def}
Y_{\ell m} = \begin{cases}
\dfrac{i}{\sqrt{2}} \left(Y_\ell^{m} - (-1)^m\, Y_\ell^{-m}\right) & \text{if}\ m < 0\\
Y_\ell^0 & \text{if}\ m=0\\
\dfrac{1}{\sqrt{2}} \left(Y_\ell^{-m} + (-1)^m\, Y_\ell^{m}\right) & \text{if}\ m > 0\, .
\end{cases}
\end{equation}
We chose to work in the galactic coordinate system. Thus, the polar angle $\theta$ corresponds to the angular distance between an object location and the dipole axis,
\begin{equation}\label{eq:cos}
\cos \theta =  \cos b_x \cos b_i \cos (l_x-l_i) + \sin b_x \sin b_i,\,
\end{equation}
where $(l_x,b_x)$ denote the Galactic coordinates of the dipole axis and $(l_i,b_i)$ those of the objects in the catalogue. However, as detailed in the next section, in our fiducial analysis, we assumed axial symmetry and retained only the $m=0$ terms, removing the dependence on $\phi$. 

We define the corresponding $\chi^2$ function per shell as
\begin{equation}
\chi^2 (\boldsymbol{p})
= \bm{\Delta \log H_{0}}\cdot \mathbf C^{-1}\cdot \bm{\Delta \log H_{0}}^{\mathsf T}
\end{equation}
where the vector $\bm{\Delta \log H_{0}} = \log H_{0,i} - \log H_{0} $, with $i = 1, \ldots, \mathcal{N}$, with $\mathcal{N}$ the number of objects in each shell; and $\boldsymbol{p} \equiv \{a_{00}, a_{10}, a_{20}, a_{30}, l_x, b_x\}$ is the parameter vector. We then explore the parameter space by using a Markov Chain Monte Carlo (MCMC) code implemented in \texttt{Wolfram Mathematica}, based on a sampling of $\sim2 \cdot 10^5$ (after cutting for burn-in and updating the covariance matrix every $\sim 2.5 \cdot 10^4$ steps), imposing only flat uninformative priors on the galactic coordinates of the dipole axis, namely $l_x \in [0^{\circ},360^{\circ})$ and $b_x \in [-90^{\circ},90^{\circ})$, and testing convergence using the method described in \cite{Dunkley:2004sv}. Finally, to assess the statistical relevance of any multipole-order model against a simple monopole signal (namely, constant $H_0$ within each shell), we also compute the Bayes factor $\mathcal{B}_{ij}$, where $\mathcal{M}_j$ denotes the reference model against which model $\mathcal{M}_i$ is tested. In our case, the reference model will be the monopole-only $H_0$ signal. The Bayes factor is defined as the ratio of the Bayesian evidences of the two models, 
$\mathcal{Z} = \int \mathcal{L}(\boldsymbol{p}) \, \pi (\boldsymbol{p}) \, \mathrm{d}\boldsymbol{p} \, ,$
where $\mathcal{L}$ is the likelihood and $\pi$ is the prior function. We evaluate the Bayesian evidence using our own \texttt{Wolfram Mathematica} implementation of the nested sampling algorithm introduced in \cite{Mukherjee:2005wg} onto the outputs of our MCMC runs. Because this algorithm is stochastic, we run it $100$ times to reduce statistical noise. This procedure yields a distribution of evidence values, from which we report the median value in our tables. Model selection is then assessed according to the Jeffreys scale \citep{Jeffreys:1939xee}: $\ln \mathcal{B}^{i}_{j} < 0$, evidence against model $\mathcal{M}_i$; $0 < \ln \mathcal{B}^{i}_{j} < 1$, inconclusive evidence; $1 < \ln \mathcal{B}^{i}_{j} < 2.5$, substantial evidence in favor of $\mathcal{M}_i$; $2.5 < \ln \mathcal{B}^{i}_{j} < 5$, strong evidence in favor of $\mathcal{M}_i$; and $\ln \mathcal{B}^{i}_{j} > 5$, decisive evidence in favor of $\mathcal{M}_i$.

{\renewcommand{\tabcolsep}{1.mm}
{\renewcommand{\arraystretch}{1.5}
\begin{table*}[!ht]
\begin{minipage}{0.9\textwidth}
\caption{Results from fitting the Hubble constant anisotropy signal with a multipole expansion. The galactic angular coordinates $(l^{dip}_g,b^{dip}_{g})$ of the (maximum of) dipole direction are in degrees; the Hubble constant $H_0$ (monopole) and the harmonic coefficients for the dipole, $a_{10}$, the quadrupole, $a_{20}$, and the octupole, $a_{30}$, are in km s$^{-1}$ Mpc$^{-1}$. We also report the values of the maximum $(H^{+}_{0})$ and of the minimum $(H^{-}_{0})$ in the signal, both from data (column 2) and from the model (column 9), the values of the minimum of the $\chi^2$ (and of the reduced $\chi^2$ for both scenarios between parenthesis), and the Bayes Factor, $\mathcal{M}_{i}$ and $\mathcal{M}_{j}$, where the monopole-only model is the reference scenario. Italic fonts indicate the values at the minimum $\chi^2$ when the posteriors are not Gaussian, noisy, or unconstrained.}
\label{tab:results_full}
\resizebox*{\textwidth}{!}{
\begin{tabular}{c cc c ccc c c }
\toprule

$\mu$ &  
$l^{dip}_g$ & $b^{dip}_g$ & $H_0$ &
$a_{10}$ & $a_{20}$ & $a_{30}$ & $\chi^2$ & $\ln \mathcal{B}^{i}_{j}$ \\

\midrule

\multicolumn{9}{c}{{\footnotesize{\textbf{CF4}}}} \\

\multirow{ 4}{*}{$[31; 36]$} & $-$ & $-$ & $76.46^{+0.09}_{-0.10}$ & $-$ & $-$ & $-$ & $15113\, (1.1)$ & $0$ \\

&  $299^{+2}_{-2}$ & $2^{+2}_{-2}$ & $76.74^{+0.10}_{-0.11}$ & $9.7^{+0.4}_{-0.4}$ & $-$ & $-$ & $14489\, (1)$ & $310$ \\

& $305^{+2}_{-2}$ & $10^{+2}_{-2}$ & $76.83^{+0.11}_{-0.11}$ & $9.7^{+0.4}_{-0.4}$ & $3.6^{+0.4}_{-0.4}$ & $-$ & $14411\, (1)$ & $350$ \\

& $303^{+2}_{-2}$ & $8^{+2}_{-2}$ & $76.89^{+0.11}_{-0.11}$ & $10.1^{+0.4}_{-0.5}$ & $3.6^{+0.4}_{-0.4}$ & $1.1^{+0.4}_{-0.4}$ & $14405\, (1)$ & $352$ \\

\addlinespace[4pt]

\multicolumn{9}{c}{{\footnotesize{\textbf{CF4$_{pec}$}}}} \\

\multirow{ 4}{*}{$[31; 36]$} & $-$ & $-$ & $76.45^{+0.10}_{-0.10}$ & $-$ & $-$ & $-$ & $12236\, (0.9)$ & $0$ \\

&  $282^{+9}_{-9}$ & $17^{+8}_{-7}$ & $76.44^{+0.11}_{-0.11}$ & $2.9^{+0.4}_{-0.4}$ & $-$ & $-$ & $12178\, (0.9)$ & $76$ \\

& $282^{+9}_{-9}$ & $19^{+8}_{-7}$ & $76.43^{+0.11}_{-0.10}$ & $2.7^{+0.4}_{-0.4}$ & $-0.56^{+0.37}_{-0.36}$ & $-$ & $12176\, (0.9)$ & $89$ \\

& $303^{+2}_{-2}$ & $8^{+2}_{-2}$ & $76.45^{+0.10}_{-0.11}$ & $2.8^{+0.4}_{-0.4}$ & $-0.52^{+0.40}_{-0.39}$ & $0.39^{+0.40}_{-0.42}$ & $12174\, (0.9)$ & $89$ \\

\hline

$z$ &  
$l^{dip}_g$ & $b^{dip}_g$ & $H_0$ &
$a_{10}$ & $a_{20}$ & $a_{30}$ & $\chi^2$ & $\ln \mathcal{B}^{i}_{j}$ \\

\midrule

\multicolumn{9}{c}{{\footnotesize{\textbf{CF4}}}} \\

\multirow{ 4}{*}{$[0.03; 0.06]$} & $-$ & $-$ & $75.49^{+0.12}_{-0.12}$ & $-$ & $-$ & $-$ & $10047\, (0.9)$ & $0$ \\

&  $299^{+4}_{-4}$ & $-14^{+3}_{-3}$ & $75.84^{+0.13}_{-0.13}$ & $6.6^{+0.5}_{-0.6}$ & $-$ & $-$ & $9893\, (0.9)$ & $28$ \\

& $296^{+4}_{-4}$ & $-10^{+3}_{-3}$ & $76.18^{+0.15}_{-0.14}$ & $6.6^{+0.6}_{-0.6}$ & $3.1^{+0.6}_{-0.6}$ & $-$ & $9866\, (0.9)$ & $29$ \\

& $296^{+4}_{-4}$ & $-10^{+5}_{-4}$ & $76.19^{+0.17}_{-0.18}$ & $6.5^{+0.7}_{-0.8}$ & $3.1^{+0.6}_{-0.6}$ & $-0.10^{+0.91}_{-1.01}$ & $9866\, (0.9)$ & $29$ \\

\addlinespace[4pt]

\multicolumn{9}{c}{{\footnotesize{\textbf{CF4$_{pec}$}}}} \\

\multirow{ 4}{*}{$[0.03; 0.06]$} & $-$ & $-$ & $75.16^{+0.12}_{-0.12}$ & $-$ & $-$ & $-$ &  $9371\, (0.9)$ & $0$ \\

&  $291^{+9}_{-9}$ & $-14^{+6}_{-7}$ & $75.33^{+0.14}_{-0.14}$ & $3.4^{+0.6}_{-0.6}$ & $-$ & $-$ & $9332\, (0.9)$ & $18$ \\

& $299^{+8}_{-8}$ & $-13^{+6}_{-6}$ & $75.57^{+0.15}_{-0.15}$ & $3.4^{+0.6}_{-0.6}$ & $1.9^{+0.6}_{-0.6}$ & $-$ & $9322\, (0.9)$ & $23$ \\

& $303^{+7}_{-9}$ & $-17^{+9}_{-5}$ & $75.64^{+0.17}_{-0.20}$ & $3.0^{+0.7}_{-0.7}$ & $1.9^{+0.6}_{-0.6}$ & $-0.99^{+0.73}_{-1.26}$ & $9319\, (0.9)$ & $24$ \\

\bottomrule

\end{tabular}}
\end{minipage}
\end{table*}}}

{\renewcommand{\tabcolsep}{1.mm}
{\renewcommand{\arraystretch}{1.5}
\begin{table*}[!ht]
\begin{minipage}{0.9\textwidth}
\caption{Same as Table~\ref{tab:results_full} but for single radial shells in $\mu$ using the CF4 catalogue.}
\label{tab:results_CF4_mu}
\resizebox*{\textwidth}{!}{
\begin{tabular}{c cc c ccc c c }

\toprule

$\mu$ &  
$l^{dip}_g$ & $b^{dip}_g$ & $H_0$ &
$a_{10}$ & $a_{20}$ & $a_{30}$ & $\chi^2$ & $\ln \mathcal{B}^{i}_{j}$ \\

\midrule

\multicolumn{9}{c}{{\footnotesize{\textbf{CF4}}}} \\

\multirow{ 4}{*}{$[31; 32]$} & $-$ & $-$ & $76.23^{+0.58}_{-0.56}$ & $-$ & $-$ & $-$ & $1094\, (2.0)$ & $0$ \\

&  $319^{+5}_{-5}$ & $16^{+3}_{-3}$ & $74.40^{+0.47}_{-0.46}$ & $25^{+2}_{-2}$ & $-$ & $-$ & $877\, (1.6)$ & $107$ \\

& $316^{+3}_{-4}$ & $13^{+2}_{-2}$ & $76.26^{+0.57}_{-0.57}$ & $25^{+2}_{-2}$ & $21^{+2}_{-2}$ & $-$ & $755\, (1.4)$ & $168$ \\

& $316^{+4}_{-3}$ & $13^{+2}_{-2}$ & $76.23^{+0.58}_{-0.56}$ & $25^{+2}_{-2}$ & $22^{+2}_{-2}$ & $0.7^{+2.1}_{-2.1}$ & $754\, (1.4)$ & $168$ \\

\hline

\multirow{ 4}{*}{$[32; 33]$} & $-$ & $-$ & $76.71^{+0.35}_{-0.35}$ & $-$ & $-$ & $-$ & $1914\, (1.8)$ & $0$ \\

& $295^{+4}_{-4}$ & $-4^{+3}_{-3}$ & $75.93^{+0.37}_{-0.36}$ & $21^{+1}_{-1}$ & $-$ & $-$ & $1661\, (1.5)$ & $125$ \\

& $305^{+2}_{-3}$ & $11^{+2}_{-2}$ & $75.94^{+0.39}_{-0.37}$ & $16^{+1}_{-1}$ & $18^{+1}_{-1}$ & $-$ & $1498\, (1.4)$ & $207$ \\

& $305^{+3}_{-3}$ & $11^{+2}_{-2}$ & $75.96^{+0.38}_{-0.38}$ & $16^{+1}_{-1}$ & $19^{+1}_{-1}$ & $1.3^{+1.4}_{-1.4}$ & $1498\, (1.4)$ & $207$ \\

\hline

\multirow{ 4}{*}{$[33; 34]$} & $-$ & $-$ & $77.99^{+0.26}_{-0.25}$ & $-$ & $-$ & $-$ & $2552\, (1.2)$ & $0$ \\

& $288^{+10}_{-10}$ & $-26^{+7}_{-7}$ & $78.22^{+0.26}_{-0.26}$ & $6.5^{+1.0}_{-1.0}$ & $-$ & $-$ & $2508\, (1.2)$ & $21$ \\

& $337^{+3}_{-3}$ & $27^{+2}_{-2}$ & $77.64^{+0.27}_{-0.28}$ & $4.9^{+1.0}_{-1.0}$ & $12^{+1}_{-1}$ & $-$ & $2357\, (1.1)$ & $96$ \\

& $334^{+3}_{-3}$ & $26^{+2}_{-2}$ & $77.55^{+0.27}_{-0.27}$ & $5.1^{+0.9}_{-0.9}$ & $12^{+1}_{-1}$ & $3.8^{+0.9}_{-0.9}$ & $2338\, (1.1)$ & $105$ \\

\hline

\multirow{ 4}{*}{$[34; 35]$} & $-$ & $-$ & $76.43^{+0.17}_{-0.17}$ & $-$ & $-$ & $-$ & $4129\, (1.0)$ & $0$ \\

& $303^{+6}_{-6}$ & $21^{+5}_{-4}$ & $76.75^{+0.19}_{-0.19}$ & $8.2^{+0.7}_{-0.7}$ & $-$ & $-$ & $3933\, (0.9)$ & $97$ \\

& $303^{+6}_{-6}$ & $21^{+6}_{-6}$ & $76.74^{+0.19}_{-0.19}$ & $8.2^{+0.7}_{-0.7}$ & $0.15^{+0.86}_{-0.87}$ & $-$ & $3932\, (0.9)$ & $96$ \\

& $302^{+6}_{-5}$ & $20^{+6}_{-5}$ & $76.68^{+0.20}_{-0.19}$ & $8.2^{+0.7}_{-0.7}$ & $0.40^{+0.84}_{-0.87}$ & $0.75^{+0.63}_{-0.64}$ & $3931\, (0.9)$ & $97$ \\

\hline

\multirow{ 4}{*}{$[35; 36]$} & $-$ & $-$ & $76.07^{+0.15}_{-0.15}$ & $-$ & $-$ & $-$ & $5366\, (1.0)$ & $0$ \\

& $299^{+4}_{-4}$ & $-2^{+3}_{-3}$ & $76.40^{+0.16}_{-0.16}$ & $9.5^{+0.7}_{-0.7}$ & $-$ & $-$ & $5170\, (0.9)$ & $97$ \\

& $297^{+4}_{-4}$ & $-5^{+3}_{-4}$ & $76.62^{+0.23}_{-0.22}$ & $9.6^{+0.7}_{-0.7}$ & $1.4^{+1.0}_{-0.9}$ & $-$ & $5167\, (0.9)$ & $98$ \\

& $302^{+6}_{-5}$ & $20^{+6}_{-5}$ & $76.41^{+0.18}_{-0.17}$ & $11^{+1}_{-1}$ & $1.1^{+0.7}_{-0.7}$ & $4.3^{+0.8}_{-0.8}$ & $5138\, (0.9)$ & $112$ \\

\bottomrule

\end{tabular}}
\end{minipage}
\end{table*}}}

{\renewcommand{\tabcolsep}{1.mm}
{\renewcommand{\arraystretch}{1.5}
\begin{table*}[!ht]
\begin{minipage}{0.9\textwidth}
\caption{Same as Table~\ref{tab:results_full} but for single radial shells in $\mu$ using the CF4$_{pec}$ catalogue.}
\label{tab:results_CF4pec_mu}
\resizebox*{\textwidth}{!}{
\begin{tabular}{c cc c ccc c c }

\toprule

$\mu$ &  
$l^{dip}_g$ & $b^{dip}_g$ & $H_0$ &
$a_{10}$ & $a_{20}$ & $a_{30}$ & $\chi^2$ & $\ln \mathcal{B}^{i}_{j}$ \\

\midrule

\multicolumn{9}{c}{{\footnotesize{\textbf{CF4$_{pec}$}}}} \\

\multirow{ 4}{*}{$[31; 32]$} & $-$ & $-$ & $78.36^{+0.53}_{-0.53}$ & $-$ & $-$ & $-$ & $461\, (0.9)$ & $0$ \\

&  $\mathit{63}$ & $\mathit{-7}$ & $78.59^{+0.54}_{-0.54}$ & $-0.6^{+4.9}_{-4.5}$ & $-$ & $-$ & $455\, (0.8)$ & $1.5$ \\

& $\mathit{57}$ & $\mathit{-10}$ & $79.12^{+0.68}_{-0.68}$ & $0.6^{+6.8}_{-5.9}$ & $3.6^{+2.8}_{-4.0}$ & $-$ & $451\, (0.8)$ & $3.0$ \\

& $\mathit{46}$ & $\mathit{-8}$ & $79.38^{+0.74}_{-0.67}$ & $0.7^{+4.9}_{-4.6}$ & $3.4^{+3.0}_{-4.1}$ & $-1.6^{+3.9}_{-6.7}$ & $448\, (0.8)$ & $4.5$ \\

\hline

\multirow{ 4}{*}{$[32; 33]$} & $-$ & $-$ & $77.26^{+0.38}_{-0.38}$ & $-$ & $-$ & $-$ & $1070\, (1.0)$ & $0$ \\

& $\mathit{232}$ & $\mathit{-15}$ & $77.25^{+0.39}_{-0.40}$ & $0.2^{+1.9}_{-1.9}$ & $-$ & $-$ & $1068\, (1.0)$ & $0.09$ \\

& $\mathit{66}$ & $\mathit{-7}$ & $77.37^{+0.44}_{-0.40}$ & $0.02^{+1.65}_{-1.62}$ & $-0.2^{+3.3}_{-3.1}$ & $-$ & $1064\, (1.0)$ & $1.5$ \\

& $\mathit{74}$ & $\mathit{-7}$ & $77.44^{+0.43}_{-0.42}$ & $0.07^{+1.79}_{-1.75}$ & $1.0^{+3.3}_{-4.1}$ & $-0.04^{+2.01}_{-2.13}$ & $1062\, (1.0)$ & $1.9$ \\

\hline

\multirow{ 4}{*}{$[33; 34]$} & $-$ & $-$ & $78.00^{+0.27}_{-0.27}$ & $-$ & $-$ & $-$ & $2002\, (0.9)$ & $0$ \\

& $\mathit{232}$ & $\mathit{-32}$ & $78.00^{+0.28}_{-0.28}$ & $0.03^{+1.10}_{-1.08}$ & $-$ & $-$ & $2002\, (0.9)$ & $-0.2$ \\

& $\mathit{163}$ & $\mathit{-51}$ & $77.95^{+0.29}_{-0.29}$ & $0.08^{+1.10}_{-1.18}$ & $0.12^{+1.41}_{-1.49}$ & $-$ & $1999\, (0.9)$ & $0.002$ \\

& $\mathit{297}$ & $\mathit{-37}$ & $78.03^{+0.29}_{-0.31}$ & $0.1^{+1.2}_{-1.2}$ & $0.03^{+1.22}_{-1.29}$ & $0.5^{+2.5}_{-3.3}$ & $1993\, (1.0)$ & $2.8$ \\

\hline

\multirow{ 4}{*}{$[34; 35]$} & $-$ & $-$ & $76.21^{+0.18}_{-0.18}$ & $-$ & $-$ & $-$ & $3685\, (0.9)$ & $0$ \\

& $294^{+21}_{-21}$ & $44^{+17}_{-15}$ & $76.21^{+0.19}_{-0.19}$ & $2.9^{+0.6}_{-0.6}$ & $-$ & $-$ & $3657\, (0.9)$ & $13$ \\

& $297^{+21}_{-24}$ & $41^{+20}_{-16}$ & $76.22^{+0.21}_{-0.23}$ & $2.8^{+0.6}_{-0.6}$ & $-0.17^{+0.78}_{-0.76}$ & $-$ & $3656\, (0.9)$ & $13$ \\

& $291^{+22}_{-22}$ & $46^{+16}_{-16}$ & $76.18^{+0.23}_{-0.21}$ & $3.0^{+0.6}_{-0.6}$ & $-0.06^{+0.72}_{-0.77}$ & $-0.70^{+0.67}_{-0.72}$ & $3654\, (0.9)$ & $13$ \\

\hline

\multirow{ 4}{*}{$[35; 36]$} & $-$ & $-$ & $75.79^{+0.15}_{-0.15}$ & $-$ & $-$ & $-$ & $4947\, (0.8)$ & $0$ \\

& $290^{+7}_{-8}$ & $15^{+7}_{-6}$ & $776.62^{+0.23}_{-0.22}$ & $4.9^{+0.7}_{-0.7}$ & $-$ & $-$ & $4884\, (0.8)$ & $30$ \\

& $289^{+8}_{-8}$ & $19^{+6}_{-6}$ & $75.64^{+0.16}_{-0.16}$ & $4.3^{+0.7}_{-0.7}$ & $-1.8^{+0.6}_{-0.6}$ & $-$ & $4875\, (0.8)$ & $35$ \\

& $290^{+7}_{-7}$ & $13^{+6}_{-5}$ & $75.71^{+0.16}_{-0.16}$ & $5.0^{+0.8}_{-0.8}$ & $-1.6^{+0.6}_{-0.6}$ & $1.7^{+0.7}_{-0.8}$ & $4868\, (0.9)$ & $38$ \\

\bottomrule

\end{tabular}}
\end{minipage}
\end{table*}}}

{\renewcommand{\tabcolsep}{1.mm}
{\renewcommand{\arraystretch}{2}
\begin{table*}[!ht]
\begin{minipage}{0.9\textwidth}
\caption{Same as Table~\ref{tab:results_full} but for single radial shells in $z$ using the CF4 catalogue.}
\label{tab:results_CF4_z}
\resizebox*{\textwidth}{!}{
\begin{tabular}{c cc c ccc c c }

\toprule

$z$ &
$l^{dip}_g$ & $b^{dip}_g$ & $H_0$ &
$a_{10}$ & $a_{20}$ & $a_{30}$ & $\chi^2$ & $\ln \mathcal{B}^{i}_{j}$ \\

\midrule

\multicolumn{9}{c}{{\footnotesize{\textbf{CF4}}}} \\

\multirow{ 4}{*}{$[0.03; 0.04]$} & $-$ & $-$ & $75.36^{+0.18}_{-0.18}$ & $-$ & $-$ & $-$ & $3985\, (0.97)$ & $0$ \\

& $306^{+5}_{-5}$ & $-15^{+4}_{-4}$ & $75.75^{+0.19}_{-0.19}$ & $7.8^{+0.8}_{-0.9}$ & $-$ & $-$ & $3893\, (0.95)$ & $45$ \\

& $294^{+5}_{-5}$ & $-9^{+3}_{-4}$ & $76.25^{+0.21}_{-0.22}$ & $7.3^{+0.9}_{-0.9}$ & $4.8^{+0.9}_{-1.0}$ & $-$ & $3869\, (0.9)$ & $56$ \\

& $294^{+5}_{-5}$ & $-12^{+5}_{-4}$ & $76.34^{+0.24}_{-0.25}$ & $6.5^{+1.2}_{-1.2}$ & $4.8^{+1.0}_{-1.0}$ & $-1.4^{+1.3}_{-1.6}$ & $3868\, (0.9)$ & $57$ \\

\hline

\multirow{ 4}{*}{$[0.04; 0.05]$} & $-$ & $-$ & $75.51^{+0.21}_{-0.21}$ & $-$ & $-$ & $-$ & $1070\, (0.9)$ & $0$ \\

& $299^{+8}_{-8}$ & $-21^{+6}_{-6}$ & $75.87^{+0.23}_{-0.23}$ & $6.4^{+0.9}_{-0.9}$ & $-$ & $-$ & $3263\, (0.9)$ & $27$ \\

& $299^{+8}_{-7}$ & $-19^{+7}_{-7}$ & $75.97^{+0.23}_{-0.25}$ & $6.5^{+0.9}_{-0.9}$ & $1.2^{+1.0}_{-1.0}$ & $-$ & $3261\, (0.9)$ & $27$ \\

& $298^{+8}_{-8}$ & $-18^{+7}_{-7}$ & $75.96^{+0.25}_{-0.25}$ & $6.4^{+1.1}_{-1.1}$ & $1.3^{+1.0}_{-1.0}$ & $-0.25^{+1.09}_{-1.10}$ & $3262\, (0.9)$ & $27$ \\

\hline

\multirow{ 4}{*}{$[0.05; 0.06]$} & $-$ & $-$ & $75.66^{+0.22}_{-0.22}$ & $-$ & $-$ & $-$ & $2742\, (0.9)$ & $0$ \\

& $283^{+11}_{-11}$ & $-7^{+8}_{-10}$ & $75.93^{+0.29}_{-0.29}$ & $4.8^{+1.2}_{-1.2}$ & $-$ & $-$ & $2722\, (0.9)$ & $9$ \\

& $289^{+10}_{-11}$ & $-4^{+11}_{-10}$ & $76.16^{+0.38}_{-0.38}$ & $4.8^{+1.3}_{-1.3}$ & $1.9^{+1.4}_{-1.4}$ & $-$ & $2721\, (0.9)$ & $9$ \\

& $295^{+9}_{-10}$ & $2^{+9}_{-9}$ & $76.04^{+0.36}_{-0.36}$ & $6.3^{+1.9}_{-1.9}$ & $1.9^{+1.3}_{-1.4}$ & $2.1^{+1.5}_{-1.6}$ & $2718\, (0.9)$ & $10$ \\

\bottomrule

\end{tabular}}
\end{minipage}
\end{table*}}}

{\renewcommand{\tabcolsep}{1.mm}
{\renewcommand{\arraystretch}{2}
\begin{table*}[!ht]
\begin{minipage}{0.9\textwidth}
\caption{Same as Table~\ref{tab:results_full} but for single radial shells in $\mu$ using the CF4$_{pec}$ catalogue.}
\label{tab:results_CF4pec_z}
\resizebox*{\textwidth}{!}{
\begin{tabular}{c cc c ccc c c }

\toprule

$z$ &
$l^{dip}_g$ & $b^{dip}_g$ & $H_0$ &
$a_{10}$ & $a_{20}$ & $a_{30}$ & $\chi^2$ & $\ln \mathcal{B}^{i}_{j}$ \\

\midrule

\multicolumn{9}{c}{{\footnotesize{\textbf{CF4$_{pec}$}}}} \\

\multirow{ 4}{*}{$[0.03; 0.04]$} & $-$ & $-$ & $74.86^{+0.19}_{-0.19}$ & $-$ & $-$ & $-$ & $3576\, (0.9)$ & $0$ \\

& $305^{+14}_{-16}$ & $-11^{+11}_{-12}$ & $74.97^{+0.18}_{-0.20}$ & $2.9^{+0.9}_{-1.0}$ & $-$ & $-$ & $3562\, (0.9)$ & $6$ \\

& $304^{+13}_{-12}$ & $-15^{+10}_{-9}$ & $75.21^{+0.23}_{-0.23}$ & $2.9^{+0.9}_{-1.0}$ & $2.0^{+1.0}_{-1.0}$ & $-$ & $3555\, (0.9)$ & $9$ \\

& $117^{+8}_{-8}$ & $24^{+6}_{-5}$ & $75.43^{+0.22}_{-0.25}$ & $-1.7^{+0.9}_{-1.0}$ & $2.2^{+0.9}_{-0.9}$ & $2.6^{+1.0}_{-1.0}$ & $3547\, (0.9)$ & $13$ \\

\hline

\multirow{ 4}{*}{$[0.04; 0.05]$} & $-$ & $-$ & $75.40^{+0.21}_{-0.21}$ & $-$ & $-$ & $-$ & $3105\, (0.9)$ & $0$ \\

& $301^{+15}_{-16}$ & $-22^{+12}_{-14}$ & $75.60^{+0.25}_{-0.23}$ & $3.4^{+1.0}_{-1.0}$ & $-$ & $-$ & $3089\, (0.8)$ & $7$ \\

& $299^{+14}_{-15}$ & $-20^{+15}_{-15}$ & $75.67^{+0.25}_{-0.22}$ & $3.4^{+1.1}_{-1.0}$ & $0.8^{+1.2}_{-1.1}$ & $-$ & $3088\, (0.8)$ & $7$ \\

& $133^{+13}_{-21}$ & $8^{+13}_{-13}$ & $75.75^{+0.24}_{-0.28}$ & $-2.3^{+2.1}_{-1.6}$ & $1.3^{+1.4}_{-1.4}$ & $1.1^{+1.6}_{-2.5}$ & $3085\, (0.8)$ & $8$ \\

\hline

\multirow{ 4}{*}{$[0.05; 0.06]$} & $-$ & $-$ & $75.33^{+0.23}_{-0.23}$ & $-$ & $-$ & $-$ & $2685\, (0.9)$ & $0$ \\

& $260^{+16}_{-18}$ & $-15^{+13}_{-13}$ & $75.79^{+0.40}_{-0.39}$ & $3.8^{+1.1}_{-1.1}$ & $-$ & $-$ & $2669\, (0.9)$ & $7$ \\

& $259^{+39}_{-25}$ & $-19^{+13}_{-15}$ & $75.79^{+0.40}_{-0.39}$ & $3.2^{+1.6}_{-1.7}$ & $-0.5^{+3.3}_{-1.6}$ & $-$ & $2669\, (0.9)$ & $7$ \\

& $287^{+23}_{-28}$ & $-8^{+10}_{-12}$ & $75.84^{+0.42}_{-0.44}$ & $5.2^{+2.3}_{-2.1}$ & $1.7^{+2.6}_{-2.5}$ & $1.8^{+1.7}_{-1.6}$ & $2666\, (0.9)$ & $7$ \\

\bottomrule

\end{tabular}}
\end{minipage}
\end{table*}}}

\section{Results}
\label{sec:results}

The results of our analysis are shown in Table~\ref{tab:results_full} for the full ranges, both in distance modulus $\mu$ and in redshift $z$ for both CF4 and CF4$_{pec}$ samples. For each sample, the table shows the results obtained by varying the multipole order of Eq.~(\ref{eq:spherical_harm}) from the monopole-only to the full octupole case. The following quantities are reported: the direction of the maximal signal in the anisotropy; the multipole moments up to the third order; and the Bayes factor, which is always calculated with respect to the corresponding monopole-only scenario.

The first comment is related to the choice to truncate the harmonic expansion at the octupole ($\ell = 3$). As shown in Table~\ref{tab:results_full}, the octupole contributes much less to the global statistical significance of the fit with respect to the dipole and quadrupole, being in most of the cases consistent with zero at $1\sigma$, thus providing no additional useful information. Furthermore, as expected, we verified that adding higher-order terms does not affect the estimation of lower-order components, further justifying the cut for higher multipoles in the expansion. Moreover, although not reported in the table, we have also examined the decision to consider only axially symmetric components ($m=0$). When $m \neq 0$ is included, the MCMC can become trapped very often in a local minimum with large $m\neq0$ and low $m=0$. Only running very long chains we consistently recover $m=0$ dominance with negligible contribution from the $m\neq0$ components. Since the $m\neq0$ components do not improve the fit, the azimuthally symmetric case is adopted as the standard scenario. {An equivalent conclusion was also drawn in \cite{Kalbouneh:2022tfw}.}

In the case of the CF4 data, we can see how the dipole direction varies slightly when we consider the cut in $\mu$ or in $z$: the longitude is strongly constrained to $l_g$ in the range $[292^{\circ}, 307^{\circ}]$ at the $1\sigma$ level, while larger variations affect the latitude. In the case of $\mu$, the latitude lies in the northern hemisphere, while in the case of $z$, it lies in the southern hemisphere. It is important to note that the data distribution differs in the two cases, as illustrated in Fig.~\ref{fig:HL_plot}. This variation could potentially be attributed to the distinct surveys from which the data were collected. 

The monopole, $H_0$, is slightly larger in the case of $\mu$ shells, but the difference is not statistically significant. Focusing on higher multipoles reveals that the dipole is much stronger in the $\mu$ shells than in the $z$ shells, the quadrupole is equivalent, and the octupole is practically absent in the $z$ shells. Nevertheless, its statistical weight is negligible in both cases. Looking at the values of the $\chi^2$ and the Bayes factor, bearing in mind that the absence of a covariance matrix may underestimate the errors and thus increase the former, we can conclude that the anisotropy signal is strongly favoured statistically in both cases compared to a constant Hubble constant scenario. The preference for directional variation in the Hubble constant cannot be attributed solely to cosmic variance; it must be physically based.

In Table~\ref{tab:results_CF4_mu}, we present the results of considering shells in $\mu$, with the aim of detecting possible variations in the dipole distance, as theorised by models such as those behind Eq.~(\ref{eq:h_angle}) in the Introduction. Examining the multipoles reveals that the octupole is subdominant to the dipole and quadrupole in most cases, except for shells $[33;34]$ and $[35;36]$, where it is comparable to and larger than the dipole and quadrupole, respectively. Examining the values of the Bayes Factor reveals that, except for those two shells, its contribution does not statistically improve the fit. The quadrupole is as large as the dipole for $\mu < 34$, while at larger distances it is highly suppressed, being consistent with zero at $1-2\sigma$. We can also observe that adding the quadrupole dramatically alters the main anisotropy direction, even changing hemisphere in latitude in the second, third, and fifth shells. The longitude, however, is much more stable, with only the third shell showing slightly more variation. Once again, as in the general case, the Bayes factors indicate that the anisotropy signal is strong and that a simple monopole fit is highly unfavourable compared to a more complicated model involving at least a dipole and a quadrupole.

\begin{figure*}[!ht]
\centering
\includegraphics[width=8.5cm]{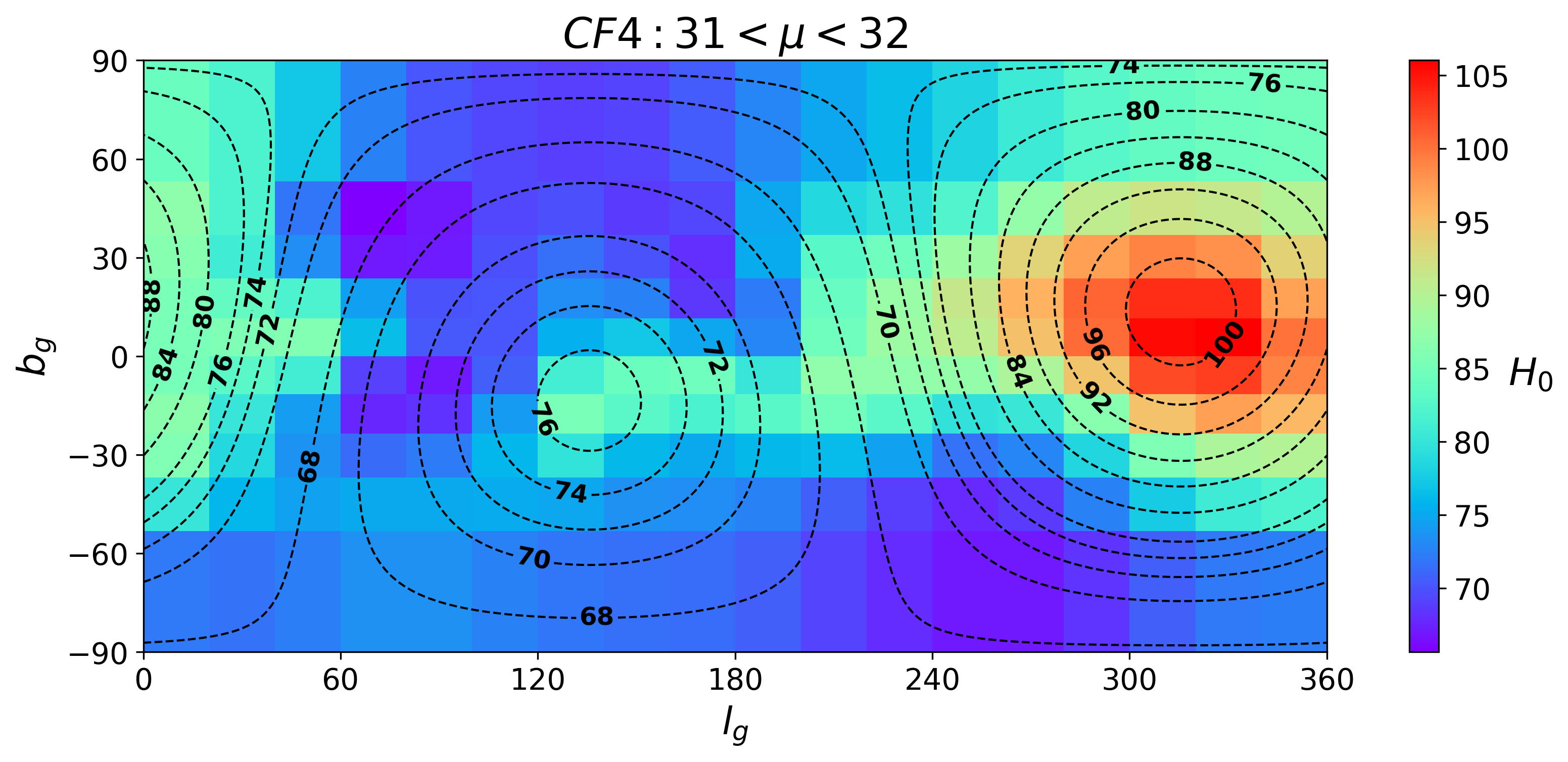}~
\includegraphics[width=8.5cm]{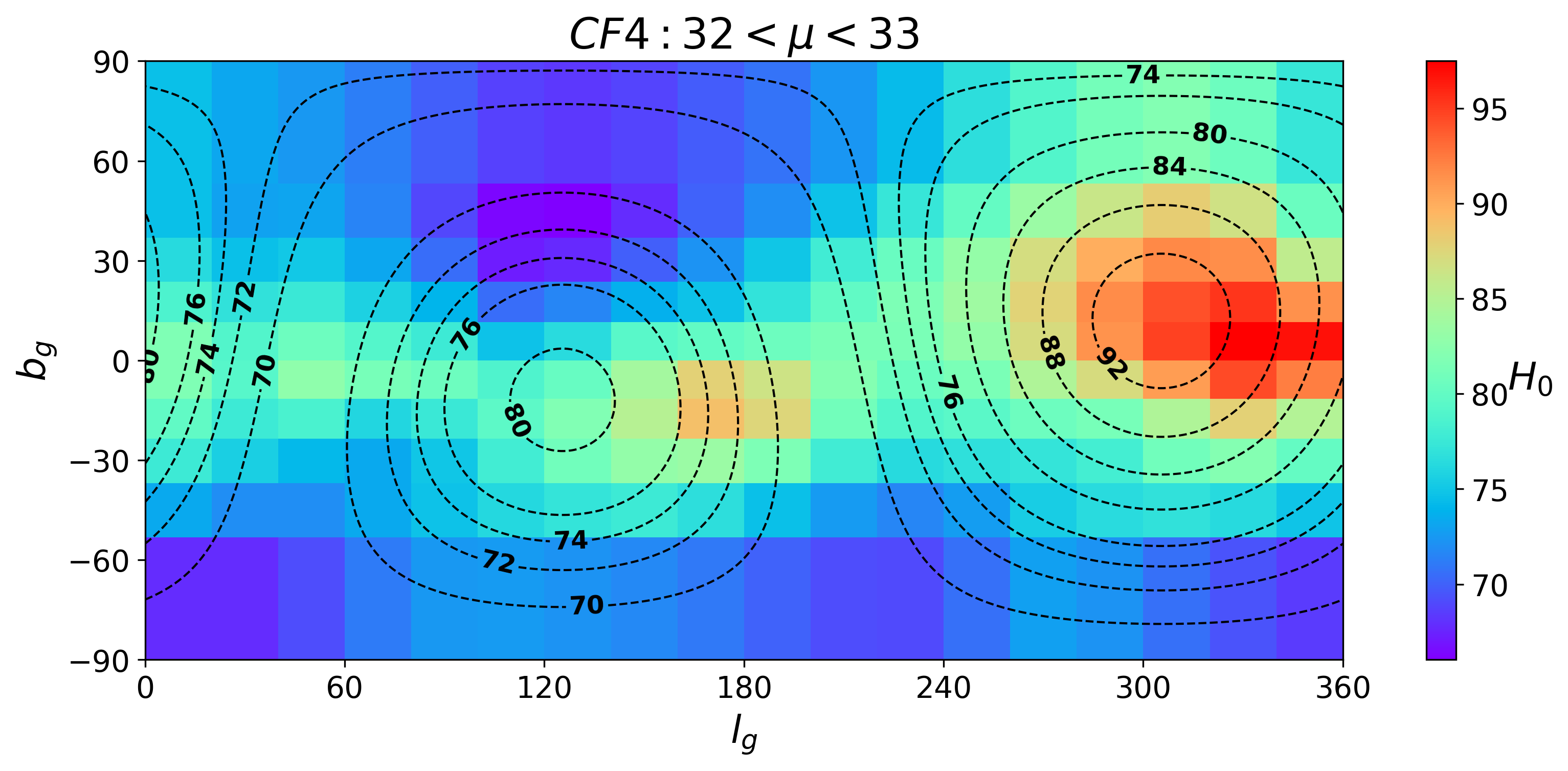}\\
~~~\\
\includegraphics[width=8.5cm]{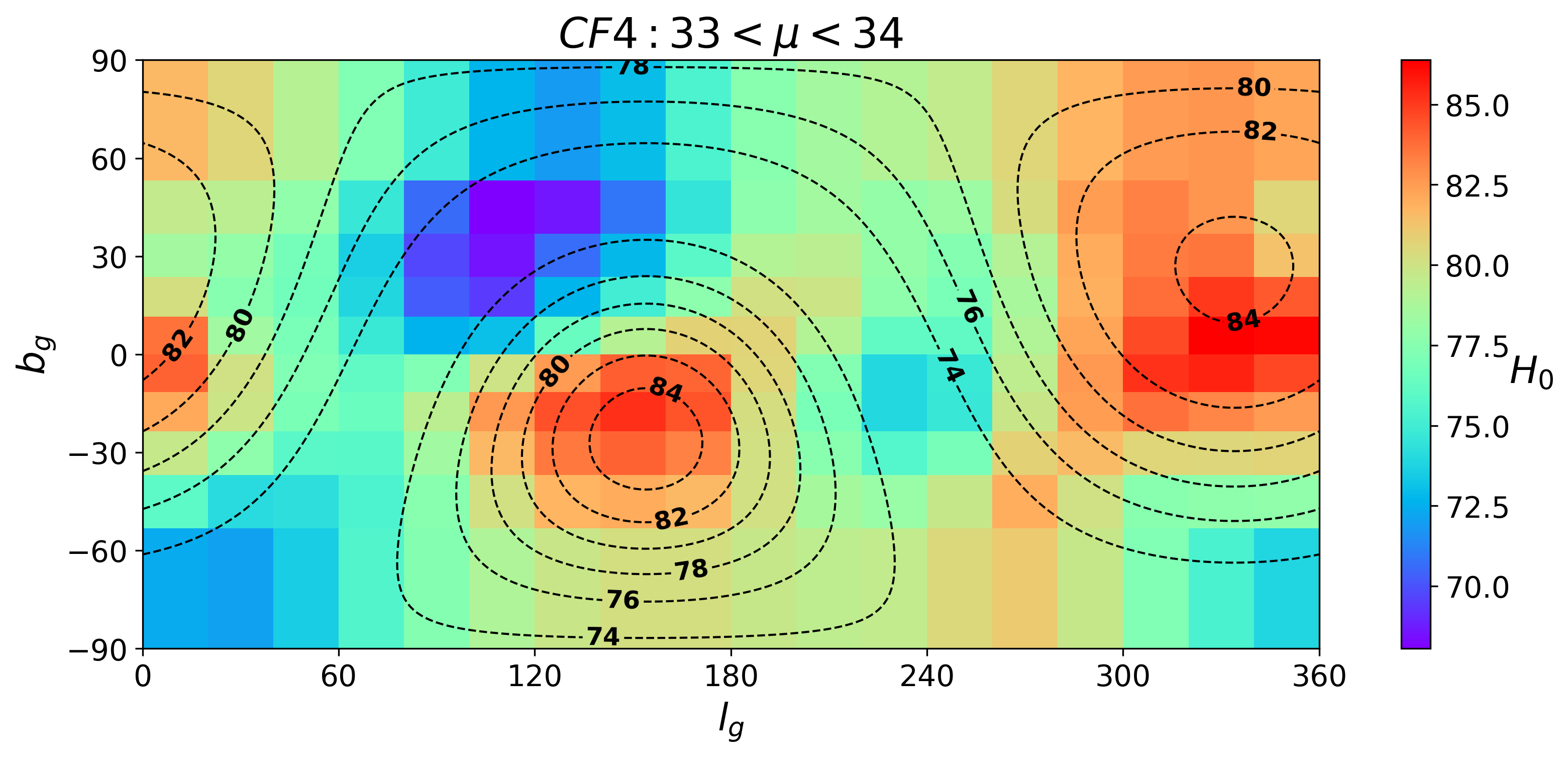}~
\includegraphics[width=8.5cm]{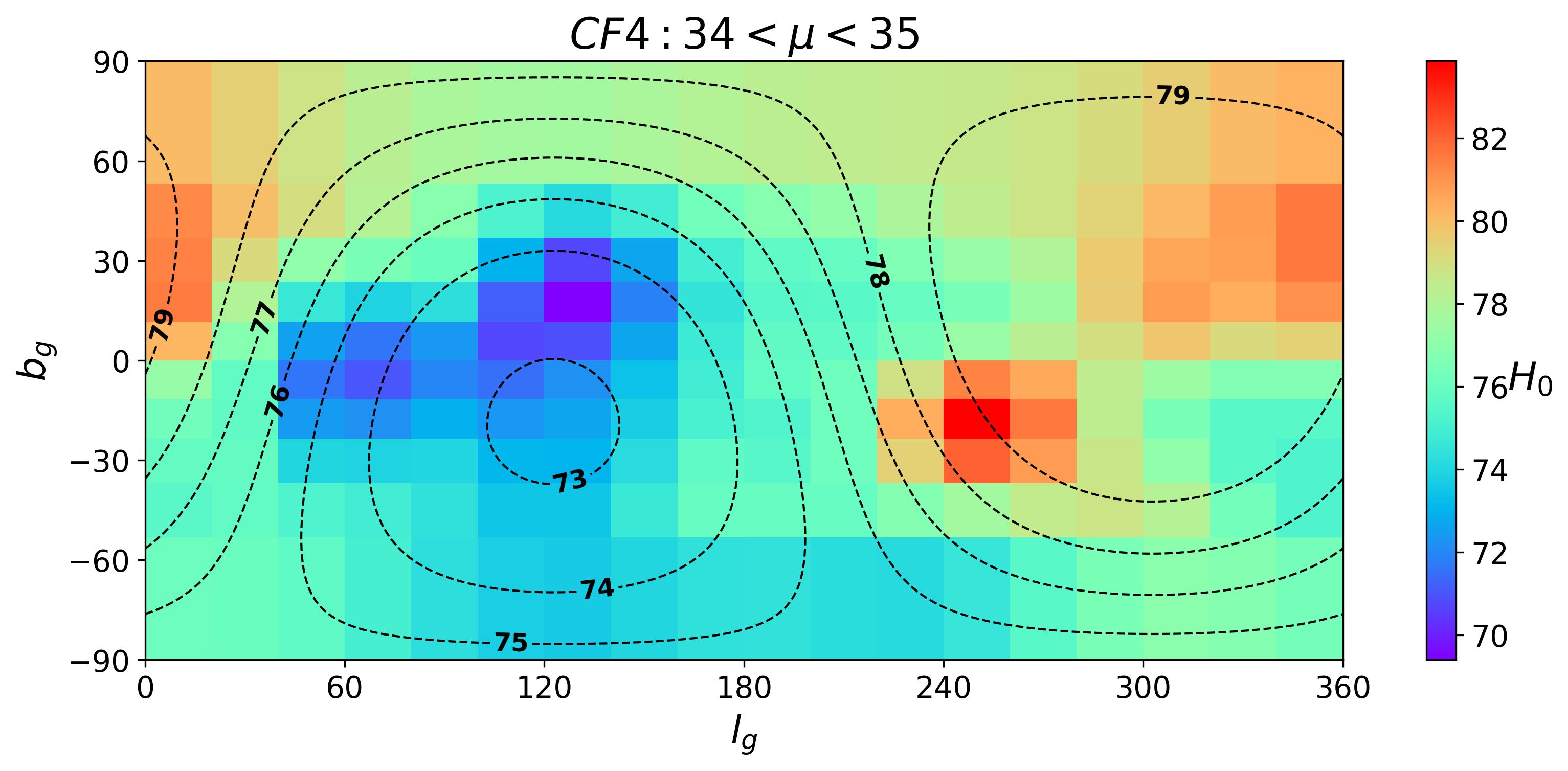}\\
~~~\\
\includegraphics[width=8.5cm]{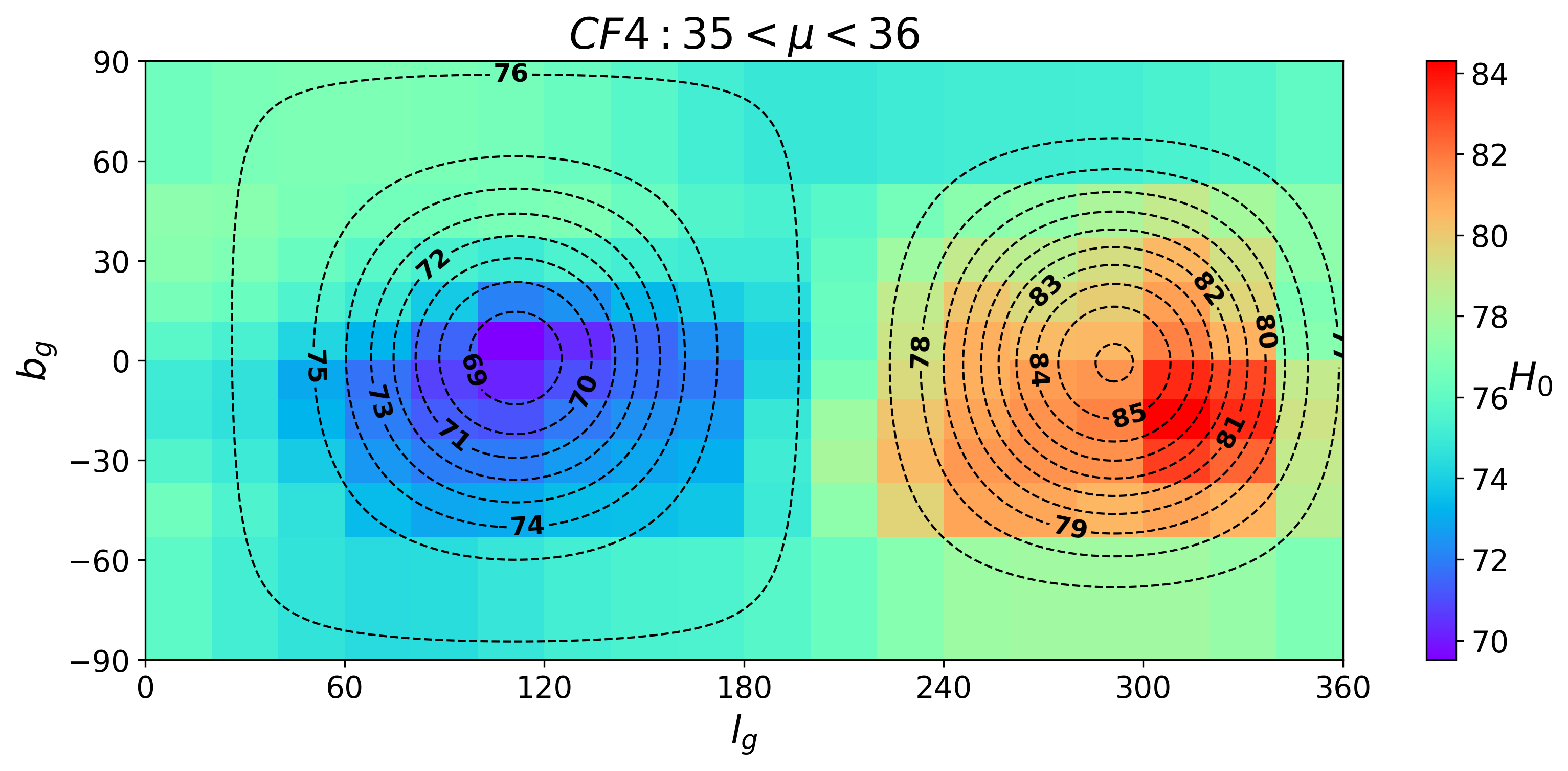}~
\includegraphics[width=8.5cm]{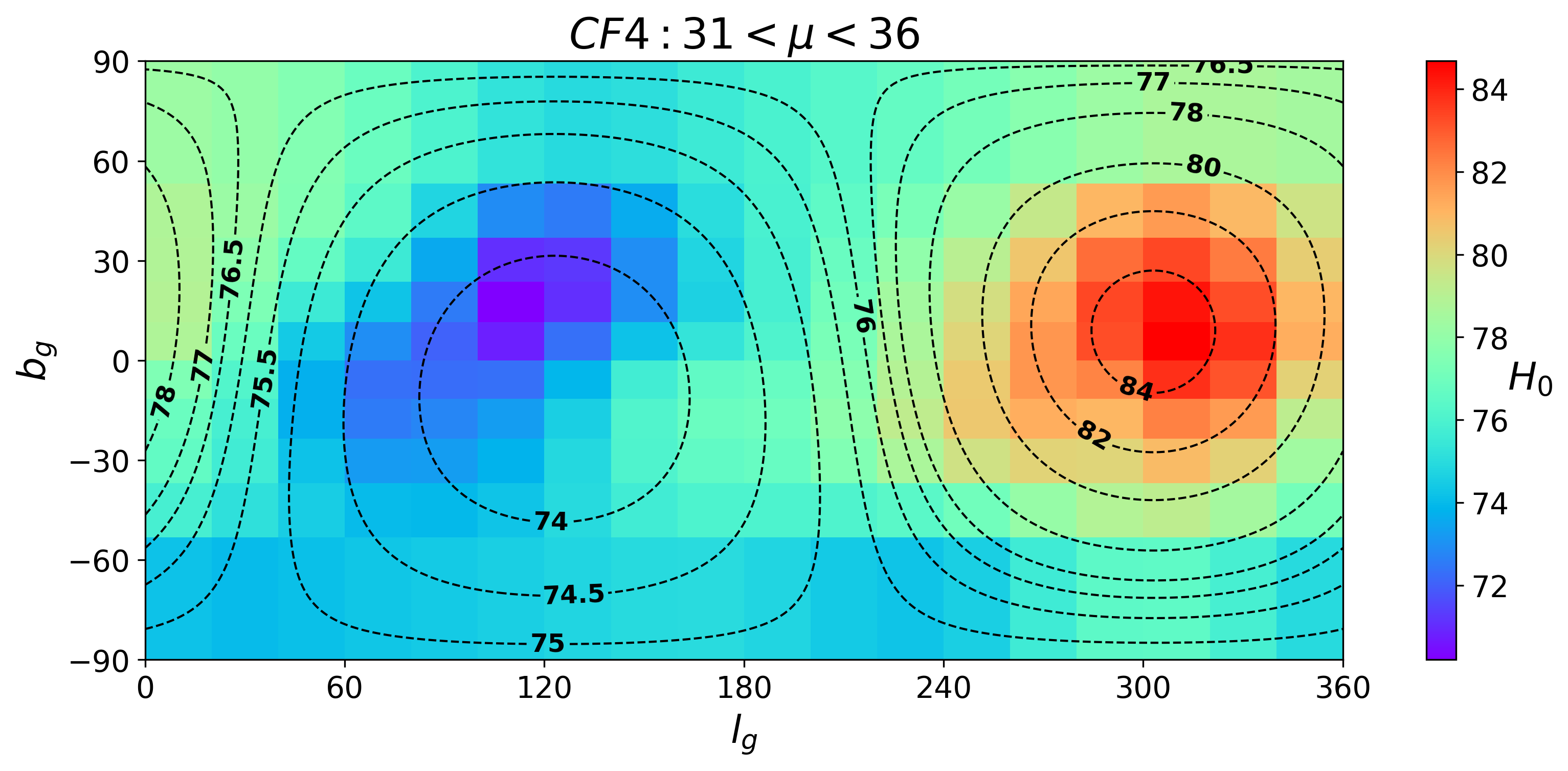}
\caption{Radial+Angular signal per radial shells in Galactic coordinate system and CMB frame for the CF4 sample. Coordinates: galactic longitude $l_g$, from $0^{\circ}$ to $360^{\circ}$; galactic latitude $b_g$, from $-90^{\circ}$ to $90^{\circ}$. The anisotropy signal in the Hubble constant from the data is shown as a colour gradient ranging from red, which indicates higher values, to violet, which indicates lower values. The black dashed lines represent the best fit to the signal, from the harmonic expansion up to the octupole.}\label{fig:Radial_Angular_shells_CF4}
\end{figure*}

Fig.~\ref{fig:Radial_Angular_shells_CF4} visually illustrates the effectiveness of the multipole fit. Representing tens of thousands of different values for the measured $H_0$ would have made the figures unreadable and dominated the map with local fluctuations, hiding the large-scale behaviours that we are mainly interested in. These behaviours can be more easily detected through the averaging procedure. The optimal configuration was found by tessellating the sky into equal solid-angle patches, with a spacing of $\Delta l_g = 20^{\circ}$ in galactic longitude and $0.2$ in $\sin b_g$, spanning $l_g \in [0^{\circ}, 360^{\circ}]$ and $b_g \in [-90^{\circ}, 90^{\circ}]$.  Each patch covers a solid angle of $\Delta \Omega = 0.0698$ sr $\approx 229$ deg$^2$. The final grid consists of the centres of each of the patches, yielding $180$ preferred directions around which the averaged Hubble constant was computed. The procedure for constructing the spatial map of $H_0$ was as follows:
\begin{enumerate}
    \item select data from the chosen CF4 subsample within a given radial interval in $\mu$ or $z$;
    \item choose one direction from the angular grid defined above;
    \item compute the average $\langle \log H_0 \rangle$ and its uncertainty for that direction starting from Eqs.~(\ref{eq:average_H0})~-~(\ref{eq:average_sH0})~-~(\ref{eq:weight_stat}) and (\ref{eq:int_err}), and adding a geometric (angular) weight, $w_{\theta}$, as Gaussian smoothing based on the angular separation from the selected direction. Thus we have
    \begin{equation}\label{eq:average_H0_geo}
    \langle \log H_0 \rangle =
    {\frac{\sum_{i=1}^N w_{\theta,i} w_i \log H_{0,i}}{\sum_{i=1}^N w_{\theta,i} w_i}}\, ,
    \end{equation}
     \begin{equation}\label{eq:average_sH0_geo}
    \sigma_{\langle \log H_0 \rangle}^2
    = {\frac{\sum_{i=1}^N w^{2}_{\theta,i} w_i} {\sum_{i=1}^N w_{\theta,i} w_i}}\, .
    \end{equation}
    We use a Gaussian in $\theta$, centered at $0$ with dispersion equal to the square root of the solid angle spanned by each cell, $\sigma_{\theta} = 15.14^{\circ}$. Note that the intrinsic scatter per shell is still obtained from Eq.~(\ref{eq:int_err}), with no geometrical weighting applied;
    \item repeat steps 2–3 for all directions in the angular grid;
    \item repeat steps 1–4 for each radial shell.
\end{enumerate}
The final angular variation maps of $H_0$ are shown in colour in each shell, ranging from lower values in blue to higher values in red. The dashed black lines show the theoretical best fit up to octupole from Eq.~(\ref{eq:spherical_harm}). For clarity, the colour scale differs between plots, although the range of variation is shown in the legends. Overall, the model accurately captures the anisotropic signal. The signal maximum is located very accurately, except in the fourth shell. It is also evident that the octupole model can describe more intricate features, such as the emergence of a secondary hot peak in shells with $\mu \in [31,34]$, after which it disappears. The cause of this deviation is unclear, but it may be related to gravitational structures in that region. This deviation is certainly not due to a lack of data, given that these shells contain enough objects to provide good statistics.

When we move on to consider the catalogue of peculiar velocities, CF4$_{pec}$, as shown in Table~\ref{tab:results_CF4pec_mu}, the scenario changes radically. Firstly, we can see that for the first three lower shells, the direction of the dipole is much more weakly constrained (italic font indicates that the posteriors are not Gaussian and span a range as large as the possible interval from the priors). There also seems to be no correlation with the CF4 case, nor with each other. No clear pattern emerges. This is also clearly reflected by the values of the multipole coefficients, which are all consistent with zero. Finally, the Bayes Factors are much lower than in the CF4, lying within the range of inconclusive or substantial evidence in favour of multipole models over monopole-only ones. However, we must once again note that our analysis is based on the possibility of underestimating the errors due to the missing CF4 covariance matrix.

\begin{figure*}[!ht]
\centering
\includegraphics[width=8.5cm]{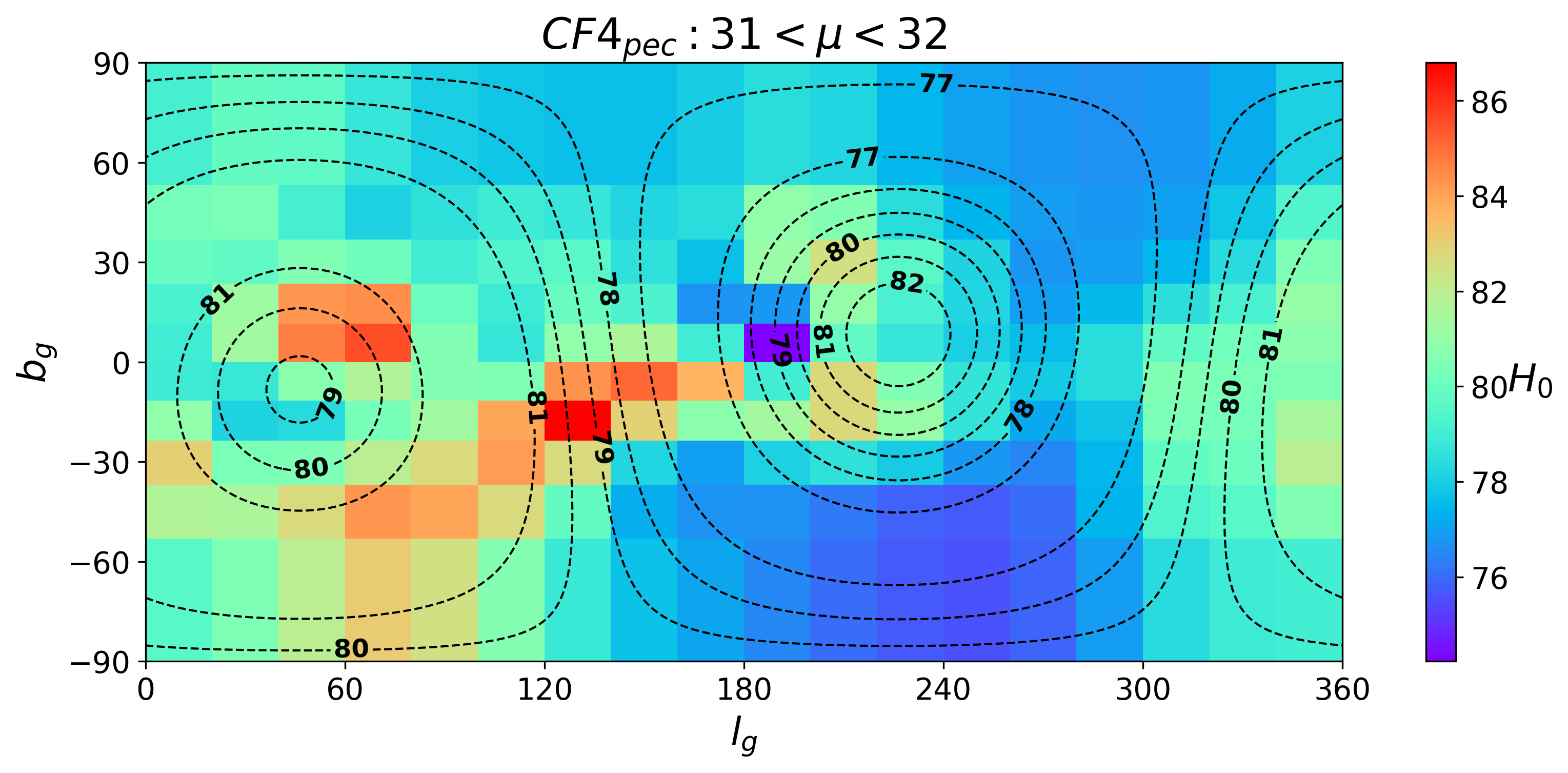}~
\includegraphics[width=8.5cm]{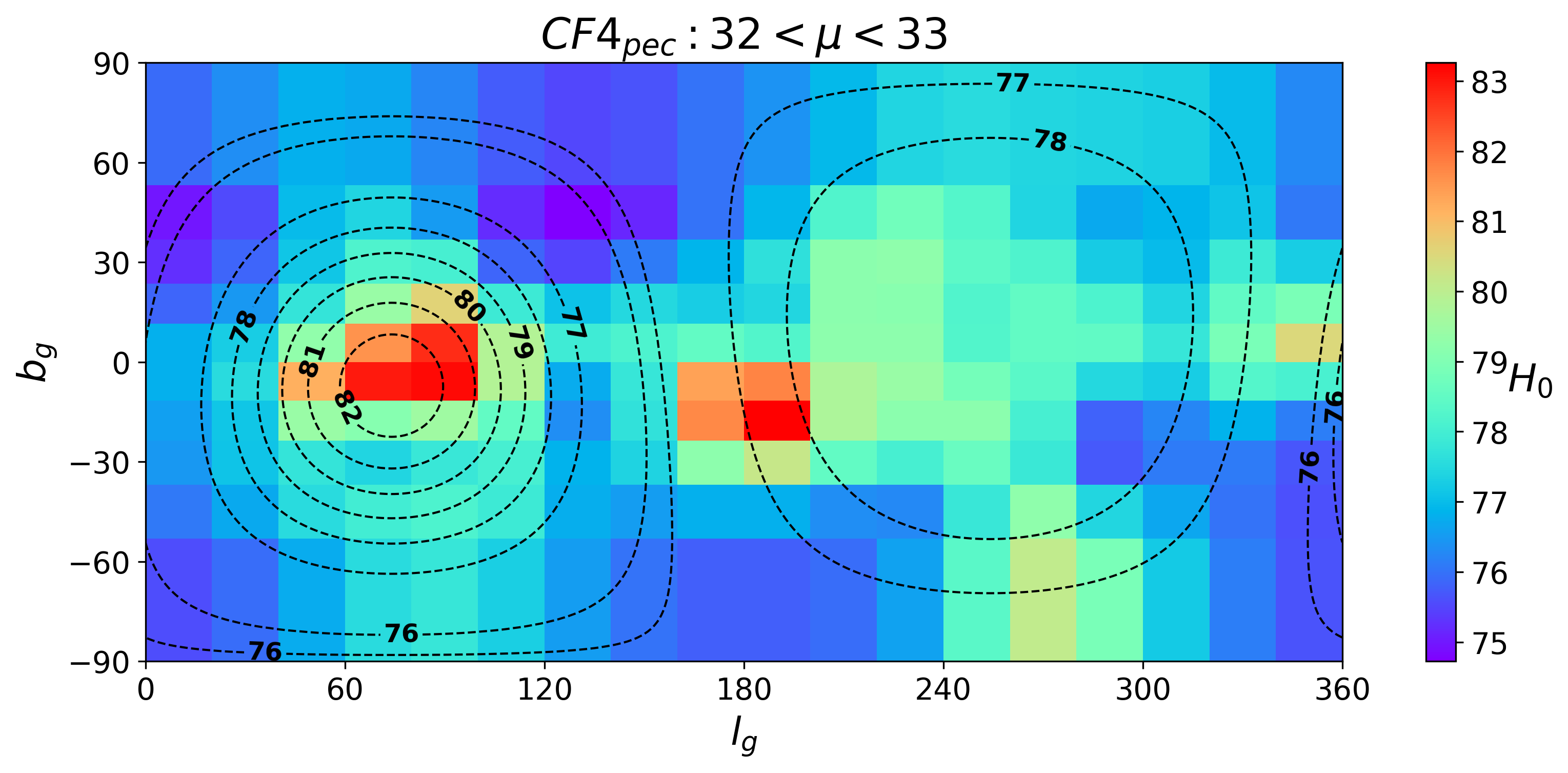}\\
~~~\\
\includegraphics[width=8.5cm]{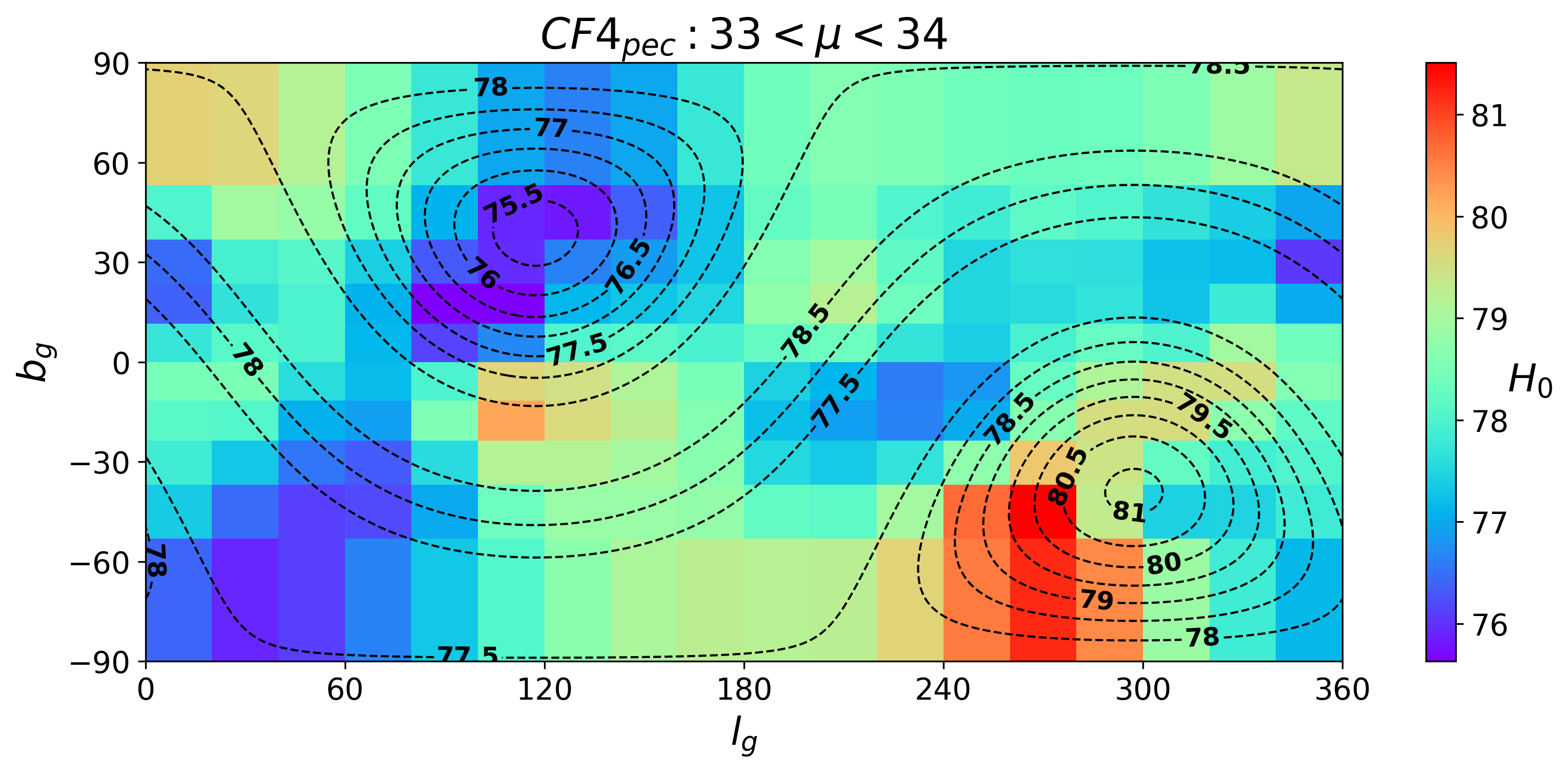}~
\includegraphics[width=8.5cm]{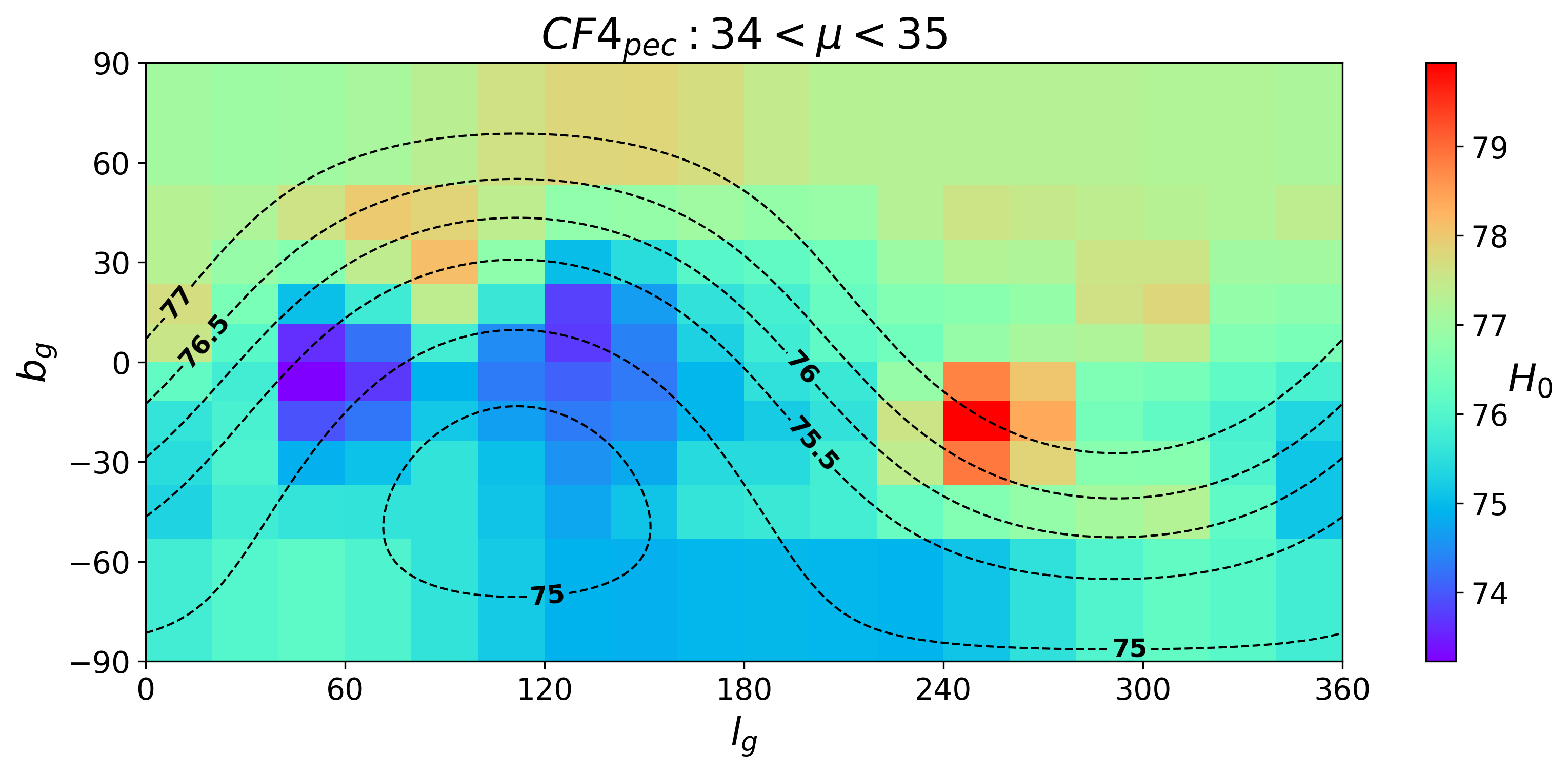}\\
~~~\\
\includegraphics[width=8.5cm]{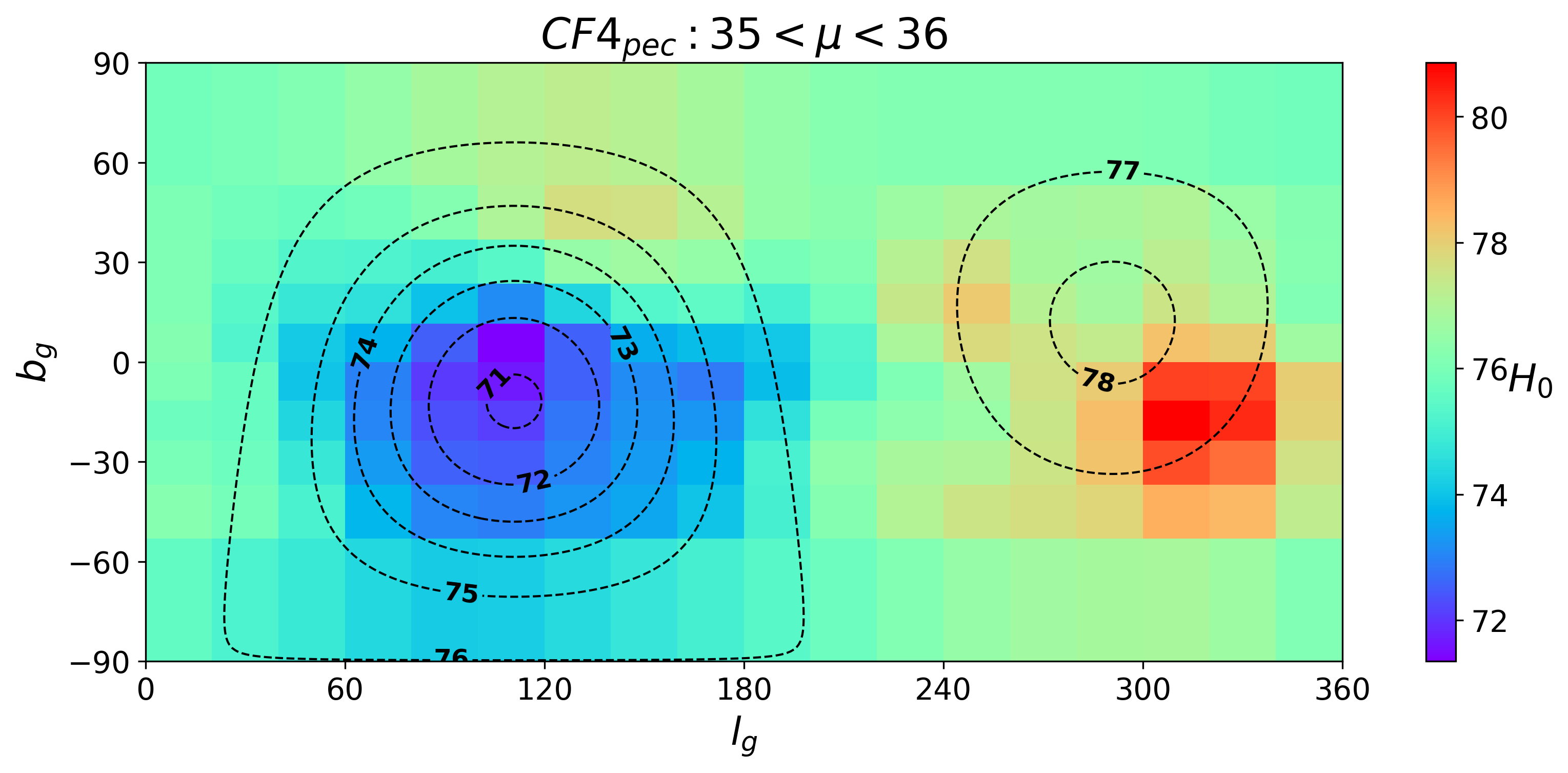}~
\includegraphics[width=8.5cm]{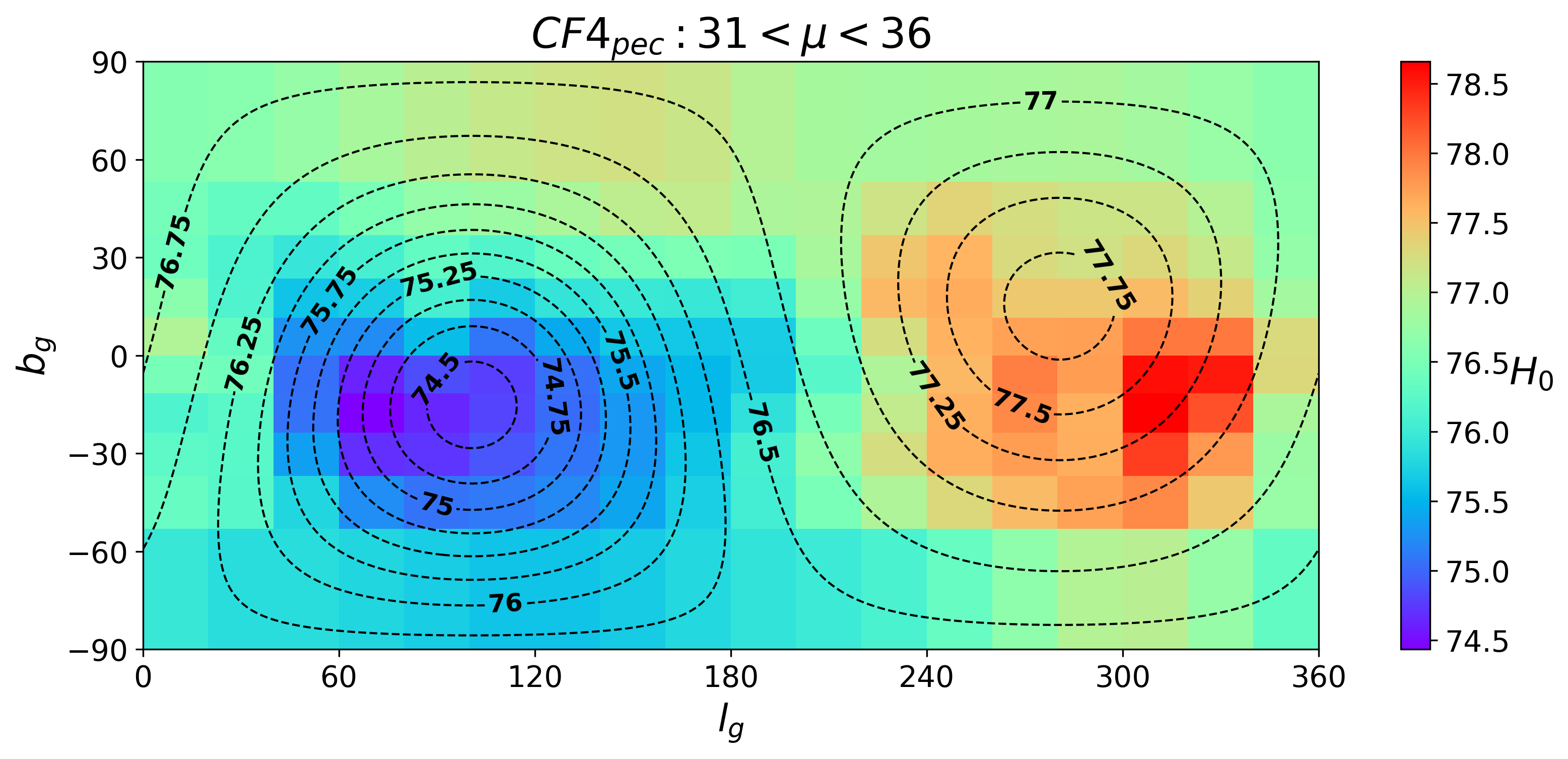}
\caption{Radial+Angular signal per radial shells in Galactic coordinate system and CMB frame for the CF4$_{pec}$ sample. Coordinates: galactic longitude $l_g$, from $0^{\circ}$ to $360^{\circ}$; galactic latitude $b_g$, from $-90^{\circ}$ to $90^{\circ}$. The anisotropy signal in the Hubble constant from the data is shown as a colour gradient ranging from red, which indicates higher values, to violet, which indicates lower values. The black dashed lines represent the best fit to the signal, from the harmonic expansion up to the octupole.}\label{fig:Radial_Angular_shells_CF4_pec}
\end{figure*}

The situation is completely different for $\mu>34$: there is a strong preference for an aligned anisotropy signal, as demonstrated by the Bayes Factors and the coefficients of the multipoles, particularly the dipole. The quadrupole and octupole are weaker, and in many cases are even consistent with zero. The direction of the preferential anisotropy alignment also shows strong regularity and is well constrained. Although the errors are larger as expected, since the signal remaining after peculiar velocity corrections is somewhat lower (as shown by the absolute values of the multipole coefficients), there is a substantial difference compared to the CF4 case, primarily in the latitude direction for the $[34;35]$ shell, while for the$[35;36]$ shell there is a disagreement at the $2\sigma$ level. All these considerations are visualised in Fig.~\ref{fig:Radial_Angular_shells_CF4_pec}, which shows that it is generally more difficult to reconstruct a clean signal, primarily due to the greatly reduced anisotropy amplitude. Given that these two shells are the most densely populated, it is not surprising that the full-range results shown in Table~\ref{tab:results_full} are consistent with them.

\begin{figure*}[!ht]
\centering
\includegraphics[width=8.5cm]{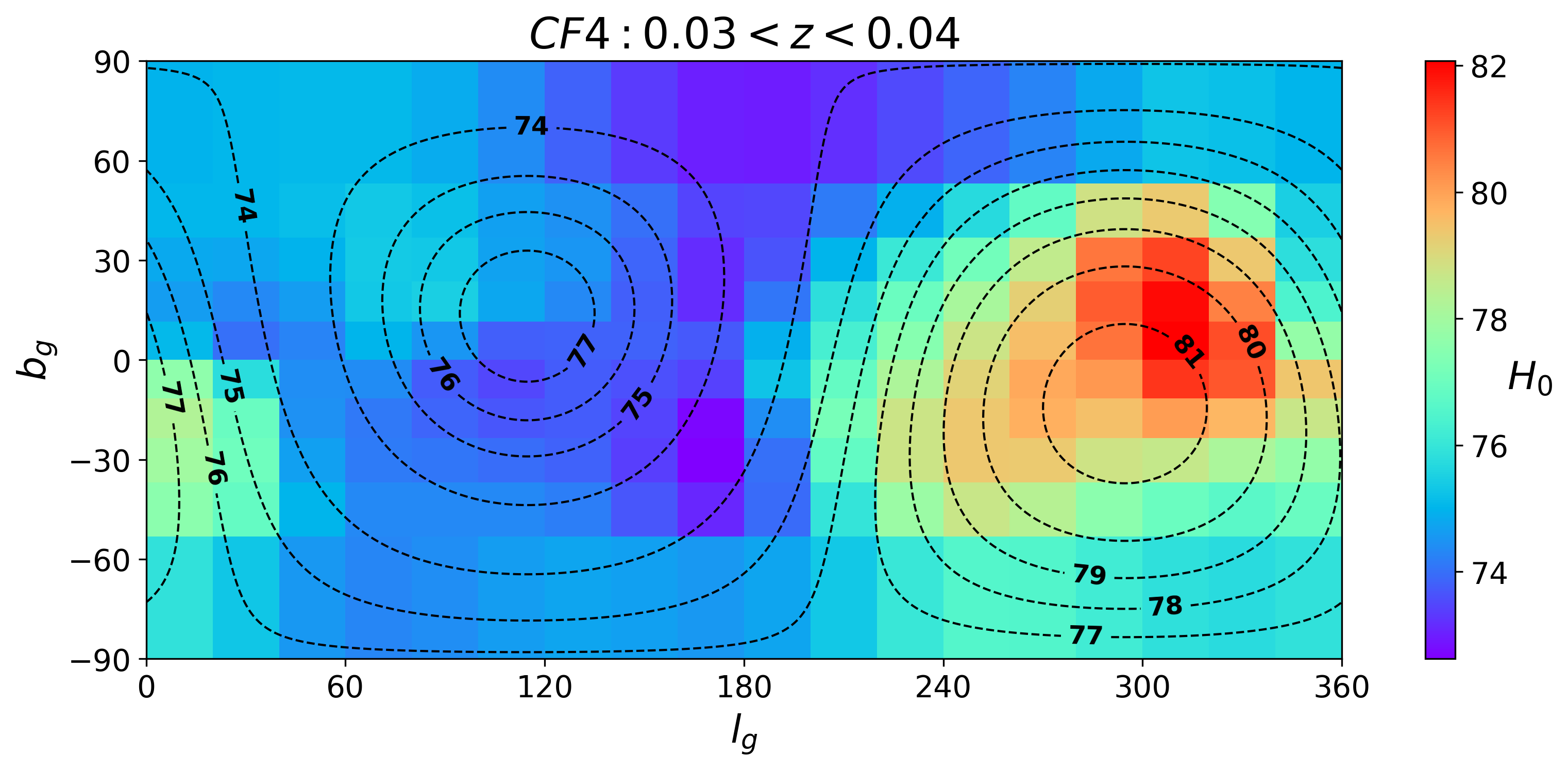}~
\includegraphics[width=8.5cm]{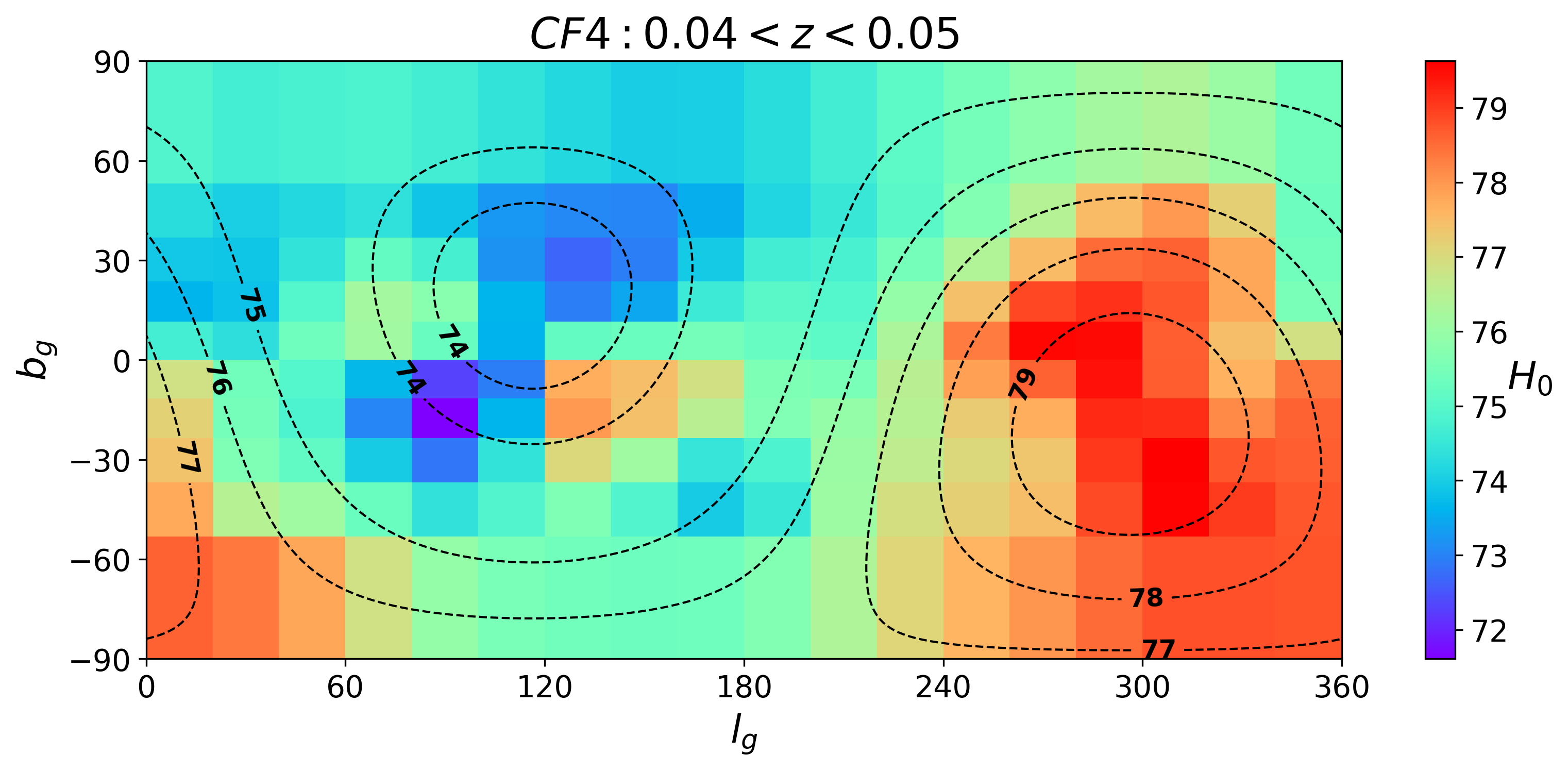}\\
~~~\\
\includegraphics[width=8.5cm]{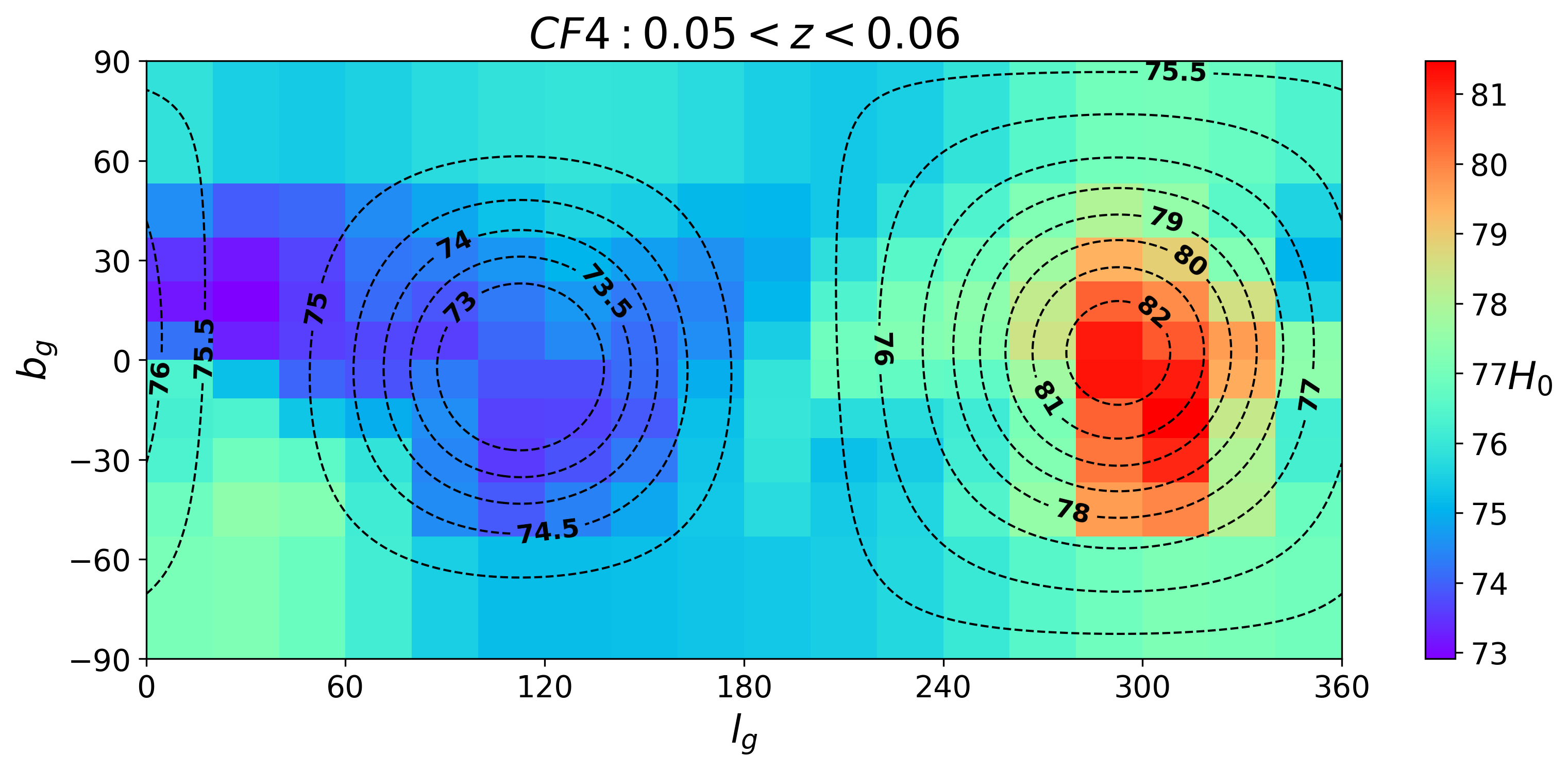}~
\includegraphics[width=8.5cm]{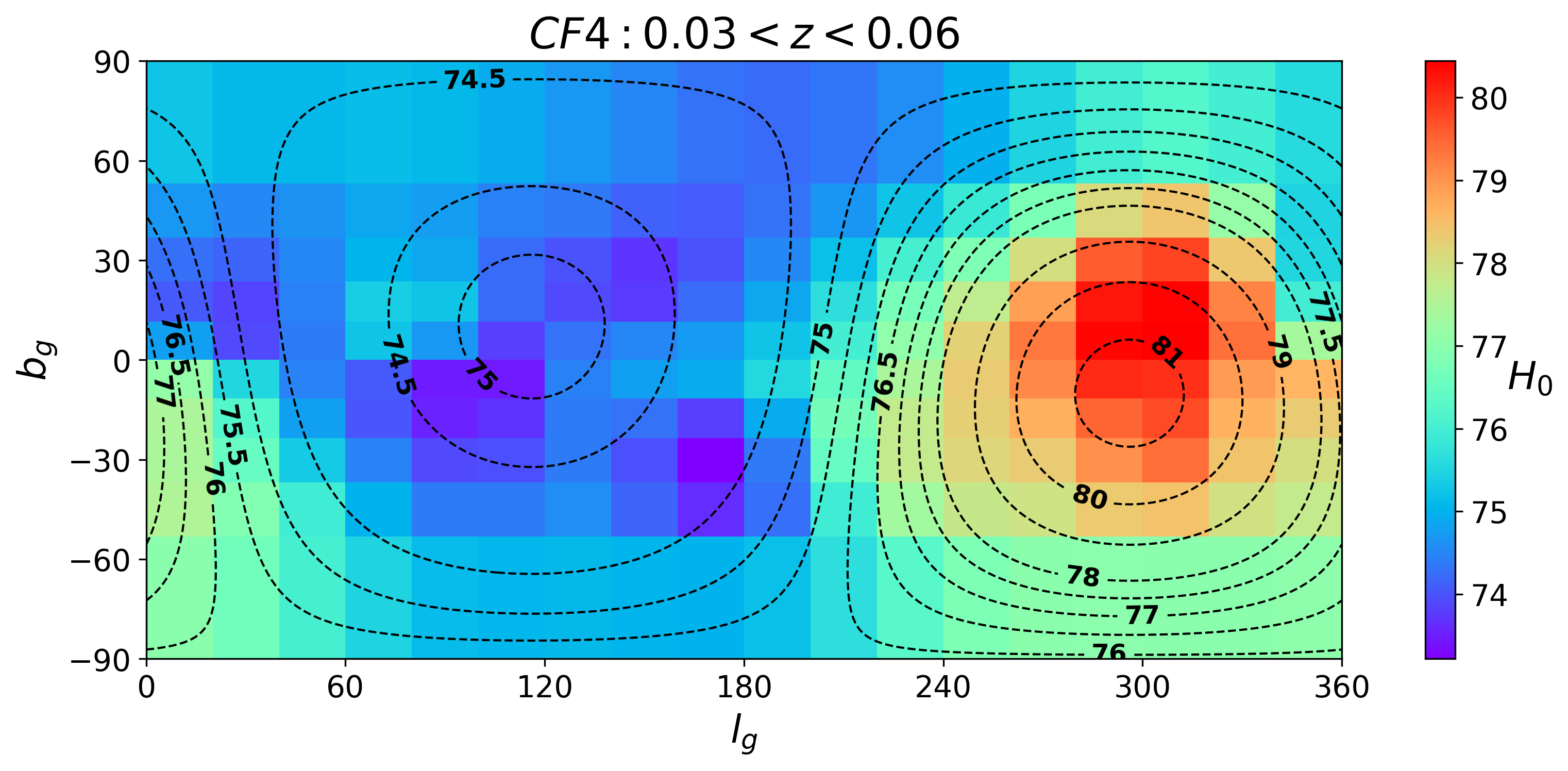}\\
~~~\\
\includegraphics[width=8.5cm]{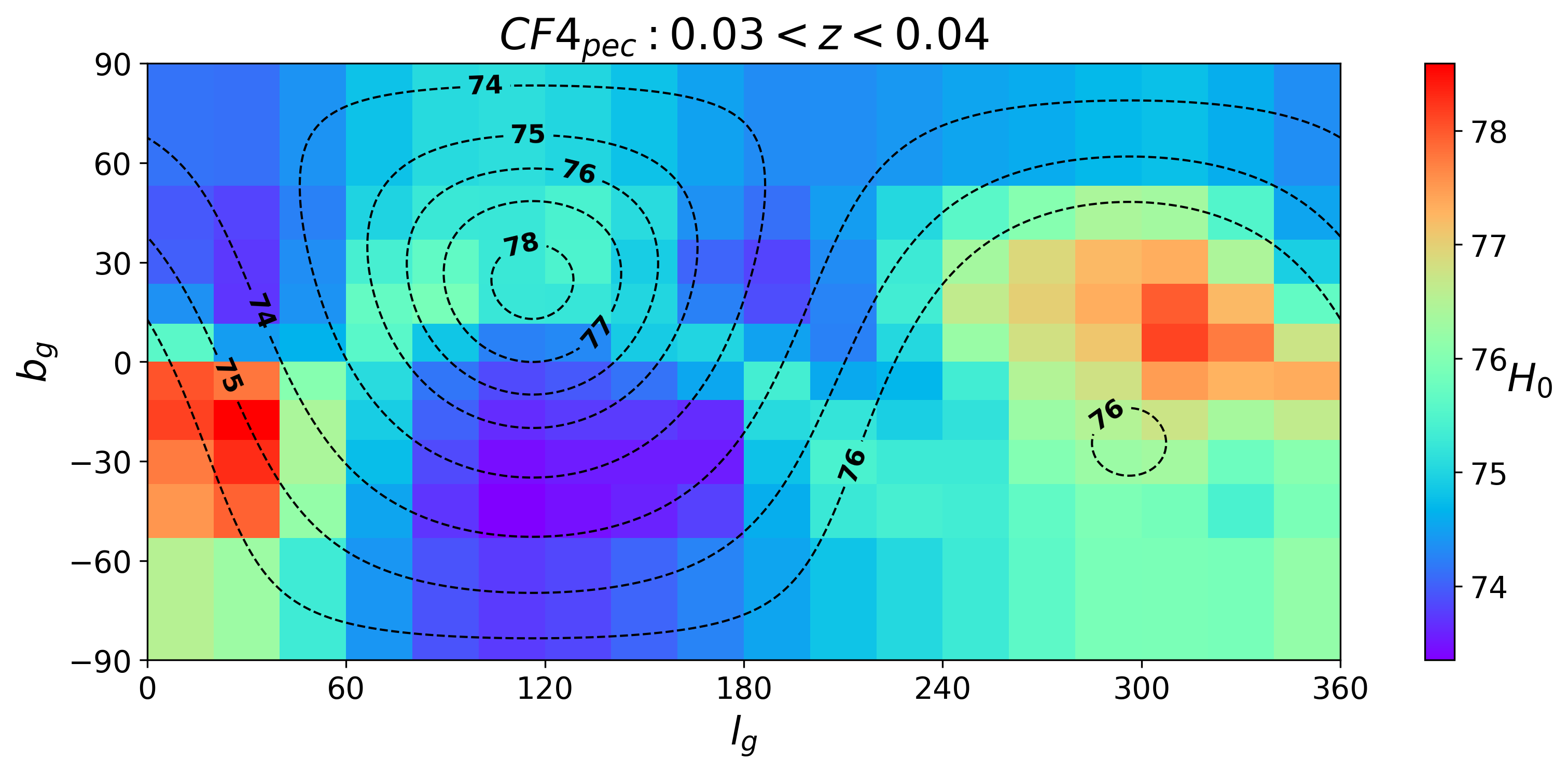}~
\includegraphics[width=8.5cm]{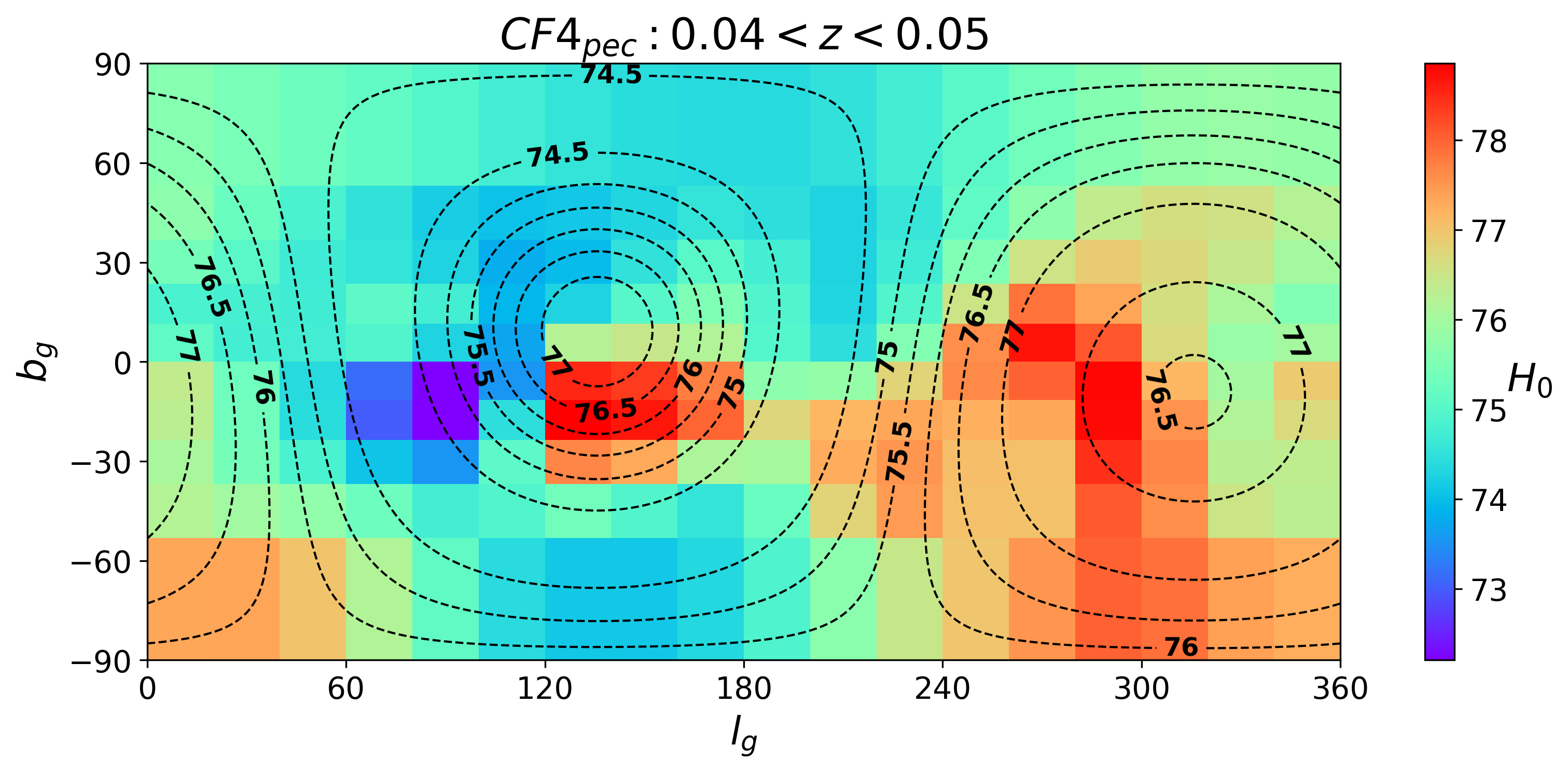}\\
~~~\\
\includegraphics[width=8.5cm]{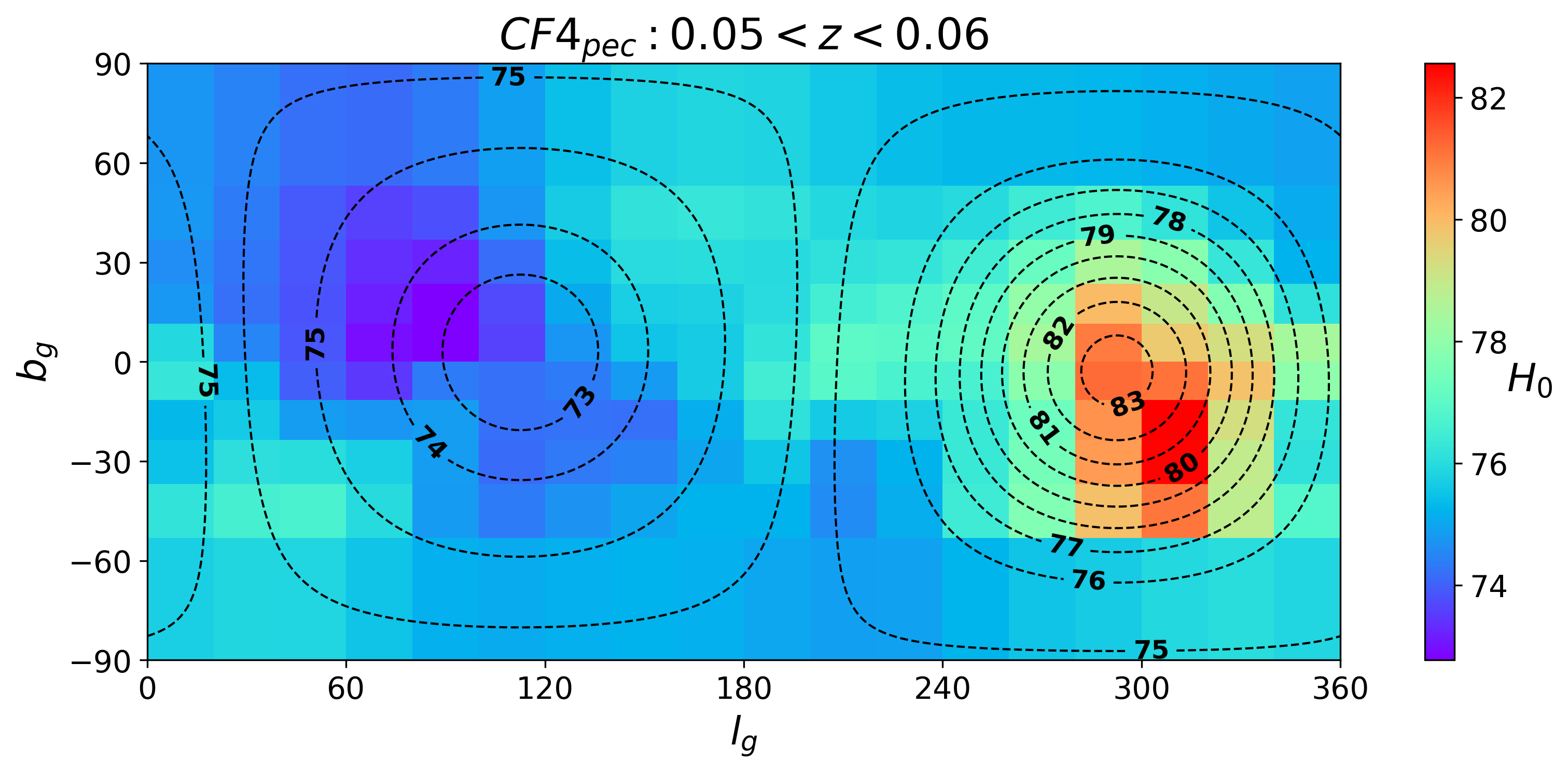}~
\includegraphics[width=8.5cm]{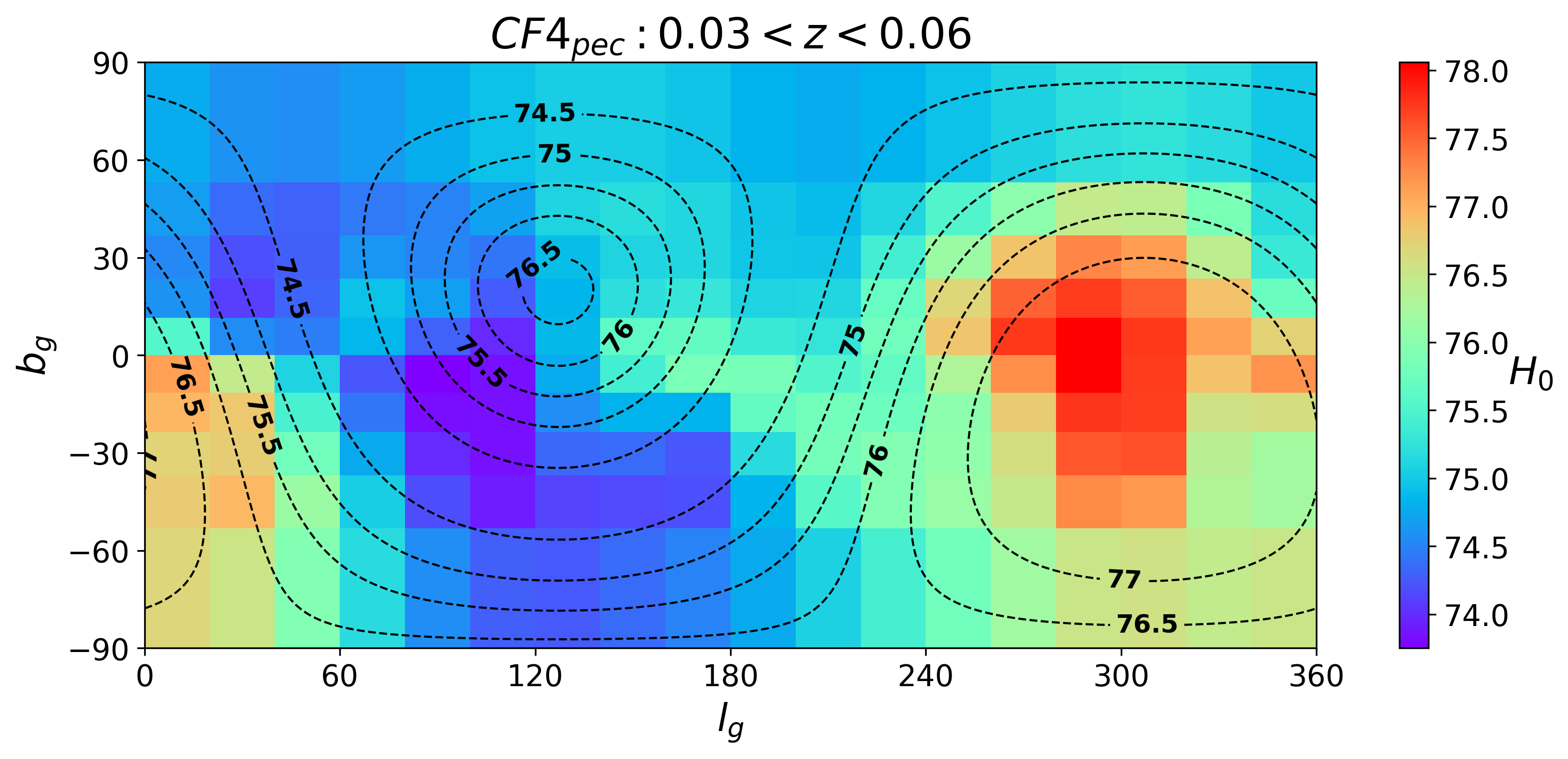}
\caption{Radial+Angular signal per radial shells in Galactic coordinate system and CMB frame for the CF4$_{pec}$ sample. Coordinates: galactic longitude $l_g$, from $0^{\circ}$ to $360^{\circ}$; galactic latitude $b_g$, from $-90^{\circ}$ to $90^{\circ}$. The anisotropy signal in the Hubble constant from the data is shown as a colour gradient ranging from red, which indicates higher values, to violet, which indicates lower values. The black dashed lines represent the best fit to the signal, from the harmonic expansion up to the octupole.}\label{fig:Radial_Angular_shells_CF4_Z}
\end{figure*}

In Tables~\ref{tab:results_CF4_z} and \ref{tab:results_CF4pec_z} and in Fig.~\ref{fig:Radial_Angular_shells_CF4_Z} we show the results when considering shells in redshift. Generally speaking, most of the considerations made for the $\mu$ shell still apply. When considering binning in redshift, note that this corresponds approximately to large $\mu$, as it can be seen from Fig.~\ref{fig:HL_plot}, where the vertical (horizontal) dashed lines represent the ranges we have considered for our analysis while binning in $z$ ($\mu$). The qualitative considerations are largely unchanged: the dipole direction is consistent with the previous binning in $\mu$ in the longitude, while the latitude is considerably different, switching hemisphere; the dipole dominates, and both the quadrupole and octupole are consistent with zero for $z>0.04$. This is also evident from the values of the Bayes factor, which show no substantial change when large multipole components are added. Finally, when peculiar velocities are considered, although the dipole correction is still relevant, its importance is greatly reduced, and the Bayes factors do not indicate a strong statistical preference with respect to a constant $H_0$ in these shells. We also note that the addition of the octupole seems to improve the fit, albeit by dramatically changing the direction of the anisotropy axis in the first two shells. However, we should again remark that the signal is much weaker in this case and that any statistically strong conclusions should be avoided.

\section{Discussion and Conclusions}
\label{sec:discussion}

In this work, we have investigated the presence of anisotropies in the Hubble constant using the Cosmicflows-4 catalogue, with the aim not only of detecting angular variations in $H_0$, but also of clarifying how such results depend on the statistical formulation of the Hubble--Lema\^{i}tre relation, on the internal consistency of the data sample, and on the treatment of peculiar velocities. A distinctive aspect of our analysis with respect to much of the existing literature is threefold. First, we formulate the Hubble--Lema\^{i}tre relation directly in logarithmic form using distance moduli, thereby preserving the Gaussian nature of the measured uncertainties. Second, the working subsamples are not chosen a priori, but are identified through internal consistency tests of the catalogue, including the depth dependence of $\langle \log H_0 \rangle$ and the behaviour of residual skewness and kurtosis. This leads to the conservative ranges $\mu \in [31,36]$ and $z \in [0.03,0.06]$, where the Hubble flow is stable and selection effects are minimized. Third, we explicitly compare results obtained using observed radial velocities (CF4) with those obtained after peculiar-velocity corrections (CF4$_{pec}$), thus separating observationally driven anisotropies from model-dependent reconstructions.

Within these validated ranges, we find that the CF4 sample exhibits a statistically significant anisotropic signal, dominated by a dipole and strongly favoured over a monopole-only model by Bayesian evidence. This raises two main questions: is the dipole consistent with previous findings in the literature? Secondly, can it be linked to a known kinematic component?

To address the first question, we first remind that when considering the full samples, there is no significant difference in the direction of the anisotropy before and after corrections for peculiar velocities. As shown in Tables~\ref{tab:results_full} the detected dipole never aligns with the CMB dipole, with Eq.~(\ref{eq:cos}) giving an angular distance of giving $\cos \theta = 0.62^{+0.02}_{-0.02}$ ($\theta = 52^{\circ}{}^{+2^{\circ}}_{-2^{\circ}}$) for $\mu \in [31,36]$, and $\cos \theta = 0.43^{+0.07}_{-0.07}$ ($\theta = 65^{\circ}{}^{+4^{\circ}}_{-5^{\circ}}$) for $z \in [0.03,0.06]$.

Our found anisotropy direction can be compared with \cite{Kalbouneh:2022tfw} and \cite{Kalbouneh:2025jnp}, who analyzed Cosmicflows-3 and Pantheon SNeIa (the former) and Cosmicflow-4 and Pantheon+ SNeIa (the latter) under the assumption $v_{pec} \ll v^{obs}_{CMB}$, respectively to the first and last rows in Table~I of \cite{Kalbouneh:2022tfw} and to Table~2 of \cite{Kalbouneh:2025jnp}, showing a substantial agreement in the dipole direction. Of all the probes cited in the introduction section, our detection is consistent with most of them. However, in some cases, such as quasars, the errors are too large to reliably assess consistency.  A more comprehensive and varied comparison certainly deserves further study in the near future.

As mentioned in the introduction, the objective of this study is also to examine the potential distance dependence of the dipole, as defined by Eq.~(\ref{eq:h_angle}) and representing a signature of differential cosmological expansion. Our results in Tables~\ref{tab:results_CF4_mu}~-~\ref{tab:results_CF4pec_mu}~-~\ref{tab:results_CF4_z} and ~-~\ref{tab:results_CF4pec_z} clearly show that no such trend is detectable. While the dipole decreases along the first three shells, it then increases again. The inability to disentangle this effect from cosmic variance or the sky scanning of the surveys in question makes it difficult to draw any firm conclusions.

Another relevant comparison is the possible link between the detected dipole and large-scale gravitational structures. In the standard cosmological framework, and in the CMB frame, no dipole should arise beyond the scale of homogeneity due to large-scale isotropy. Yet local inhomogeneities can break this symmetry and indeed within the distance range probed by the Cosmicflows-4 sample, we find the structures listed in \cite{Valade:2024riw}.

Thus, when using uncorrected radial velocities, a dipole aligned with the dominant local bulk flows from large structures is expected. We estimate the corresponding bulk flow by interpreting the dipole signal as a bulk motion. Although this approach is less precise than detailed reconstructions of the local gravitational field (e.g. \cite{Watkins:2023rll,Whitford:2023oww,Hoffman:2023pac}), which are however model dependent, it provides a reasonable approximation. The bulk flow associated with the dipole is given by
\begin{equation}
v^{dip}_{bulk} \equiv \langle v \rangle = \frac{(H^{max}_0 - H^{min}_{0})}{2} \langle d \rangle\, = \frac{a_{10}}{2} \langle d \rangle\, ,
\end{equation}
where $\langle \ldots \rangle$ denotes the weighted average on a shell. In our nominal range we find, for $\mu \in [31,36]$, $|v^{dip}_{bulk}| = 298^{+13}_{-13}$ km s$^{-1}$ at $\langle d \rangle \sim 44\,h^{-1}$ Mpc, closely aligned with the Shapley supercluster, while for $z \in [0.03,0.06]$, $|v^{dip}_{bulk}| = 298^{+13}_{-13}$ km s$^{-1}$. In fact, a more detailed analysis \cite{Courtois:2017mrq} has revealed that the location of the dominant attractor in the local universe region covered by Cosmicflows-4 lies in the direction $(306^{\circ},22^{\circ})$, which is within $1\sigma$ of our estimate for the longitude, while a larger deviation is for the latitude. As shown in Fig.~9 of \cite{Qin:2021tak} and Fig.~9 and Table~2 of \cite{Duangchan:2025uzj}, our estimate is in perfect agreement with many earlier measurements and theoretical estimations from the standard cosmological model. 

Instead, for the redshift binning, with $z \in [0.03,0.06]$, we find $|v^{dip}_{bulk}| = 564^{+47}_{-45}$ km s$^{-1}$ at $\langle d \rangle \sim 120\,h^{-1}$ Mpc. This value is clearly more difficult to frame into a standard cosmological picture.

The key question, then, is whether any dipole should remain after the radial recessional velocities have been corrected for peculiar motions \textit{within a standard cosmological framework}. As such corrections are meant to eliminate all anisotropies, one might wonder what would be the origin of a residual signal, if there were one. The central result of this work emerges from the comparison with the peculiar-velocity-corrected catalogue. Once peculiar velocities are subtracted, the anisotropy amplitude is strongly suppressed, particularly at lower distances, and only a weak residual signal remains, with an inferred bulk flow of $\sim 87$ km s$^{-1}$ in the nominal range. This behaviour indicates that the dominant contribution to the anisotropy observed in CF4 is associated with local kinematics rather than with a genuine anisotropy of the background cosmological expansion. Interestingly, as we have discussed in the previous section, most of the contribution to this possible bulk flow comes from the more distant shells, which contain more data; the lower ones are statistically consistent with zero velocity. We stress that the CF4$_{pec}$ velocities are themselves model-dependent reconstructions, obtained within a $\Lambda$CDM framework, and therefore the suppression of the signal should be interpreted in this context.

A second objective of this work was to test whether the dipole exhibits a systematic dependence on distance, as expected in scenarios of differential expansion such as those described by Eq.~(\ref{eq:h_angle}). Our analysis of radial shells in both $\mu$ and $z$ shows no robust evidence for a monotonic evolution of the dipole amplitude. Although variations are observed across shells, they are not consistent with a simple decreasing trend with distance, and are difficult to disentangle from cosmic variance and survey geometry effects. This absence of a clear radial behaviour weakens the interpretation of the observed anisotropy as a signature of large-scale deviations from isotropic expansion.

\begin{figure*}
\centering
\includegraphics[width=8.5cm]{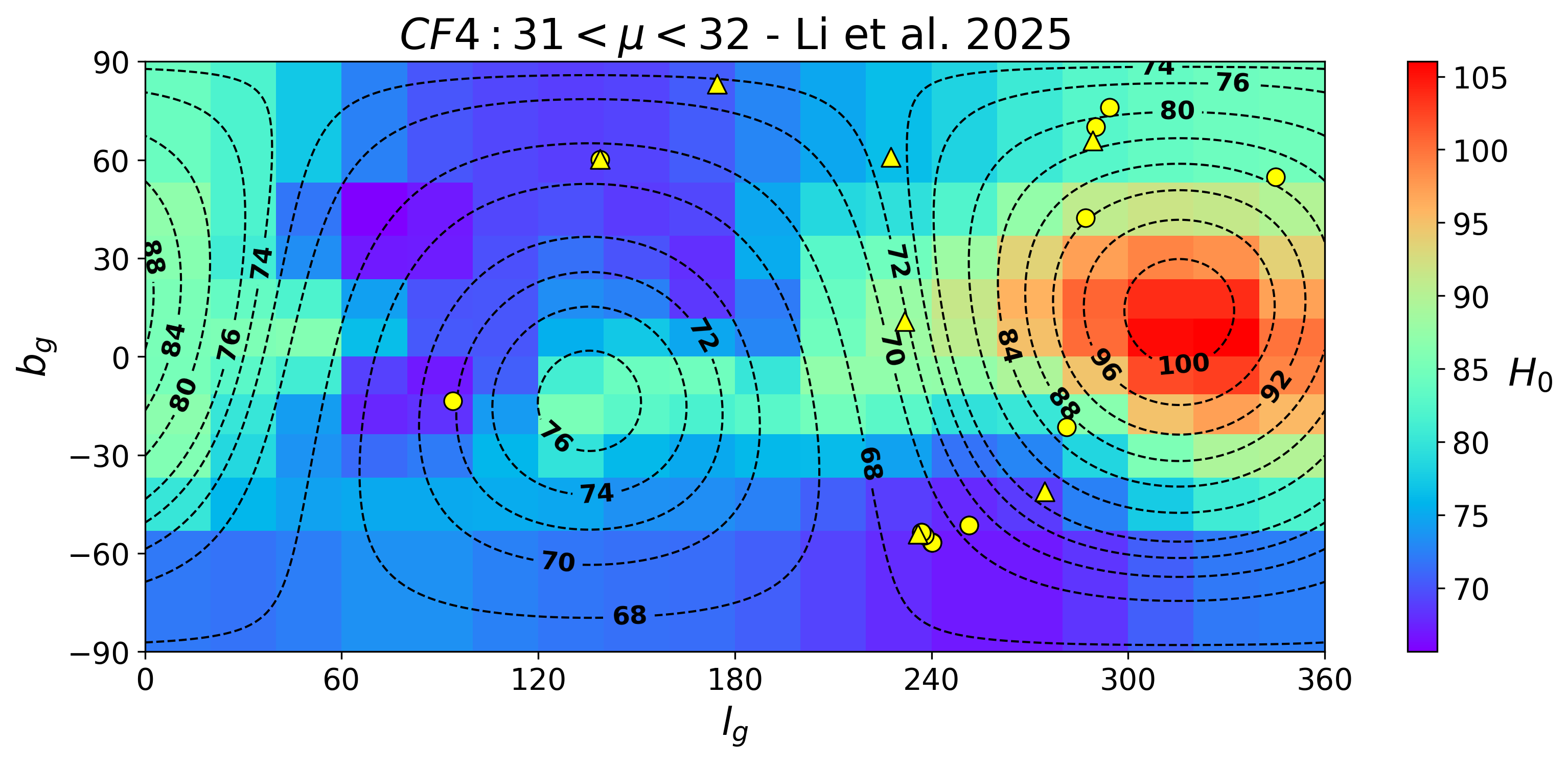}~~
\includegraphics[width=8.5cm]{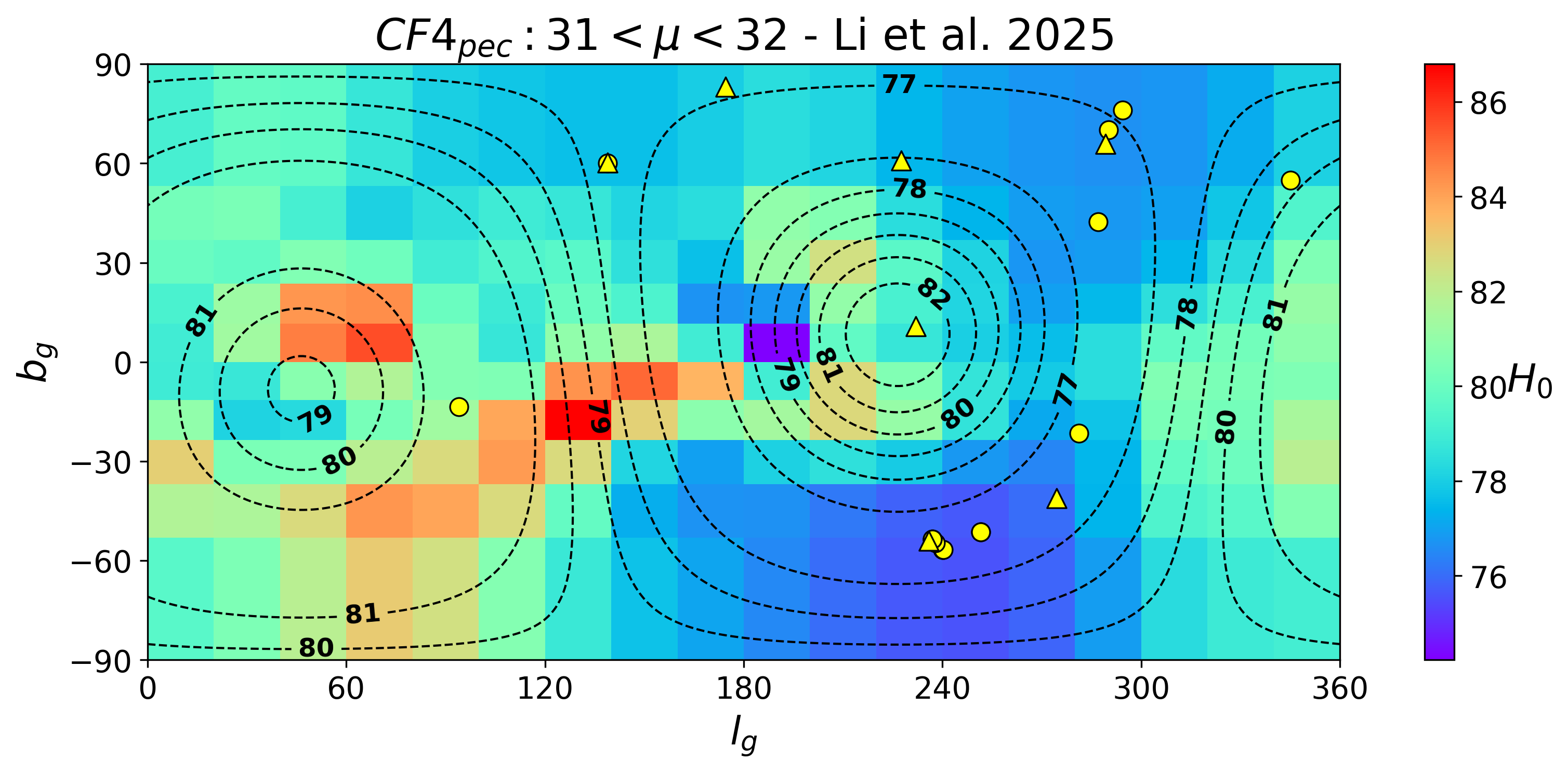}\\
\includegraphics[width=8.5cm]{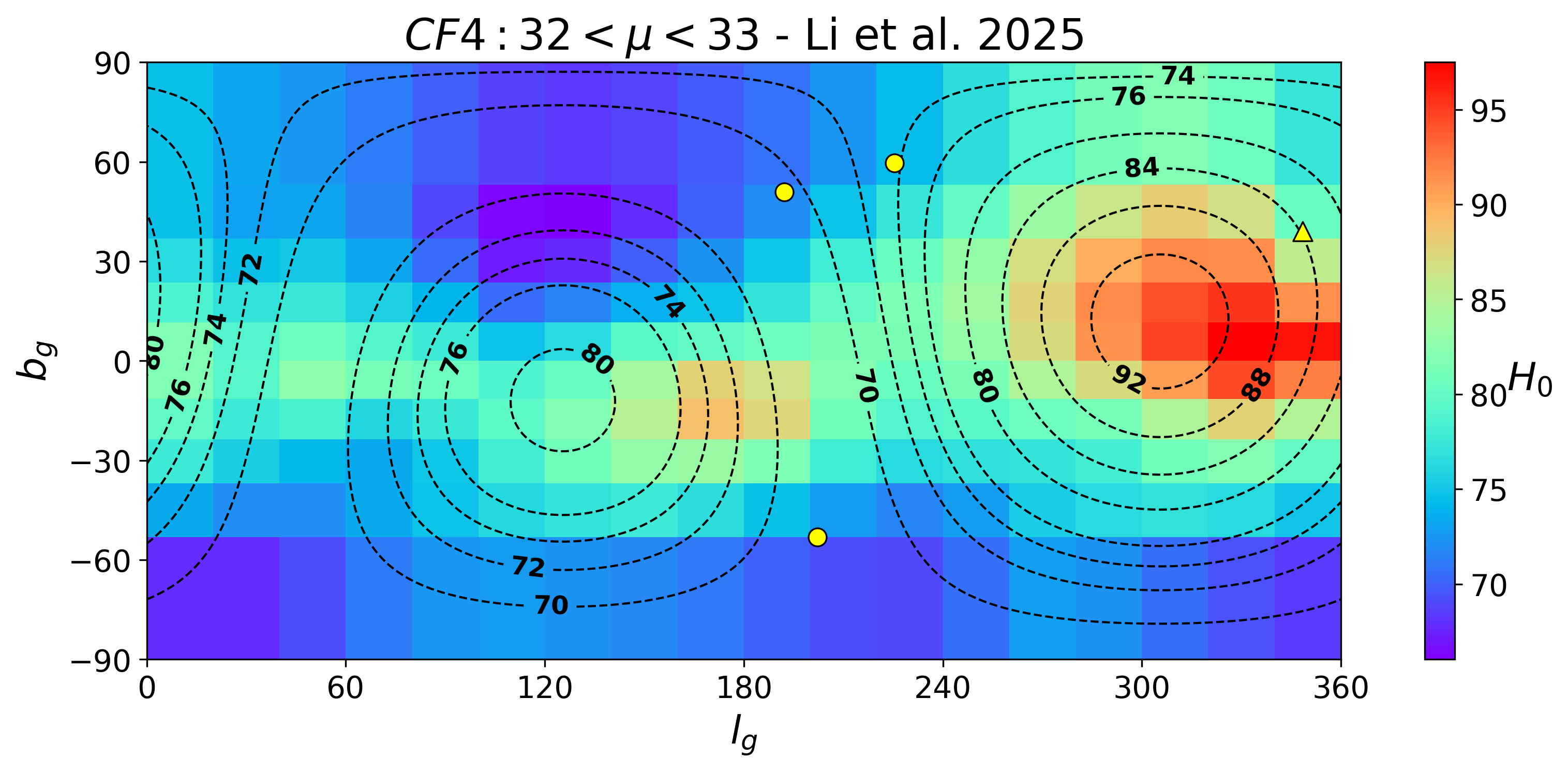}~~
\includegraphics[width=8.5cm]{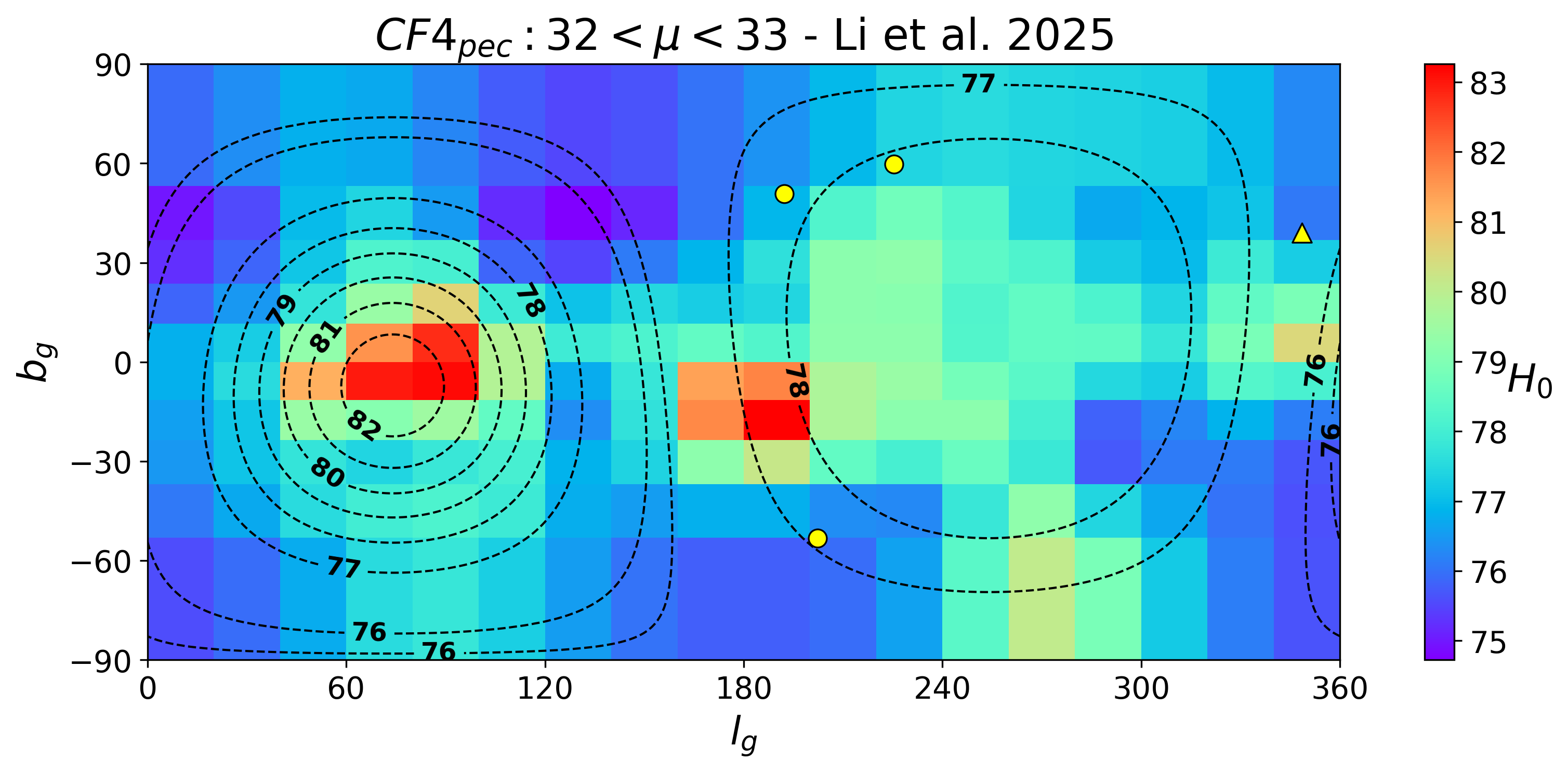}
\caption{Radial+Angular signal per radial shells in Galactic coordinate system and CMB frame for the CF4 sample (left) and the CF4$_{pec}$ (right) sample highlighting the positions of SNeIa from \textit{CCHP} (circles) and \textit{SH0ES} (triangles) as reported in \cite{Li:2025lfp}.}\label{fig:trends_Li_1}
\end{figure*}

Finally, we examined whether the detected anisotropy could have a significant impact on the Hubble tension. Figure~\ref{fig:trends_Li_1} shows the $35$ host galaxies from Table~$3$ of \cite{Li:2025lfp}, the largest current sample of galaxies with \textit{HST} and \textit{JWST} TRGB measurements. Circles mark the hosts used by the \textit{CCHP} team to estimate $H_0$, whose value differs from that of \textit{SH0ES} and is not in tension with \textit{Planck} estimation; triangles indicate hosts excluded from \textit{CCHP}. According to \textit{SH0ES} team \cite{Li:2025lfp}, these excluded hosts reconcile the two estimates, implying a selection bias in \textit{CCHP}. Interpreting the role of the detected anisotropy in this context is not straightforward. In the CF4 sample (left panels of Fig.~\ref{fig:trends_Li_1}), which reflects the imprint of peculiar velocities on $H_0$, the hosts discarded by \textit{CCHP} do not appear to be preferentially concentrated in a ``hot'' or ``cold'' region of the detected anisotropy signal. This suggests that including them would not significantly alter the estimation of $H_0$ from \textit{CCHP}. It should be noted that the CCHP team does not apply peculiar-velocity corrections, which are known to increase $H_0$ \cite{Li:2025lfp}. The situation does not seem to change in the CF4$_{pec}$ case, although we observe that most of the hosts are located near the ``cold'' regions.This implies that both the \textit{CCHP} and \textit{SH0ES} samples could actually be underestimating the cosmological value of $H_0$, thus exacerbating the Hubble tension.

\begin{figure*}
\centering
\includegraphics[width=8.5cm]{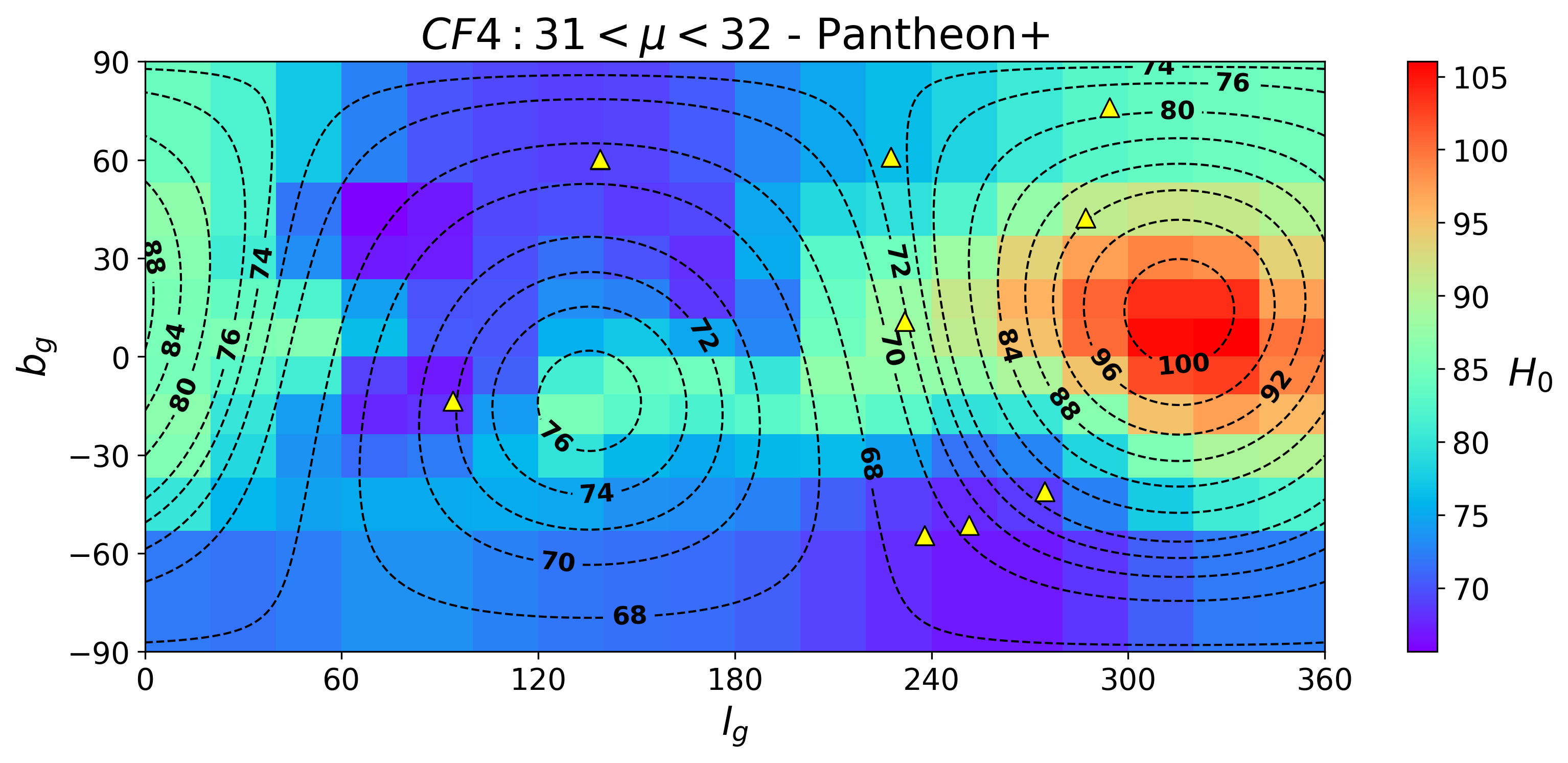}~~
\includegraphics[width=8.5cm]{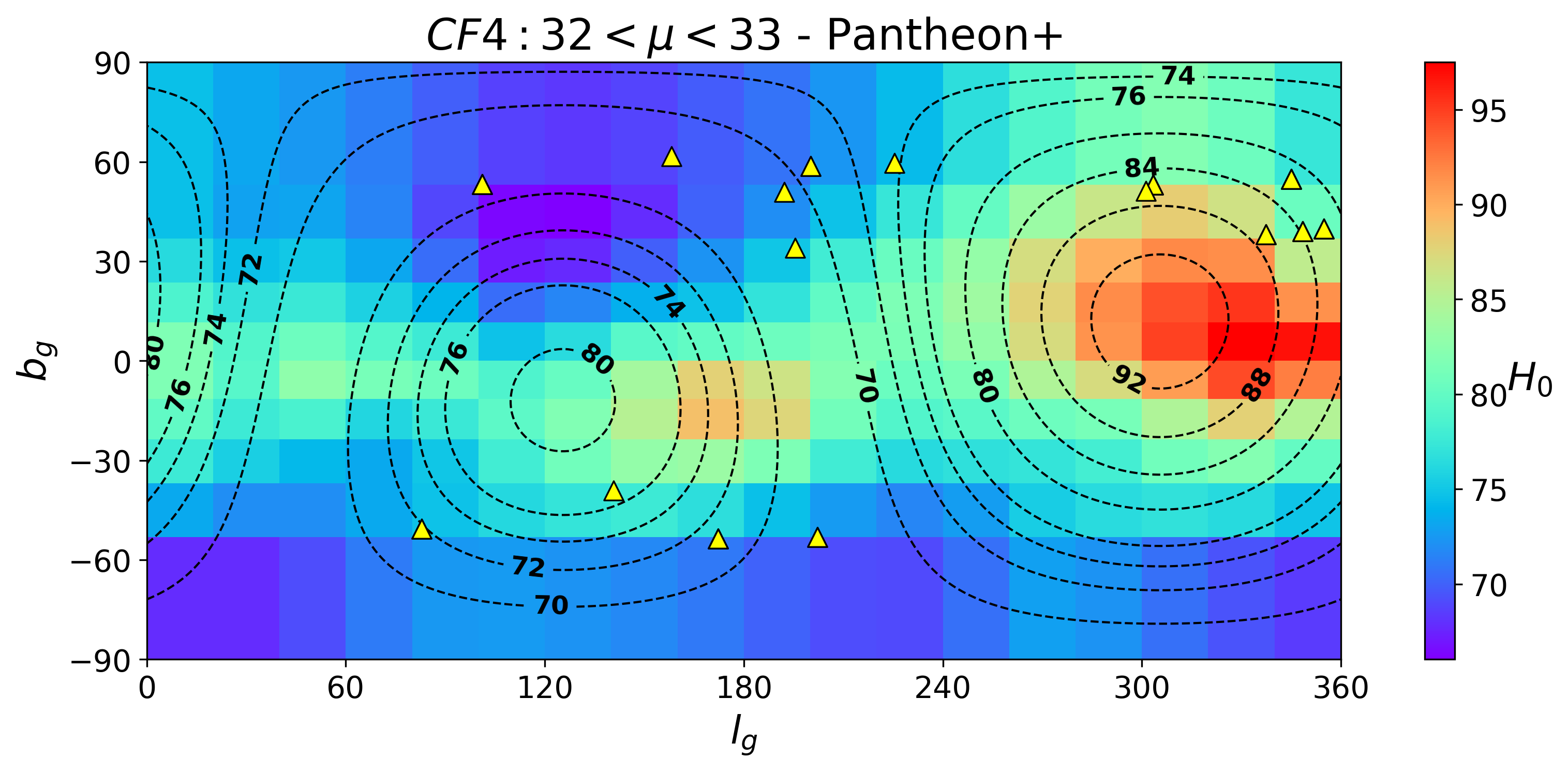}\\
\includegraphics[width=8.5cm]{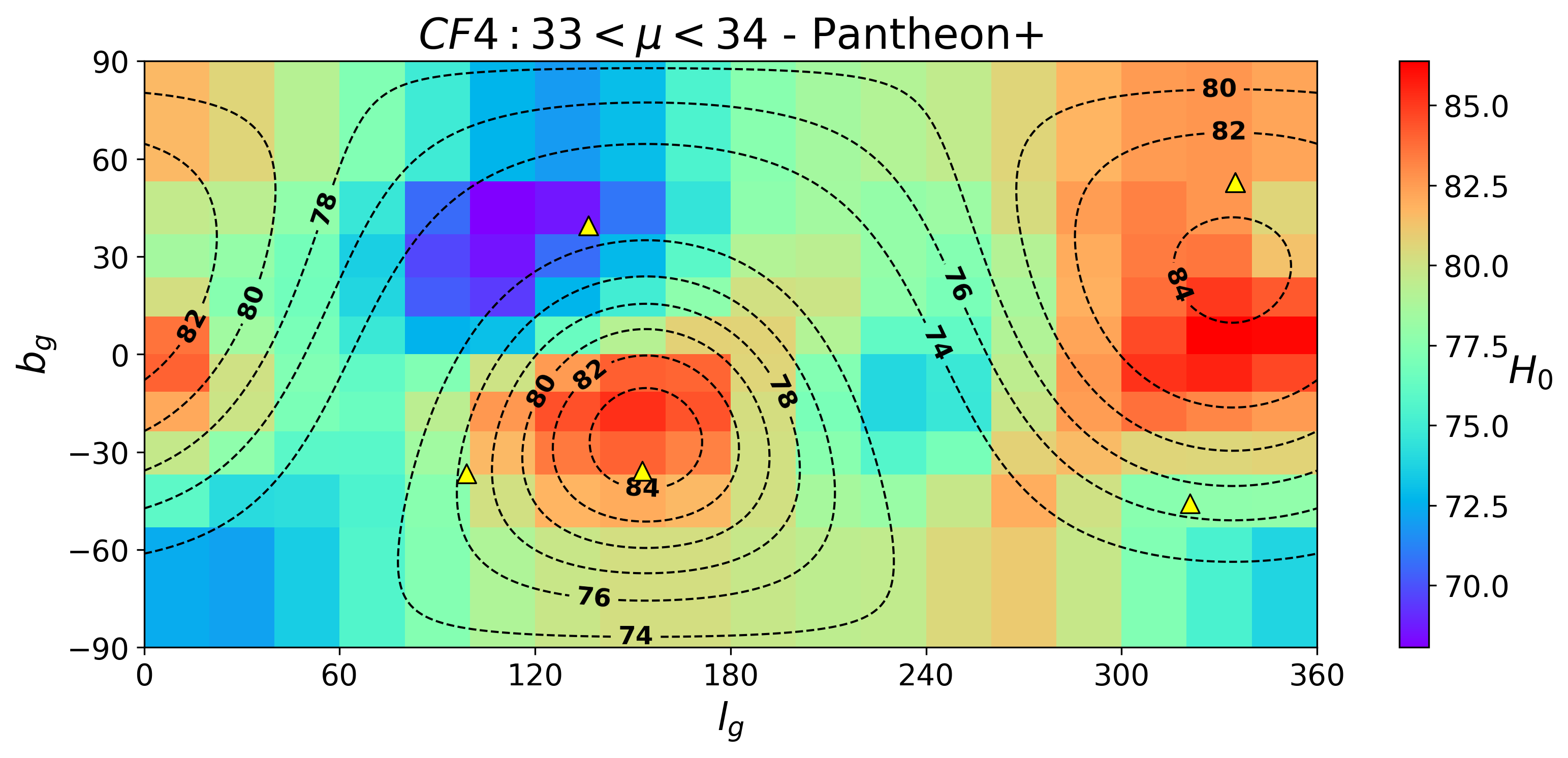}~~
\includegraphics[width=8.5cm]{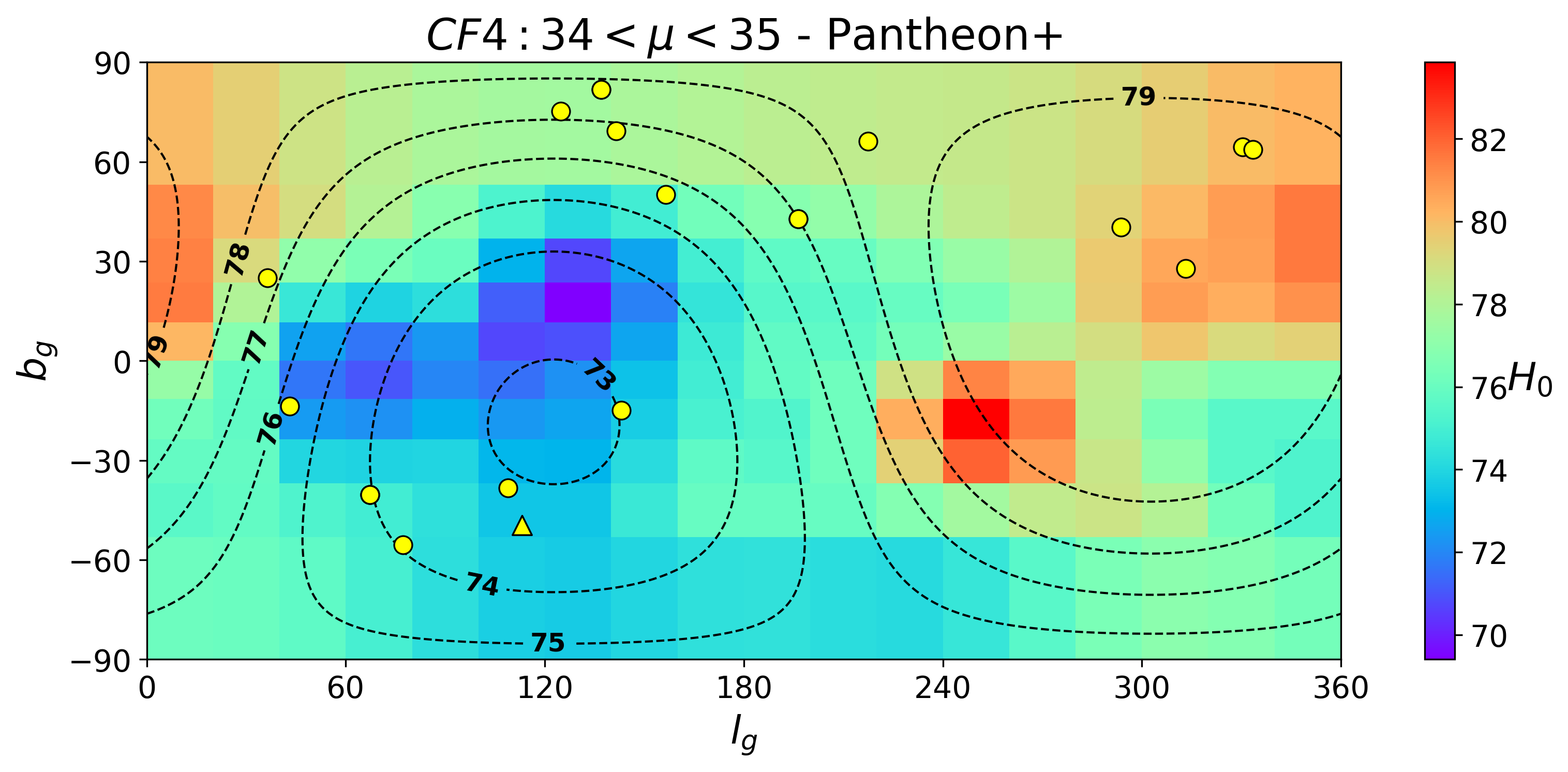}\\
\includegraphics[width=8.5cm]{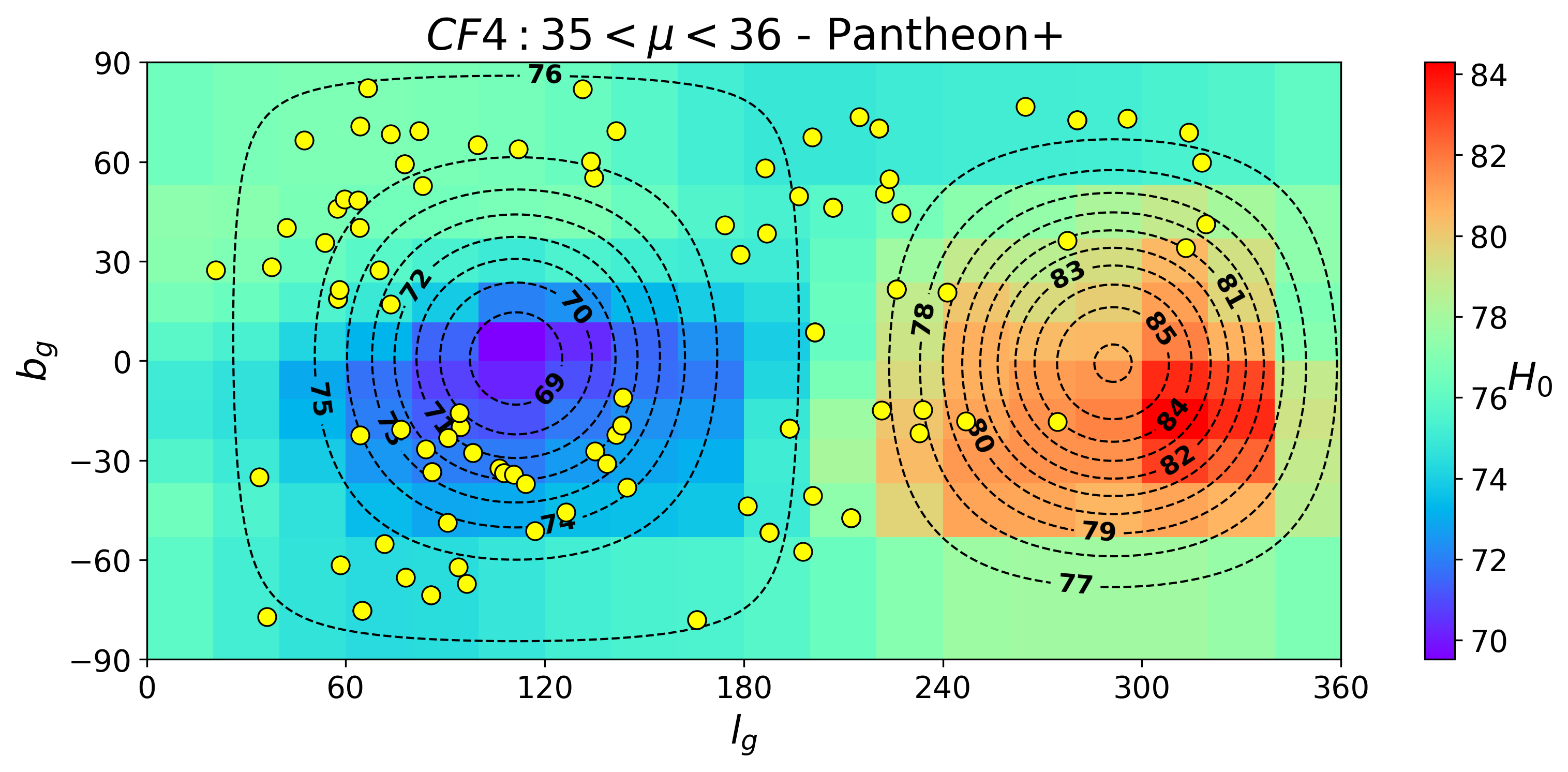}
\caption{Radial+Angular signal per radial shells in Galactic coordinate system and CMB frame for the CF4 sample highlighting the positions of SNeIa from calibrator hosts (triangles) and from the Hubble flow (circles) as reported in the Pantheon+ catalogue.}
\label{fig:trends_Pantheon_1}
\end{figure*}

\begin{figure*}
\centering
\includegraphics[width=8.5cm]{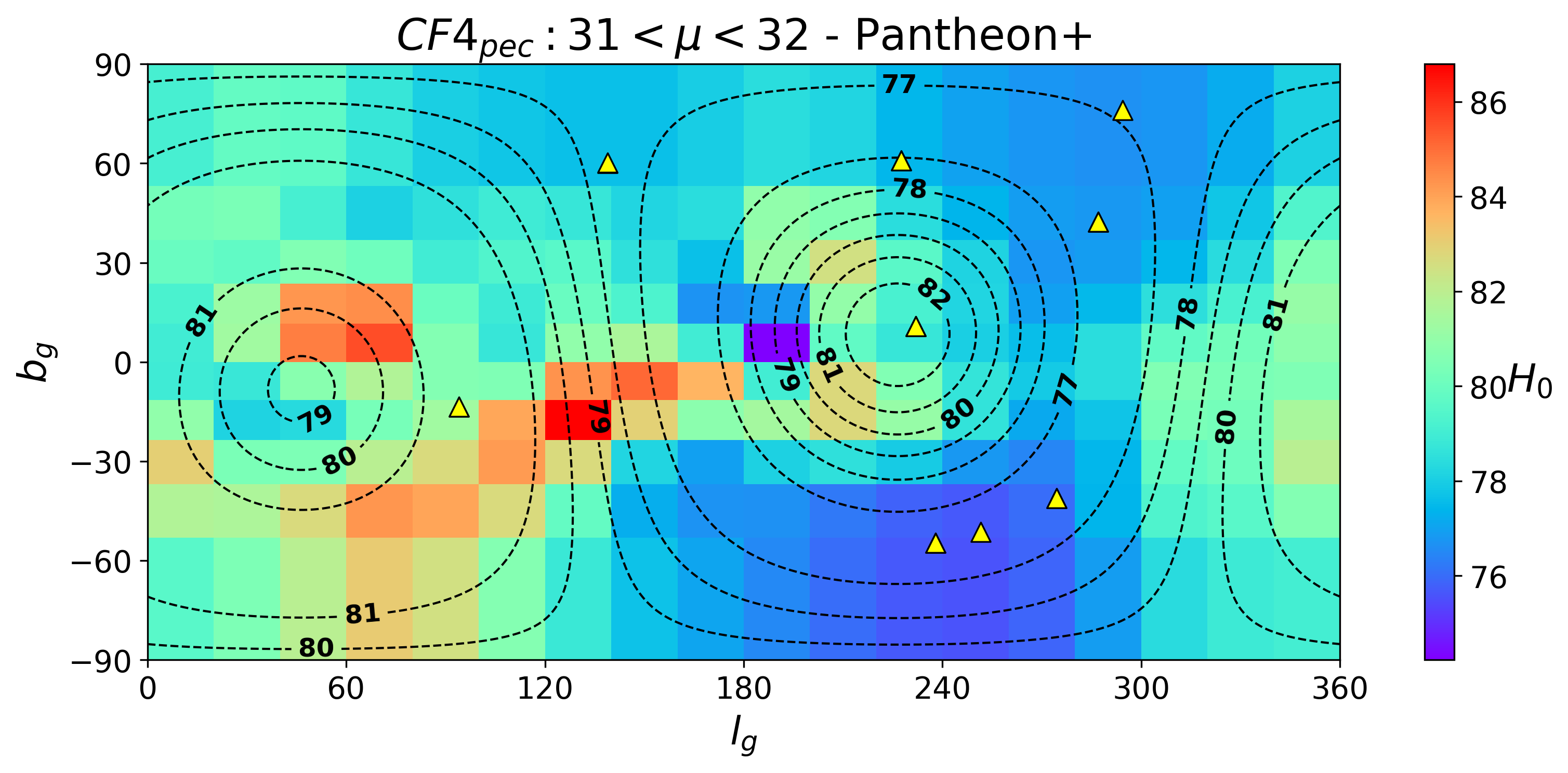}~~
\includegraphics[width=8.5cm]{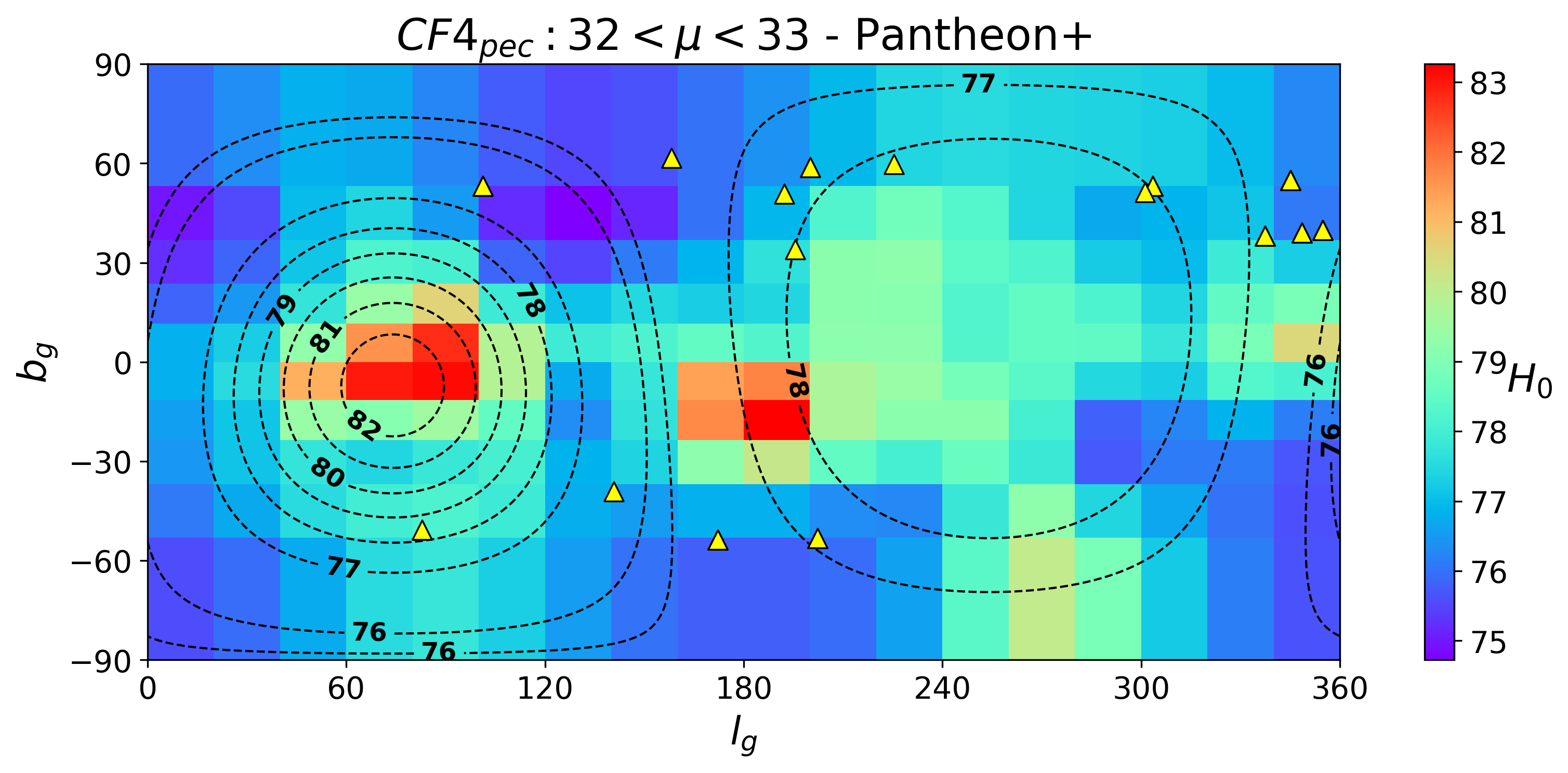}\\
\includegraphics[width=8.5cm]{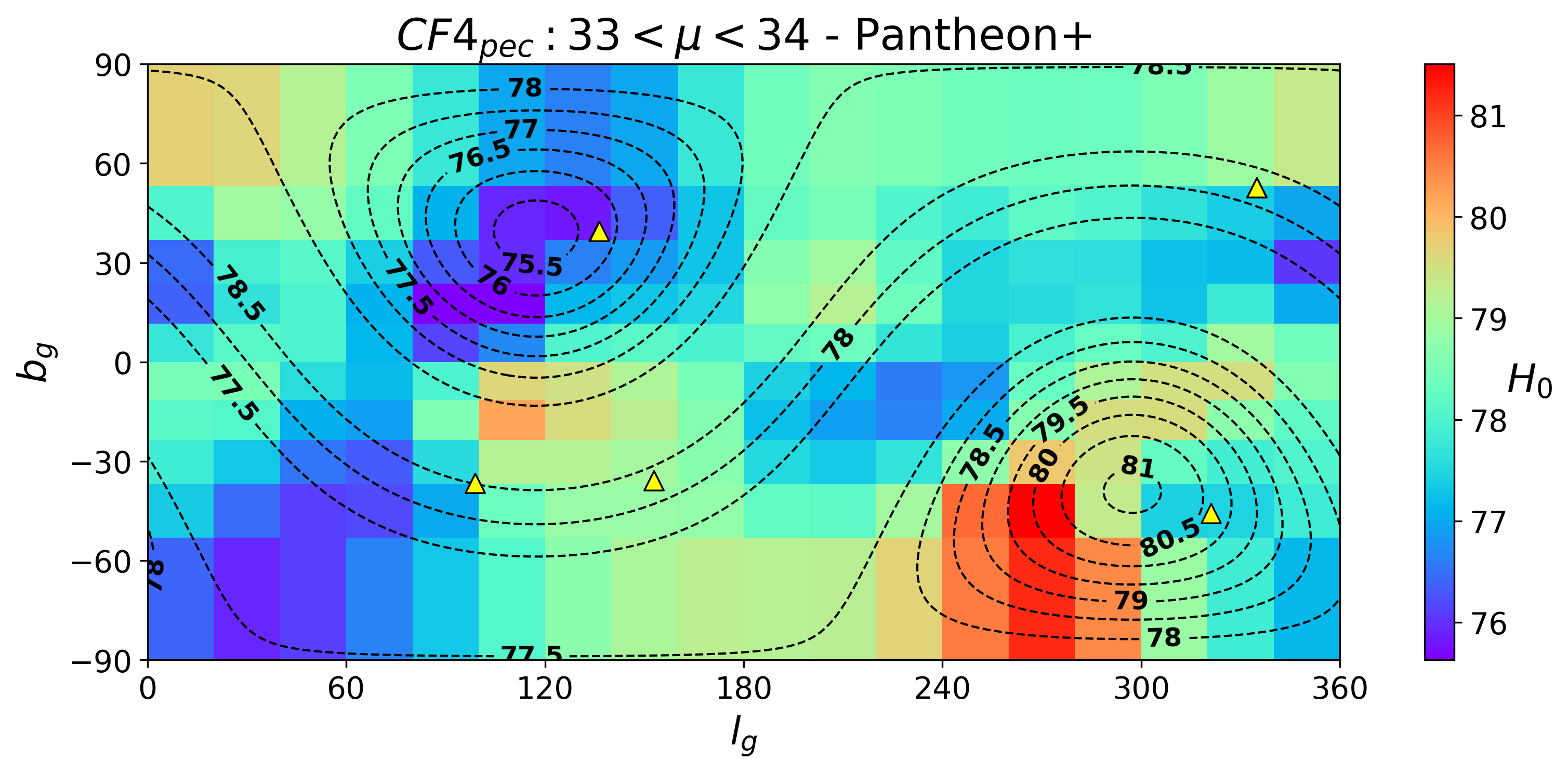}~~
\includegraphics[width=8.5cm]{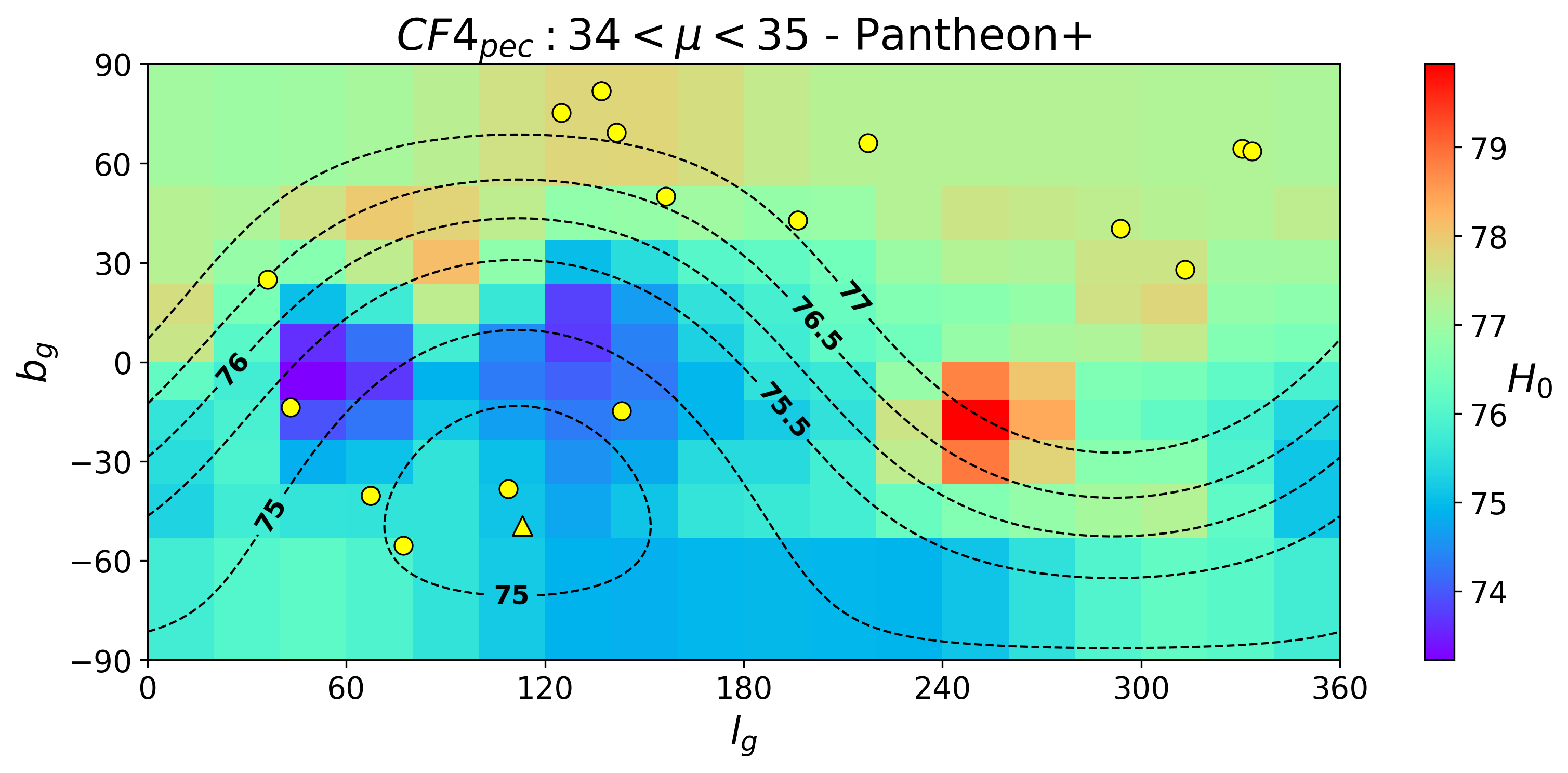}\\
\includegraphics[width=8.5cm]{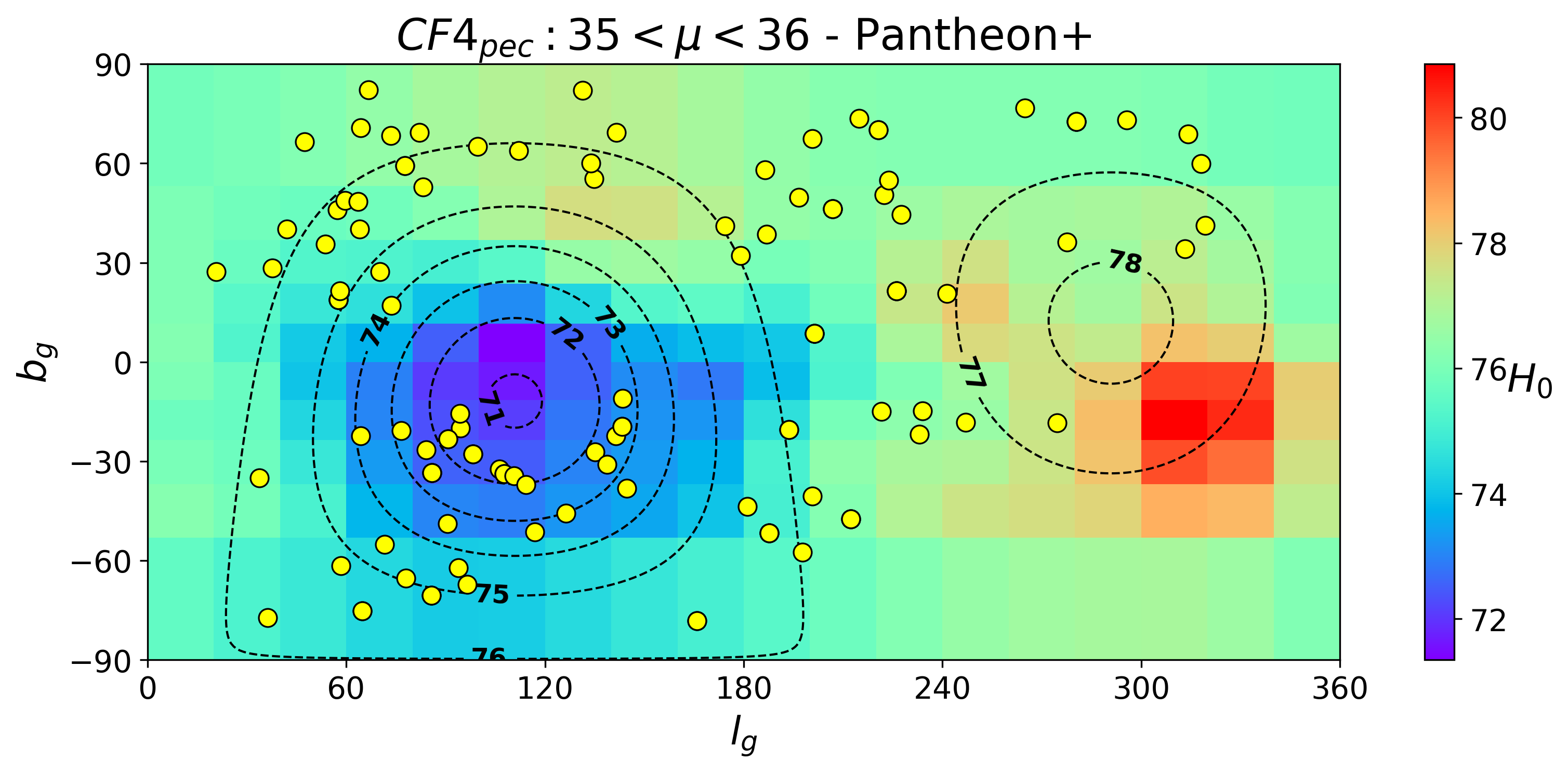}
\caption{Radial+Angular signal per radial shells in Galactic coordinate system and CMB frame for the CF4$_{pec}$ sample highlighting the positions of SNeIa from calibrator hosts (triangles) and from the Hubble flow (circles) as reported in the Pantheon+ catalogue.}
\label{fig:trends_Pantheon_2}
\end{figure*}

Taking a broader perspective, it is important to recall that these hosts are part of the sample which is used by \textit{SH0ES} to provide a powerful constraint on the magnitude zero-point of SNeIa. This calibration constrains $H_0$ when combined with Hubble-flow SNeIa, which are the only ones able to trace the cosmic expansion. Figures~\ref{fig:trends_Pantheon_1} and \ref{fig:trends_Pantheon_2} show the spatial distribution of Pantheon+ Hubble-flow SNe~Ia (circles) and calibration hosts (triangles). Once again, no clear and decisive asymmetry appears relative to the background anisotropy. However, we should point out that most of the Hubble-flow SNe Ia seem to populate the ``cold'' region of the anisotropy more densely, once again suggesting that the real value of $H_0$ may be underestimated, although this is unlikely to be the cause of the Hubble tension.

In summary, our results suggest that the anisotropy of the Hubble constant detected in the Cosmicflows-4 catalogue is real at the level of local flows, but is largely suppressed once peculiar velocities are accounted for, and does not exhibit the radial behaviour expected from large-scale anisotropic expansion. This highlights the importance of combining statistically consistent formulations, internally validated samples, and a careful distinction between observed and reconstructed quantities when interpreting anisotropy signals in cosmological data. Future surveys with improved sky coverage, homogeneous selection functions, and full covariance information will be essential to determine whether any residual anisotropy persists beyond the local kinematic regime and can be robustly linked to new cosmological physics.

\section*{Acknowledgements}

VS warmly thanks Yehuda Hoffmann for clarifications regarding the Cosmicflows-4 data, Christian Marinoni for important insights into anisotropic cosmography, and Geraint F. Lewis for discussion. This article is based upon work from the COST Action CosmoVerse CA21136, supported by COST (European Cooperation in Science and Technology). JBJ and DB acknowledge support from grants PID2021-122938NB-I00 and PID2024-158938NB-I00 funded by MICIU/AEI/10.13039/501100011033 and by “ERDF A way of making Europe”, the Project SA097P24 funded by Junta de Castilla y Le\'on and the research visit grant PRX23/00530.

\section*{Data Availability}

The CF4 data is publicly available at \url{https://edd.ifa.hawaii.edu/dfirst.php}. The peculiar velocity catalogue, CF4$_{pec}$, is available upon reasonable request sent to AV.

\bibliographystyle{apsrev4-2}
\bibliography{biblio.bib}

\end{document}